\documentclass[preprint,12pt]{elsarticle}
\journal{Journal of Molecular Liquids}

\usepackage{amsmath}
\usepackage{amssymb}
\usepackage{graphicx}
\usepackage{natbib}
\biboptions{numbers,sort&compress}
\usepackage{csquotes}
\usepackage{siunitx}
\usepackage{hyperref}
\usepackage{enumitem}
\usepackage{physics}
\usepackage{caption}
\usepackage{subcaption}
\usepackage{booktabs}
\usepackage{multirow}
\usepackage{cancel}
\usepackage{bm}
\usepackage[capitalise]{cleveref}
\usepackage{color}
\usepackage[toc,page]{appendix}
\usepackage{cleveref}

\DeclareSIUnit\Molar{\textsc{M}}
\DeclareSIUnit\ergs{ergs}
\DeclareSIUnit\esu{esu}
\DeclareSIUnit\statV{stat\textsc{V}}
\DeclareSIUnit\cm{\centi\metre}

\begin{document}
	\begin{frontmatter}
	
	
	\title{Contact theorems for electrolyte-filled hollow charged nanoparticles: Non-linear osmotic pressure in confined electrolytes}
	
	\author{M. Lozada-Cassou\corref{cor1}}
	\ead{marcelolcmx@ier.unam.mx}
	\author{S. A. Rivera-Cerecero}
	
	\address{Instituto de Energías Renovables, Universidad Nacional Autónoma de México, Priv. Xochicalco S/N, Temixco, Morelos 62580, México.}
	\cortext[cor1]{Corresponding author}
	
	\begin{abstract}
		\small
		Analytical expressions for the osmotic pressure of electrolytes confined within electrolyte-filled, hollow, charged nanoparticles are studied using contact theorems for cavity nanoshells immersed in a low-concentration electrolyte. Three shell geometries are considered: planar, cylindrical, and spherical. The nanoparticles are modeled as charged cavities of internal radius \( R \) and wall thickness \( d \), and are assumed to be at infinite dilution in the surrounding bulk electrolyte. Numerical calculations of the osmotic pressure are presented as a function of several model parameters. For cylindrical and spherical shells, the osmotic pressure exhibits absolute maxima as a function of shell size. This behavior arises from the competition between the violation of the local electroneutrality condition (VLEC) within the shell cavity and the nonlinear profile of the effective electric field. In contrast, the osmotic pressure of the planar (slit-shell) geometry is a monotonic, nonlinear, decreasing function of cavity width. The results are analyzed in terms of the steric and electrostatic (Maxwell stress tensor) contributions appearing in the corresponding contact theorems. Because the analysis is restricted to low surface charge densities and low electrolyte concentrations, the electric double layers (EDLs) inside and outside the shells are obtained from analytical solutions of the linearized Poisson--Boltzmann equation. We report two novel phenomena: \textit{confinement charge reversal} (CCR) and \textit{confinement overcharging} (CO). These arise naturally from the topology of the system. By construction, the confined and bulk electrolytes are maintained at the same chemical potential. These findings have implications for the design of synthetic nanocapsules and ion-selective membranes.
	\end{abstract}
\end{frontmatter}

\section{Introduction}\label{introduction}

Inorganic~\cite{Nie-inorganic-vesicles-2012} and organic~\cite{Biophysics-book-1983,Avinoam-membranes-1994,Evans-Wennerstrom-1999} vesicles are of considerable relevance in science~\cite{Bailey-Langmuir-1990,Hille-Ion-channels-book-2001,Tuszynski-Introduction-Molecular-Biophysics-book-2003,Bohinc_2008} and technology~\cite{bezadhi-encapsulation-2016,Bohinc-2018,Doan-Nguyen-nanoparticles-2023}, as in the oil and chemical industry~\cite{Contreras-oil-industry-vesicles-2019}, biology~\cite{Coffey-Biology-2023}, pharmaceutical manufacturing~\cite{Almoshari-Osmotic-Pressure-pharmaceuticals-2022}, gene therapy~\cite{Gelbart-Phys-Today-2000}, medicine~\cite{Alberts-Mol-Cell-Bio-2002,Extracellular-vesicles-collection-2023}, bulk nanocapacitors~\cite{jia-RSC-adv-2023} and energy storage~\cite{Samyabrata-overcharging-2024}. In addition, polymerization of self-assembled particles, as vesicles, has been applied to specific industrial applications, reproducing some functions of biological membranes and the development of new materials~\cite{Daoud-soft-matter-book-1999}.

One of the properties that is most relevant in macroions separation~\cite{Phys-Chem-macromolecules-Tanford-1961} and the design and appropriate function of vesicles or biological cells is their osmotic pressure~\cite{Esteki-living-cell-2021,Wennerstrom_cells_2022}, $\pi$, defined as the difference between the inside and outside pressures in the vesicles. For example, in some animal cells this pressure should be small for them to survive. However, in bacteria and plants the osmotic pressure need not to be small since their walls can support pressures in the range of 1 to 10 atm ($\approx1.013x10^6 N/m^2$)~\cite{Tuszynski-Introduction-Molecular-Biophysics-book-2003}. In vesicle aqueous solutions, the stability or fusion of vesicles depends, in addition of steric and electrostatic forces, on undulation and peristaltic forces between their membrane walls. These forces, when vesicles become within 1 nm of each other, may make them to fuse. Increasing the osmotic pressure might prevent their fusion~\cite{Israelachvili-book}. While the materials of the fluids and walls in these nanoparticle systems vary widely, and their structures are topologically diverse—referred to in the literature as core-shells, nanopores, vesicles, or cavity shells—they all share a common characteristic: they are fluid-filled, hollow, charged shells. Thus, hereinafter let us refer to these kind of hollow nanoparticles simply as shells, cavity shells or cavity nanoparticles. Additionally, while some of the results presented in this article are relatively general concerning the topology and materials of nanoparticles, we focus specifically on electrolyte-filled, hollow, charged nanoparticles with planar, cylindrical, and spherical geometries.

The osmotic pressure in shells immersed in an electrolyte solution depends on their size, the charge, $\sigma_{\scriptscriptstyle_{0}}$, wall thickness, $d$, and on the electrolyte's bulk concentration, $\rho_{\scriptscriptstyle_{0}}$, and valence, among other parameters. The pressure on a single wall can be calculated through contact theorems. These theorems, derived from general liquid theory arguments, have been applied  to planar~\cite{Henderson-Contact-JCP-1978,Jimenez_2004_Nov,Holovko-2005}, cylindrical and spherical electrodes~\cite{Bari-Contact-Mol-Phys-2015,Holovko-Contact-2023}, as well as for planar~\cite{Lozada_1984,Lozada_1990-I}, cylindrical and spherical cavity shells~\cite{Yu_1997,Aguilar_2007}. The cavity shells' contact conditions as well as other properties, such as their capacitance, reduce to those for nano-electrodes by making the internal radius equal to zero~\cite{Adrian-JML-2023}. In particular, the osmotic pressure as a function of the inner radius of shells immersed in a restricted primitive model (RPM) electrolyte has been calculated, and an interesting charge separation was reported, resembling a Van der Waals liquid-vapor phase transition equation~\cite{Yu_1997,Aguilar_2007}. In the RPM electrolyte, the ions are considered as hard spheres of diameter $a$, and the solvent is modeled only through its dielectric constant, $\varepsilon$. In those studies, the walls of the shells were assumed to have the same dielectric constant as that of the electrolyte, to avoid image forces. The above mentioned contact theorems, can also be obtained from a simple force balance between steric and electric forces on the cavity shells' walls~\cite{Yu_1997,Aguilar_2007}. In this paper, we outline these derivations for prompt reference and congruity of our results discussion.

To calculate the osmotic pressure of shells through contact theorems, the electrical double layer (EDL) formed inside and outside of the shells is needed, i.e., the electrolyte contact concentration next to the shell's walls, along with its associated electrical field profile. Under the action of an external electric field, a bulk electrolyte becomes an inhomogeneous fluid with an uneven distribution of counter-ions and co-ions next to the electrode. This is the EDL~\cite{Hansen-book-2013}, which by itself is relevant in the study of colloids~\cite{Verwey_TheoryStabilityLyophobicColloids_1948,Israelachvili-book}, biological systems~\cite{Evans-Wennerstrom-1999}, biomaterials~\cite{Bohinc_2008,Bohinc-2018}, medicine~\cite{Coffey-Biology-2023}, and oil industry~\cite{HUANG-oil-1996,Oil-Recovery-book-2019}, among others.

The EDL has been studied in the past through point-ion models with the Gouy-Chapman equation~\cite{Gouy_1910,Chapman_1913}, which is the analytical solution of the Poisson-Boltzmann (PB) equation for a planar electrode~\cite{Verwey_TheoryStabilityLyophobicColloids_1948,Hiemenz-book-1977}. The PB equation is a non-linear, inhomogeneous, second order differential equation, and its integral equation version has also been numerically solved for planar~\cite{Lozada_1982}, cylindrical~\cite{Gonzalez_1985}, and spherical~\cite{Gonzalez_1989} electrodes geometries. For the RPM, the EDL has also been studied with several other statistical mechanics equations, such as  integral equations~\cite{,Carnie_1981,Henderson_1992_FIF, Lozada_1992_FIF,Attard_1996,Croxton_1981,Henderson_1982,Bratko_1982,Vlachy-Donnan-1992}, density functional~\cite{Patra_1994,Gillespie_2005,Goel_2008,Hartel_2017,Patra_2020}, and modified Poisson-Boltzmann (MPB)~\cite{Outhwaite_1986,Bhuiyan_1993_CMT,Bhuiyan_1994}, as well as computer simulations~\cite{Bret_1984,Bratko-Vlachy-osmotic-1985,Card_1970,Torrie_1980,Degreve_1993,Goel_2008,Boda-confinement-2024}. Theoretical studies for electrolytes confined in shells have been published in the past~\cite{Lozada_1984,Kjellander-Marcelja-Chem-Phys-Letts-1986,Lozada-Cassou-Yamada-1988,Kjellander-two-plates-1988,Kjellander-Akesson-Jonsson-Marcelja-1992,Lozada_1990-I,Lozada_1990-II,Vlachy1989,Yeomans1993,Yu_1997,Vlachy2001,Grosse-2002,Henderson2005,Aguilar_2007,Peng2009,Ala-Nisila_2011,Henderson2012,Pizio2012,Spada_two_plates_2018,biagooi-Nature2020,Keshavarzi_2022,Feng-nanopores-topology-2023}. The role of confinement and dielectric constant on ion adsorption in planar, cylindrical, and spherical shells was recently investigated using Monte Carlo simulations~\cite{Boda-confinement-2024}. In that study, it was explicitly noted that ionic size effects become negligible at low electrolyte concentrations.

In references~\cite{Yu_1997,Aguilar_2007}, the osmotic pressure in shells was studied through the RPM, where the fluid was taken to be a $2:2$, $\rho_{\scriptscriptstyle_{0}}=0.971M$ electrolyte. For biological and many other systems this salt concentration is very high.  Therefore, in this paper we present results for the osmotic pressure of planar, cylindrical and spherical shells immersed in a low to very low electrolyte concentration, and a low to very low charge on the shells' walls. For these low concentrations and charges we will use the analytical expressions for the concentration and electric field profiles obtained from the linearized Poisson-Boltzmann (LPB) equation, recently reported for nanocapacitors of planar, cylindrical and spherical geometry~\cite{Adrian-JML-2023}.  The cavity nanoparticles are taken to be at infinite dilution. Hence, here we will not address to systems of self-assembled or designed arrays of cavity nanoparticles, which can have relevant application in medicine~\cite{jia-RSC-adv-2023} and supercapacitors~\cite{Supercapacitors-Book-2013}, or consider charge regulation effects~\cite{ninham-JTB-1971,Bohinc-2018}.

The Poisson--Boltzmann (PB) theory has long served as a cornerstone in the theoretical description of electrostatic interactions in both homogeneous and inhomogeneous systems, including simple fluids (bulk and confined electrolytes) and complex fluids such as colloidal dispersions. Its origins trace back to the early theoretical developments of Debye and H\"uckel (yielding the linearized PB, or LPB equation), and to the Verwey--Overbeek theory, which extended PB to interacting planar or spherical colloidal particles~\cite{Verwey_TheoryStabilityLyophobicColloids_1948}.

In the context of bulk electrolytes, classical statistical mechanics texts such as McQuarrie’s \textit{Statistical Mechanics}~\cite{McQuarrie_StatMech} emphasize that, at sufficiently low concentrations:
\begin{quote}
	``\dots the short-range interaction between ions does not play an important role, and its precise nature is unimportant,''\\
	and\\
	``\dots the linearization of the Poisson--Boltzmann equation is valid as well.''
\end{quote}

Accordingly, the LPB equation has been shown to yield both \textit{qualitative} and \textit{quantitative agreement} with predictions from \textit{integral equation theories}, \textit{density functional approaches}, \textit{modified Poisson--Boltzmann formulations}, and \textit{Monte Carlo simulations} of inhomogeneous fluids at low electrolyte concentrations~\cite{Lozada_1982,Degreve_1993,Degreve_1995,Gonzalez_2018}. Recent Monte Carlo studies on ion adsorption in shells of planar, cylindrical, and spherical geometries further confirm that ionic size effects become negligible at low concentrations, supporting the applicability of the point-ion model employed in our analysis~\cite{Boda-confinement-2024}.

Moreover, both the PB and LPB equations continue to serve as \textit{benchmark models} for validating the development of more advanced theoretical and computational methods for charged systems. Their analytical tractability and physical transparency make them invaluable tools for understanding the fundamental mechanisms that govern electrostatic interactions in soft matter and electrochemical systems.

This paper is organized as follows.  
In \cref{Theory}, we outline the derivation of the contact theorems for planar, cylindrical, and spherical shells, based on the balance between steric and electrostatic forces acting on the cavity walls. We also derive the corresponding electroneutrality conditions for these nanoparticle geometries and summarize the analytical solutions for the electric double layers (EDLs), obtained from the linearized Poisson--Boltzmann (LPB) equation.  
In \cref{Res_Disc}, we present our results for the osmotic pressure in these cavities as a function of several key model parameters. We also report two novel phenomena: \textit{confinement charge reversal} (CCR) and \textit{confinement overcharging} (CO), which are analyzed in terms of their underlying steric and electrostatic contributions.  
Finally, in \cref{Conclusions}, we summarize the main findings of the study and outline potential directions for future work.

\section{Theory}\label{Theory}

Due to the equivalence between particle and field representations, it has long been established that theories of bulk fluids can be extended to inhomogeneous fluids by treating one of the species—typically at infinite dilution—as a large particle that acts as an external field source. Based on this general framework, various theoretical approaches have been developed for the Poisson–Boltzmann equation~\cite{Verwey_TheoryStabilityLyophobicColloids_1948}, charging parameter expansions~\cite{Kirkwood-Poirier-1954,Stillinger-Kirkwood-1960}, and the Hypernetted-Chain (HNC) integral equation formalism for planar~\cite{Lozada_1981}, cylindrical~\cite{Lozada_1983,Gonzalez_1985}, and spherical~\cite{Gonzalez_1989,Degreve_1993} electrodes.

Following the same line of reasoning, both Poisson–Boltzmann and integral equations have been formulated for charged slit-shells immersed in electrolyte solutions. Using the Born–Green–Yvon (BGY) hierarchy, an exact expression for the mean force between the slit-shell walls was derived in~\cite{Lozada_1984}. This result constitutes a contact theorem for the slit-shell geometry. To compute this force, the electric double layer (EDL) profiles inside and outside the slit are required.

Using this general approach, the Three-Point-Extension Hypernetted-Chain Mean Spherical Approximation (TPE-HNC/MSA) was applied to obtain the EDL structure near the slit-shell walls. These profiles were numerically calculated for both constant surface potential~\cite{Lozada_1990-I,Lozada_1990-II,Olivares_1995} and constant surface charge density conditions~\cite{Lozada-Cassou-PRL1996,Lozada-Cassou-1996}. When substituted into the contact theorem, the resulting osmotic pressure showed both quantitatively accurate and qualitatively consistent agreement with Monte Carlo simulations~\cite{Valleau-Ivkov-Torrie-two-plates-1991,Lozada1996}. This net pressure is interpreted as the osmotic pressure of the slit-shell.

For cylindrical and spherical shells, due to their radial symmetry, the osmotic pressure can be evaluated using only the HNC/MSA-calculated EDL concentration profiles. In previous studies on these geometries, the salt concentration was assumed to be high, and the shell walls highly charged~\cite{Lozada1996,Lozada-Cassou-1996,Yu_1997,Aguilar_2007}. In contrast, in the present study we focus on the **low-concentration, low-charge** regime.

The contact theorems for the three shell geometries with inner radius \( R \) reduce to those for their respective solid electrodes in the limit \( R \to 0 \). Similar contact theorems, based on general statistical mechanical arguments, have also been derived for planar~\cite{Henderson-Contact-JCP-1978,Henderson-contact-Theorem-1979,McQuarrie-contact-theorem-1980,Holovko-2005}, cylindrical, and spherical electrodes~\cite{Holovko-Contact-2023}.

The pressure exerted on a charged wall by an adjacent electrolyte arises from the balance between steric and electrostatic (Maxwell stress tensor) components of the mean force. For cavity shells, the **net pressure**—defined as the difference between the fluid pressure on the inner and outer walls—corresponds to the **osmotic pressure**. Because the fluid on both sides of the shell wall is at the same chemical potential~\cite{Lozada-Cassou-PRL1996}, the EDL structures inside and outside the shell are inherently correlated.

Contact theorems for planar, cylindrical, and spherical shells have been previously derived~\cite{Lozada_1984,Yu_1997,Aguilar_2007}. In the present work, we summarize these derivations using a simplified force balance framework. While such balances could be extended to arbitrary shell shapes and general molecular fluids, we restrict ourselves to a **restricted primitive model (RPM)** electrolyte in planar, cylindrical, and spherical geometries. However, numerical results are presented for a **point-ion model** with a **Stern correction**, where ionic size is only considered at contact with the shell walls~\cite{Verwey_TheoryStabilityLyophobicColloids_1948}.

This force-balance approach has also been applied to **Donnan equilibrium** systems involving semi-permeable membranes immersed in primitive model (PM) macroion solutions~\cite{Vlachy-Donnan-1992,Jimenez_2004_Nov}. In contrast to the RPM, the PM allows for **asymmetric ionic sizes**, enabling more detailed modeling of realistic systems.

\subsection{Contact Theorems}\label{contact theorems}

In this section, we summarize the derivation of contact theorems for planar, cylindrical, and spherical shells, based on a force balance between steric and electrostatic contributions acting on the shell walls~\cite{Yu_1997,Aguilar_2007}. While we consider shells immersed in a restricted primitive model (RPM) electrolyte, numerical results are presented for a point-ion model with a Stern correction~\cite{Verwey_TheoryStabilityLyophobicColloids_1948}.

\begin{figure}[htbp]
	\centering
	\begin{subfigure}{.495\textwidth}
		\centering
		\includegraphics[angle=-90,width=.95\linewidth]{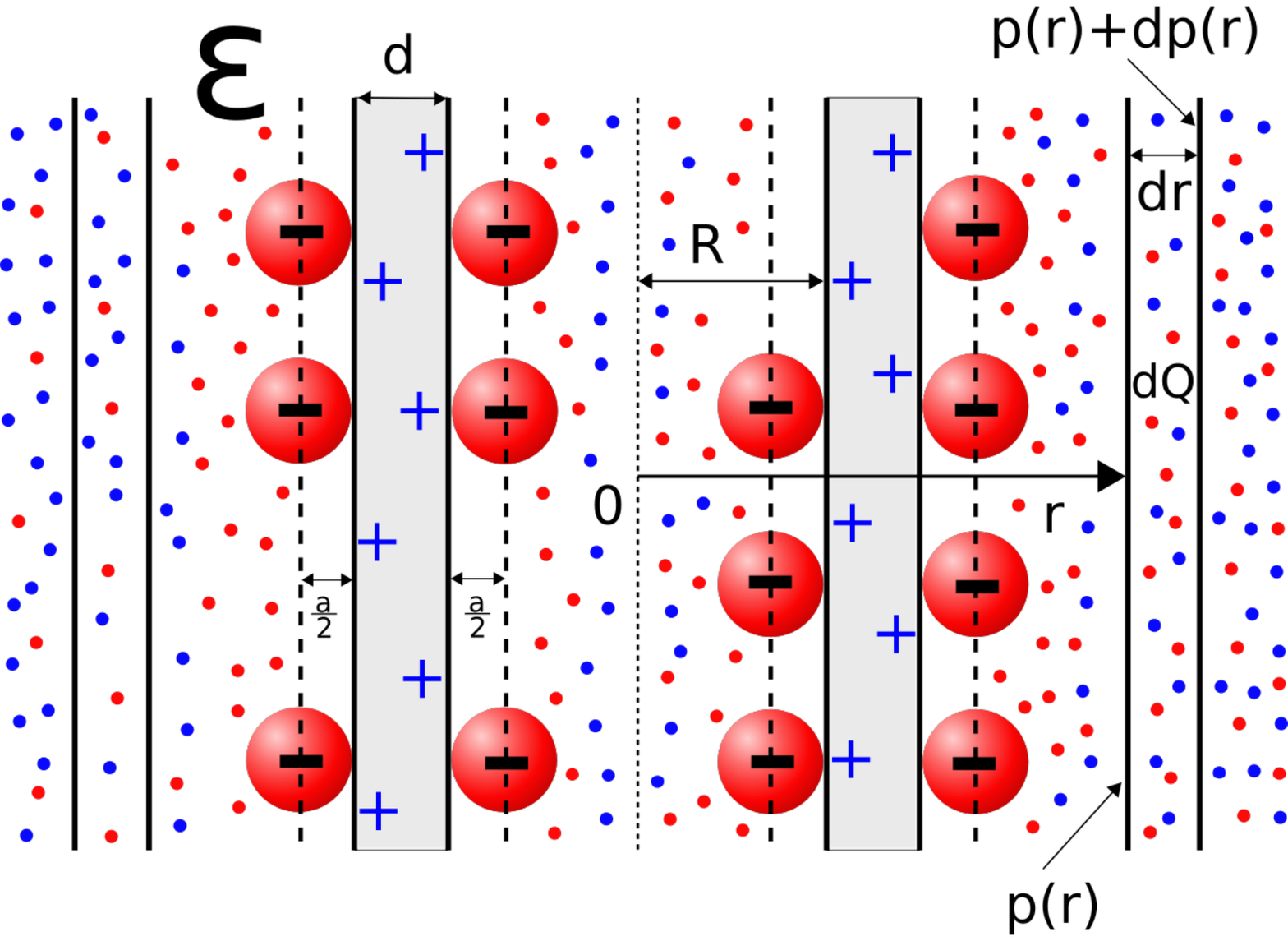}
		\caption{Slit-shell (planar geometry, 2D).}
		\label{Geometry_twoplates}
	\end{subfigure}
	\begin{subfigure}{.425\textwidth}
		\centering
		\includegraphics[angle=-90,width=\linewidth]{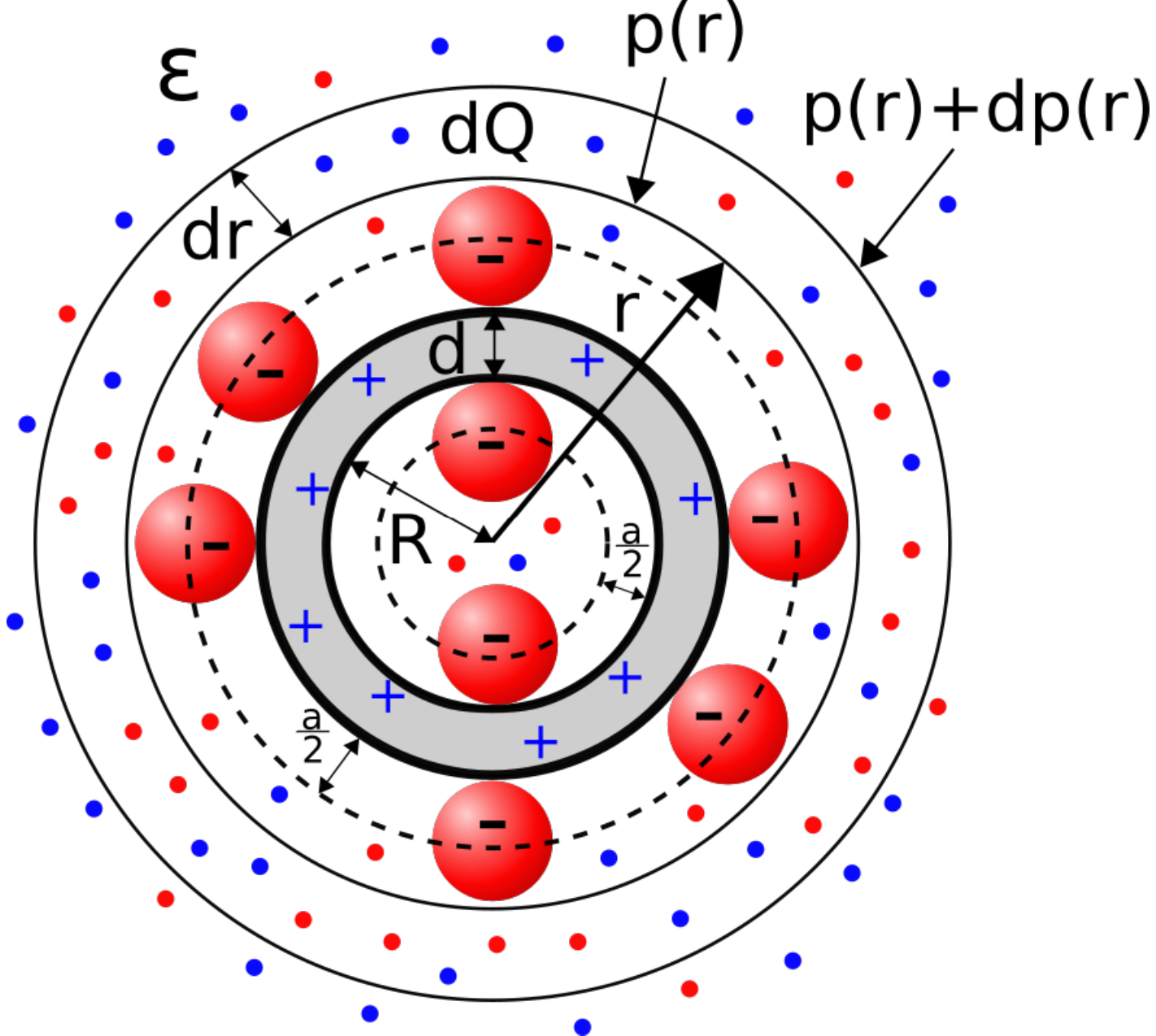}
		\caption{Spherical shell (2D cross-section).}
		\label{Geometry_sphere}
	\end{subfigure}
	
	\vspace{0.5em}
	
	\begin{subfigure}{.495\textwidth}
		\centering
		\includegraphics[width=.98\linewidth]{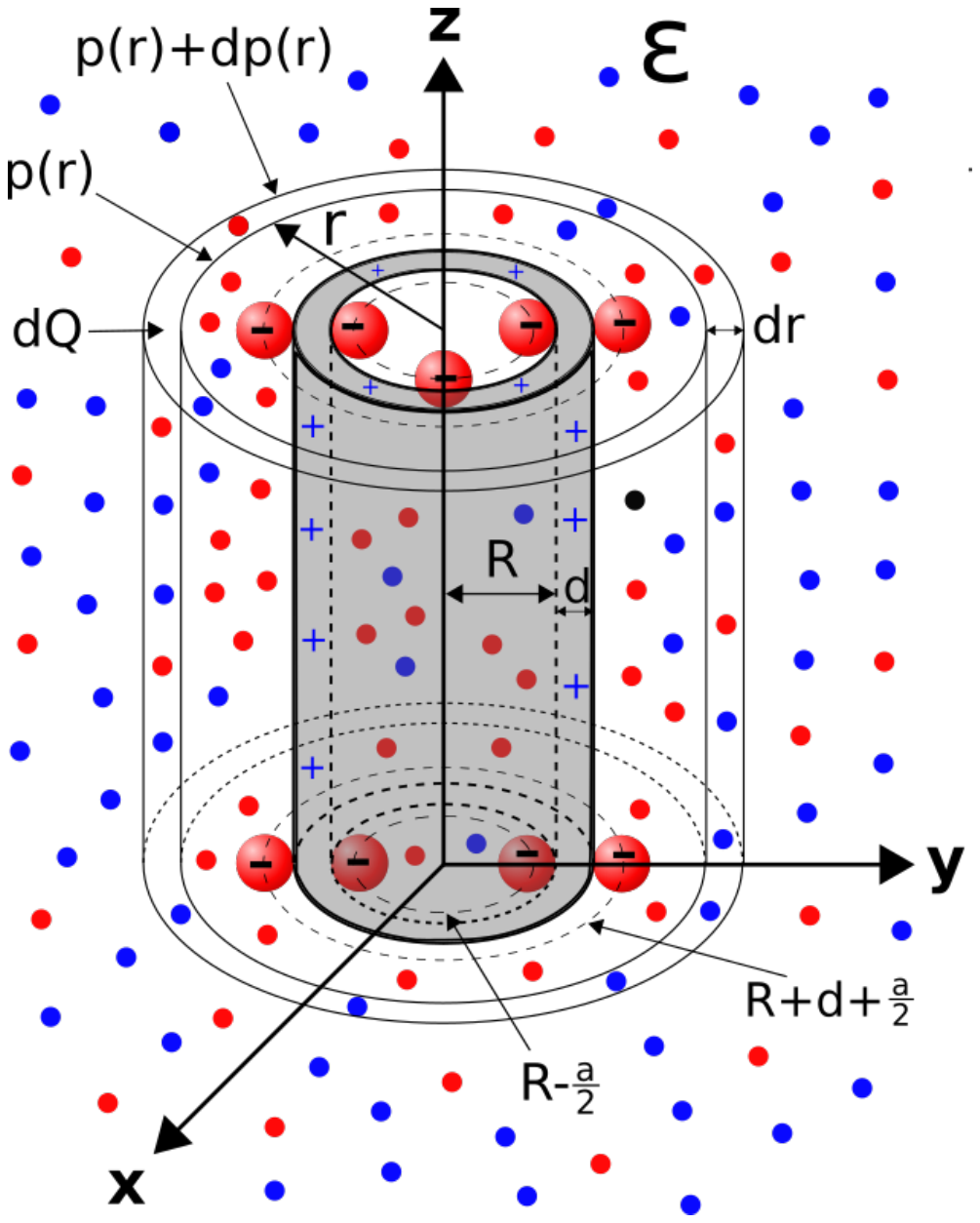}
		\caption{Cylindrical shell (3D).}
		\label{Geometry_cylinder}
	\end{subfigure}
	
	\caption{Electrolyte-filled hollow charged nanoparticles immersed in a bulk electrolyte.}
	\label{Geometry_shells}
\end{figure}

We consider three geometries: planar (slit-shell), cylindrical, and spherical, as shown in \cref{Geometry_shells}. The electrostatic potential and field obey Poisson's equation, and the contact theorems relate pressure differences across the shell to the ionic distribution and electric field at contact. 

The derivations rely on calculating the force on a thin slab of electrolyte near the shell surface, using both electrostatics and statistical mechanics. These force balances yield expressions for the pressure at contact in terms of the electric field and ion concentration profiles. In each geometry, the net pressure—defined as the difference between the inner and outer pressures on the shell—is the osmotic pressure, which we decompose into steric (entropic) and electrostatic (Maxwell stress tensor) contributions.

\subsubsection{Slit-Shell}\label{slit shell}

The simplest of the contact theorems is that corresponding to a slit-shell. In \cref{Geometry_twoplates}, a schematic representation of a slit-shell is shown. The geometrical center of the slit is located at the origin, and each plate (of thickness $d$ and infinite extent in the $y$ and $z$ directions) lies at positions $(\pm R, 0, 0)$. Due to the system's symmetry, the mean electrostatic potential (MEP), $\psi(\vec{r})$, and the electrostatic pressure, \( p_{\scriptscriptstyle{E}}(r) \), depend only on the distance $r = |x|$ from the center.

The electric field, defined as
\begin{equation}
	\vec{E}(x, y, z) = -\nabla \psi(x, y, z) = -\left( \frac{\partial \psi}{\partial x}, \frac{\partial \psi}{\partial y}, \frac{\partial \psi}{\partial z} \right),
	\label{E_field_slit}
\end{equation}
reduces, by symmetry, to a one-dimensional field:
\begin{equation}
	E(x) = -\frac{d\psi(x)}{dx},
	\label{E_slit}
\end{equation}
which satisfies the odd-parity condition:
\begin{equation}
	E(x) = -E(-x).
	\label{E_slit_odd}
\end{equation}
Consequently, $\psi(x)$ is an even function:
\begin{equation}
	\psi(r) = \psi(x) = \psi(-x),
	\label{psi_slit_even}
\end{equation}
and we consider only $x \geq 0$ without loss of generality.

We now consider a thin slab located at $x = r$ with thickness $dr$ and surface area $A$. Let \( dQ = A\, \rho_{\scriptscriptstyle{el}}(r) \, dr \) be the charge in the slab, where \( \rho_{\scriptscriptstyle{el}}(r) \) is the local charge density. The **electrostatic component of the pressure**, \( p_{\scriptscriptstyle{E}}(r) \), represents the contribution to the total pressure due to the Maxwell stress tensor, i.e., the momentum flux associated with the electric field.

Applying a force balance over this slab yields:
\begin{equation}\label{Slit-Force-Balance}
	A\left(p_{\scriptscriptstyle{E}}(r) + dp_{\scriptscriptstyle{E}}(r) \right) = A\, p_{\scriptscriptstyle{E}}(r) + E(r)\, dQ(r).
\end{equation}
Dividing through and simplifying, we obtain:
\begin{equation}
	\frac{dp_{\scriptscriptstyle{E}}(r)}{dr} = -\rho_{\scriptscriptstyle{el}}(r) \frac{d\psi(r)}{dr}.
	\label{Balance}
\end{equation}

From Poisson’s equation,
\begin{equation}
	\frac{d^2 \psi(r)}{dr^2} = -\frac{1}{\varepsilon_0 \varepsilon} \rho_{\scriptscriptstyle{el}}(r),
	\label{PEq}
\end{equation}
we can substitute into \cref{Balance}:
\begin{equation}
	\frac{dp_{\scriptscriptstyle{E}}(r)}{dr} = \varepsilon_0 \varepsilon \frac{d^2 \psi(r)}{dr^2} \frac{d\psi(r)}{dr}.
\end{equation}
By inspection,
\begin{equation}
	\frac{d}{dr} \left[ p_{\scriptscriptstyle{E}}(r) \right] = \frac{\varepsilon_0 \varepsilon}{2} \frac{d}{dr} \left[ \left( \frac{d\psi(r)}{dr} \right)^2 \right].
	\label{Balance2}
\end{equation}
Using \cref{E_slit}, and integrating \cref{Balance2} from $r_0$ to $r$,
\begin{equation}
	p_{\scriptscriptstyle{E}}(r) - p_{\scriptscriptstyle{E}}(r_0) = \frac{\varepsilon_0 \varepsilon}{2} \left[ E^2(r) - E^2(r_0) \right].
	\label{Presion}
\end{equation}

To obtain the net (osmotic) pressure on the slit walls, we must also consider the steric (entropic) component, defined as:
\begin{equation}
	p_{\scriptscriptstyle{S}}(r) = kT \sum_i \rho_i\, g_{\alpha i}(r) = kT \rho_{\alpha s}(r),
	\label{Pressure-Steric}
\end{equation}
where \( g_{\alpha i}(r) \) is the reduced concentration profile for species $i$, and

\begin{equation}
\rho_{\alpha s}(r) = \sum_i \rho_i g_{\alpha i}(r)
	\label{Ions-Total-Concentration}
\end{equation}

\noindent is the total local ion concentration.

Combining steric and electric contributions, the **total pressure** becomes:
\begin{equation}
	p_{\scriptscriptstyle{F}}(r) = p_{\scriptscriptstyle{S}}(r) - p_{\scriptscriptstyle{E}}(r).
\end{equation}

By evaluating these quantities at the contact points \( r = R - a/2 \) (inside) and \( r = R + d + a/2 \) (outside), and taking \( E(r_0) \to 0 \) at \( r_0 \to 0 \) (center) or \( r_0 \to \infty \) (bulk), we obtain the **contact theorem** for the slit-shell geometry:
\begin{equation} \label{Osmotic-plates-Net}
	\begin{split}
		p_{\scriptscriptstyle{N}}(R) &= p^{\text{in}}_{\scriptscriptstyle{F}}(R) - p^{\text{out}}_{\scriptscriptstyle{F}}(R) \\
		&= kT \left[ \rho_{\alpha s}\left(R - \frac{a}{2} \right) - \rho_{\alpha s}\left(R + d + \frac{a}{2} \right) \right] \\
		&\quad - \frac{\varepsilon_0 \varepsilon}{2} \left[ E^2\left(R - \frac{a}{2} \right) - E^2\left(R + d + \frac{a}{2} \right) \right].
	\end{split}
\end{equation}

This result decomposes the **osmotic pressure** into its **steric** and **electrostatic** components:
\begin{equation}
	p_{\scriptscriptstyle{S}}(R) = kT \left[ \rho_{\alpha s}\left(R - \frac{a}{2} \right) - \rho_{\alpha s}\left(R + d + \frac{a}{2} \right) \right]
\label{Osmotic-plates-Steric}
\end{equation}

\noindent and 

\begin{equation}
	p_{\scriptscriptstyle{E}}(R) = -\frac{\varepsilon_0 \varepsilon}{2} \left[ E^2\left(R - \frac{a}{2} \right) - E^2\left(R + d + \frac{a}{2} \right) \right].
\label{Osmotic-plates-Electric-Plates-E(r)}
\end{equation}

Although this decomposition is somewhat artificial (since in general entropic and electrostatic effects are coupled), it offers valuable insight into the physical mechanisms driving osmotic pressure under confinement.

Finally, since \( E(r) = \sigma(r)/(\varepsilon_0 \varepsilon) \), the net osmotic pressure can also be expressed in terms of effective surface charge densities at contact:
\begin{equation}
	p_{\scriptscriptstyle{N}}(R) = kT \left[ \rho_{\alpha s}\left(R - \frac{a}{2} \right) - \rho_{\alpha s}\left(R + d + \frac{a}{2} \right) \right] - \frac{1}{2 \varepsilon_0 \varepsilon} \left[ \sigma_{\text{Hi}}^2 - \sigma_{\text{Ho}}^2 \right],
\end{equation}
where \( \sigma_{\text{Hi}} = \sigma(R - a/2) \) and \( \sigma_{\text{Ho}} = \sigma(R + d + a/2) \).

\subsubsection{Cylindrical Shell}\label{Cylindrical-shell}

We now extend the force-balance argument to the case of a cylindrical shell immersed in an electrolyte, as shown in \cref{Geometry_cylinder}. Due to cylindrical symmetry, the electric field depends only on the radial coordinate \( r \).

A force balance on a cylindrical shell segment of radius \( r \), height \( z \), and thickness \( dr \), leads to the following differential equation for the **electrostatic component of the pressure**, \( p_{\scriptscriptstyle{E}}(r) \):
\begin{equation}
	2\pi (r+dr)z \left[ p_{\scriptscriptstyle{E}}(r) + dp_{\scriptscriptstyle{E}}(r) \right] = 2\pi r z p_{\scriptscriptstyle{E}}(r) + E(r)\, dQ(r),
	\label{Cylinder-force-balance}
\end{equation}
where the charge in the cylindrical shell is \( dQ(r) = 2\pi r z\, \rho_{\scriptscriptstyle{el}}(r)\, dr \), and the electric field is \( E(r) = -\frac{d\psi(r)}{dr} \).

Simplifying and dividing by \( 2\pi z \), we obtain:
\begin{equation}
	\frac{dp_{\scriptscriptstyle{E}}(r)}{dr} = -\frac{1}{r} p_{\scriptscriptstyle{E}}(r) - \rho_{\scriptscriptstyle{el}}(r) \frac{d\psi(r)}{dr}.
	\label{Cylinder-differential-equation}
\end{equation}

This equation can be solved analytically using Poisson’s equation for cylindrical symmetry:
\begin{equation}
	\rho_{\scriptscriptstyle{el}}(r) = -\varepsilon_0 \varepsilon \left( \frac{d^2 \psi(r)}{dr^2} + \frac{1}{r} \frac{d\psi(r)}{dr} \right).
	\label{Poisson-eq-Cyl}
\end{equation}

Substituting into \cref{Cylinder-differential-equation} and integrating gives the following expression for the electrostatic pressure:
\begin{equation}
	\begin{split}
		r\, p_{\scriptscriptstyle{E}}(r) = r_0\, p_{\scriptscriptstyle{E}}(r_0)
		+ \frac{\varepsilon_0 \varepsilon}{2} \left[ 
		r E^2(r) - r_0 E^2(r_0) + \int_{r_0}^{r} E^2(y)\, dy
		\right].
		\label{Balance4-Cyl}
	\end{split}
\end{equation}

Using the same approach as in the slit-shell case, and applying the appropriate boundary conditions (vanishing fields at \( r \to 0 \) and \( r \to \infty \)), the **fluid pressure** outside and inside the shell becomes:

\textbf{Outside} \( (r = R + d + a/2) \):
\begin{equation}
	\begin{split}
		p^{\text{out}}_{\scriptscriptstyle{F}}(R) &= kT\, \rho_{\alpha s}(R + d + a/2) \\
		&\quad - \frac{\varepsilon_0 \varepsilon}{2} \left[ 
		E^2(R + d + a/2)
		- \frac{1}{R + d + a/2} \int_{R + d + a/2}^{\infty} E^2(y)\, dy 
		\right].
		\label{pi-Cyl-out}
	\end{split}
\end{equation}

\textbf{Inside} \( (r = R - a/2) \):
\begin{equation}
	\begin{split}
		p^{\text{in}}_{\scriptscriptstyle{F}}(R) &= kT\, \rho_{\alpha s}(R - a/2) \\
		&\quad - \frac{\varepsilon_0 \varepsilon}{2} \left[ 
		E^2(R - a/2)
		+ \frac{1}{R - a/2} \int_0^{R - a/2} E^2(y)\, dy 
		\right].
		\label{pi-Cyl-in}
	\end{split}
\end{equation}

Thus, the **electrostatic (Maxwell) component** of the net osmotic pressure becomes:
\begin{equation}
	\begin{split}
		p_{\scriptscriptstyle{E}}(R) &= -\frac{\varepsilon_0 \varepsilon}{2} \left[
		E^2(R - a/2) - E^2(R + d + a/2)
		\right] \\
		&\quad - \frac{\varepsilon_0 \varepsilon}{2(R - a/2)} \int_0^{R - a/2} E^2(y)\, dy
		- \frac{\varepsilon_0 \varepsilon}{2(R + d + a/2)} \int_{R + d + a/2}^{\infty} E^2(y)\, dy.
		\label{Osmotic-Cyl-Electric}
	\end{split}
\end{equation}

The **steric (entropic) component** is again:
\begin{equation}
	p_{\scriptscriptstyle{S}}(R) = kT \left[
	\rho_{\alpha s}(R - a/2) - \rho_{\alpha s}(R + d + a/2)
	\right].
	\label{Osmotic-Steric-Press-Cyl}
\end{equation}

Hence, the **net osmotic pressure** for the cylindrical shell is:
\begin{equation}
	\begin{split}
		p_{\scriptscriptstyle{N}}(R) = p_{\scriptscriptstyle{S}}(R) + p_{\scriptscriptstyle{E}}(R)
		&= kT \left[
		\rho_{\alpha s}(R - a/2) - \rho_{\alpha s}(R + d + a/2)
		\right] \\
		&\quad - \frac{\varepsilon_0 \varepsilon}{2} \left[
		E^2(R - a/2) - E^2(R + d + a/2)
		\right] \\
		&\quad - \frac{\varepsilon_0 \varepsilon}{2(R - a/2)} \int_0^{R - a/2} E^2(y)\, dy \\
		&\quad - \frac{\varepsilon_0 \varepsilon}{2(R + d + a/2)} \int_{R + d + a/2}^{\infty} E^2(y)\, dy.
		\label{Osmotic-Net-Press-Cyl}
	\end{split}
\end{equation}

In the limit \( R \rightarrow 0 \), the cylindrical shell becomes a solid charged cylinder. In that case, the inner wall vanishes and the negative of \cref{Osmotic-Net-Press-Cyl} reduces to the classical contact theorem for a single charged cylindrical electrode.

\subsubsection{Spherical Shell}\label{Spherical-shell}

We now consider a spherical shell of inner radius \( R \), wall thickness \( d \), and immersed in an electrolyte. Following the same approach, a force balance on a spherical shell segment yields an expression for the electrostatic pressure \( p_{\scriptscriptstyle{E}}(r) \).

Starting with the electrostatic force balance in spherical coordinates:
\begin{equation}
	4\pi (r + dr)^2 \left[ p_{\scriptscriptstyle{E}}(r) + dp_{\scriptscriptstyle{E}}(r) \right] 
	= 4\pi r^2 p_{\scriptscriptstyle{E}}(r) + E(r)\, dQ(r),
	\label{Sphere-Force-Balance}
\end{equation}
where \( dQ(r) = 4\pi r^2\, \rho_{\scriptscriptstyle{el}}(r)\, dr \). 

Using Poisson's equation for spherical symmetry:
\begin{equation}
	\rho_{\scriptscriptstyle{el}}(r) = -\varepsilon_0 \varepsilon \left( \frac{d^2 \psi(r)}{dr^2} + \frac{2}{r} \frac{d\psi(r)}{dr} \right),
	\label{Poisson-eq-Sph}
\end{equation}
we find:
\begin{equation}
	\begin{split}
		\frac{d}{dr} \left( r^2 p_{\scriptscriptstyle{E}}(r) \right) 
		&= \varepsilon_0 \varepsilon \left[
		\frac{d}{dr} \left( r^2 \left( \frac{d\psi(r)}{dr} \right)^2 \right)
		- r^2 \frac{d^2\psi(r)}{dr^2} \frac{d\psi(r)}{dr}
		\right].
		\label{Balance2-Sph}
	\end{split}
\end{equation}

Solving this equation and applying the appropriate boundary conditions, we obtain the following expressions for the fluid pressure.

\textbf{Outside} the shell (at \( r = R + d + a/2 \)):
\begin{equation}
	\begin{split}
		p^{\text{out}}_{\scriptscriptstyle{F}}(R) &= kT\, \rho_{\alpha s}(R + d + a/2) 
		- \frac{\varepsilon_0 \varepsilon}{2} E^2(R + d + a/2) \\
		&\quad + \frac{\varepsilon_0 \varepsilon}{\left(R + d + a/2\right)^2} \int_{R + d + a/2}^{\infty} y\, E^2(y)\, dy.
		\label{Pressure-sphere-out}
	\end{split}
\end{equation}

\textbf{Inside} the shell (at \( r = R - a/2 \)):
\begin{equation}
	\begin{split}
		p^{\text{in}}_{\scriptscriptstyle{F}}(R) &= kT\, \rho_{\alpha s}(R - a/2) 
		- \frac{\varepsilon_0 \varepsilon}{2} E^2(R - a/2) \\
		&\quad - \frac{\varepsilon_0 \varepsilon}{(R - a/2)^2} \int_0^{R - a/2} y\, E^2(y)\, dy.
		\label{Pressure-sphere-in}
	\end{split}
\end{equation}

Thus, the **electrostatic (Maxwell) component** of the osmotic pressure is:
\begin{equation}
	\begin{split}
		p_{\scriptscriptstyle{E}}(R) = 
		&- \frac{\varepsilon_0 \varepsilon}{2} \left[ E^2(R - a/2) - E^2(R + d + a/2) \right] \\
		&- \frac{\varepsilon_0 \varepsilon}{(R - a/2)^2} \int_0^{R - a/2} y\, E^2(y)\, dy \\
		&- \frac{\varepsilon_0 \varepsilon}{(R + d + a/2)^2} \int_{R + d + a/2}^{\infty} y\, E^2(y)\, dy.
		\label{Osmotic-Press-Sph-Electric}
	\end{split}
\end{equation}

The **steric (entropic) component** is:
\begin{equation}
	p_{\scriptscriptstyle{S}}(R) = kT \left[
	\rho_{\alpha s}(R - a/2) - \rho_{\alpha s}(R + d + a/2)
	\right].
	\label{Osmotic-Steric-Press-Sphere}
\end{equation}

The total **osmotic pressure** is then:
\begin{equation}
	\begin{split}
		p_{\scriptscriptstyle{N}}(R) = p_{\scriptscriptstyle{S}}(R) + p_{\scriptscriptstyle{E}}(R)
		&= kT \left[ \rho_{\alpha s}(R - a/2) - \rho_{\alpha s}(R + d + a/2) \right] \\
		&\quad - \frac{\varepsilon_0 \varepsilon}{2} \left[ E^2(R - a/2) - E^2(R + d + a/2) \right] \\
		&\quad - \frac{\varepsilon_0 \varepsilon}{(R - a/2)^2} \int_0^{R - a/2} y\, E^2(y)\, dy \\
		&\quad - \frac{\varepsilon_0 \varepsilon}{(R + d + a/2)^2} \int_{R + d + a/2}^{\infty} y\, E^2(y)\, dy.
		\label{Osmotic-sphere-Net}
	\end{split}
\end{equation}

As in the other cases, in the limit \( R \rightarrow 0 \), the spherical shell reduces to a solid charged sphere, and the negative of \cref{Osmotic-sphere-Net} becomes the classical contact theorem for a single spherical electrode.


\subsubsection{Electroneutrality Condition}\label{electroneutrality condition}

For the three shell geometries, the electroneutrality condition is
\begin{equation}\label{Electroneutrality-condition-general}
	Q_{\scriptscriptstyle{0}} \left( R \right)+Q_{\scriptscriptstyle{0}} \left( R+d \right)+Q_{\scriptscriptstyle{Hi}} \left( R-a/2 \right) +Q_{\scriptscriptstyle{Ho}} \left( R+d+a/2 \right)=0,
\end{equation}
\noindent where $Q_{\scriptscriptstyle{0}} \left( R \right)$ and  $Q_{\scriptscriptstyle{0}} \left( R+d \right)$, are the electrical charges at the surfaces $A_{\gamma}(R)$, $A_{\gamma}(R+d)$, while $Q_{\scriptscriptstyle{Hi}} \left( R-a/2\right)$ and $Q_{\scriptscriptstyle{Ho}} \left( R+d+a/2\right)$ are the induced charges inside and outside of the shells, given by

\begin{equation}\label{Induced-charge-in}
	\begin{split}
		Q_{\scriptscriptstyle{Hi}} \left( R-a/2 \right)=\int\limits_{\boldsymbol{\omega_{\scriptscriptstyle_{in}}}}\rho_{el}(r)d\mathbf{V}\ = f_{\scriptscriptstyle_{\gamma}}\int_{ 0}^{R-a/2}  r^\gamma \rho_{el}(r)dr,
	\end{split}
\end{equation}
\noindent and
\begin{equation}\label{Induced-charge-out}
	\begin{split}
		Q_{\scriptscriptstyle{Ho}} \left( R+d+a/2 \right)=\int\limits_{\boldsymbol{\omega_{\scriptscriptstyle_{out}}}}\rho_{el}(r)d\mathbf{V}\ =f_{\scriptscriptstyle_{\gamma}}\int_{ R+d+a/2}^{\infty} r^\gamma\rho_{el}(r) dr.\
	\end{split}
\end{equation}

\noindent  \cref{Induced-charge-in,Induced-charge-out} are volume integrals over all the space inside, $\boldsymbol{\omega_{\scriptscriptstyle_{in}}}$, and outside, $\boldsymbol{\omega_{\scriptscriptstyle_{out}}}$, of the shells, respectively. Hence, $f_{\scriptscriptstyle_{\gamma}}$ are geometrical  constants resulting from these volume integrals in Cartesian, cylindrical or spherical coordinates; $\gamma=0,1,2$ for the planar, cylindrical and spherical geometry, respectively. Consequently, the surface charge densities profiles for $r\leq (R-a/2)$ and $r\geq(R+d+a/2)$ are 
\begin{equation}\label{Induced-charge-density-in}
	\sigma_{\scriptscriptstyle{\gamma}}(r)=\frac{1}{(r)^{\gamma}}\int_{ 0}^{r}r^\gamma \rho_{el}(r) dr,
\end{equation}
\noindent and
 \begin{equation}\label{Induced-charge-density-out}
	\sigma_{\scriptscriptstyle{\gamma}}(r)=-\frac{1}{(r)^{\gamma}}\int_{ r}^{\infty}r^\gamma \rho_{el}(r) dr.
\end{equation}

\noindent Thus, the electroneutrality condition becomes,

\begin{equation}\label{Electroneutrality-condition-general2}
	\begin{split}
	&R^{\gamma}\sigma_{\scriptscriptstyle{0}} +(R+d)^{\gamma}\sigma_{\scriptscriptstyle{0}}
	+(R-a/2)^{\gamma}\sigma_{\scriptscriptstyle{\gamma}} \left( R-a/2 \right)\\ &=(R+d+a/2)^{\gamma}\sigma_{\scriptscriptstyle{\gamma}} \left( R+d+a/2 \right).
\end{split}
\end{equation}

\noindent $\sigma_{\scriptscriptstyle{\gamma}}(R-a/2)/(\epsilon_{\scriptscriptstyle_{0}}\epsilon)$ and $\sigma_{\scriptscriptstyle{\gamma}}(R+d+a/2)/(\epsilon_{\scriptscriptstyle_{0}}\epsilon)$ are the effective electrical fields at $r=R-a/2$ and $r=R+d-a/2$, respectively. Thus, from Eq. \ref{Electroneutrality-condition-general2}, the electrical fields balance is

 \begin{equation}\label{Electrical-field-balance}
 	\begin{split}
 	R^{\gamma}E_{\scriptscriptstyle{\gamma}}(R)\mathbf{e_{r}} +(R-a/2)^{\gamma}E_{\scriptscriptstyle{\gamma}} \left( R-a/2 \right)\mathbf{e_{r}}+(R+d)^{\gamma}E_{\scriptscriptstyle{\gamma}}(R+d)\mathbf{e_{r}}\\
 	 =(R+d+a/2)^{\gamma}E_{\scriptscriptstyle{\gamma}} \left( R+d+a/2 \right)\mathbf{e_{r}}, \qquad\qquad\qquad
 	 	\end{split}
 \end{equation}

\noindent where $\mathbf{e_{r}}$ is a unitary vector along the $r$-direction. In \cref{Electrical-field-balance}, $E(R-a/2)$ is negative, whereas $E(R)$, $E(R+d)$ and $E(R+d+a/2)$ are positive, since we have chosen $\sigma_{\scriptscriptstyle{0}}$ to be a positive surface charge density. This is, \cref{Induced-charge-density-out} states that $\sigma_{\scriptscriptstyle{\gamma}}(R+d+a/2)$, the effective surface charge density at $r=R+d+a/2$,  \textit{is also equal to the negative of the induced surface charge density outside the shells}, whereas \cref{Induced-charge-density-in} states that $\sigma_{\scriptscriptstyle{\gamma}}(R-a/2)$ is equal to the \textit{induced charge inside the shells}. In reference \cite{Adrian-JML-2023} it was defined $\sigma_{\scriptscriptstyle{\gamma}}(R-a/2)$ with a negative sign in \cref{Induced-charge-density-in}. This implies a change in notation with no effect in the results presented in that or this article, other that in a change in sign of $E(r)$, for $0\leq r \leq (R-a/2)$. For the discussion given in this article, we will use the electroneutrality balance condition of Eq. \ref{Electrical-field-balance}. In the \cref{The electical double layer} we will drop the sub-index $\gamma$ and will use $\sigma_{\scriptscriptstyle{Hi}}$ for $\sigma_{\scriptscriptstyle{\gamma}}(R-a/2)$ and $\sigma_{\scriptscriptstyle{Ho}}$ for $\sigma_{\scriptscriptstyle{\gamma}}(R+d+a/2)$, as defined in \cref{Induced-charge-density-in,Induced-charge-density-out}, respectively. However, notice that $\sigma_{\scriptscriptstyle{Hi}}$ and $\sigma_{\scriptscriptstyle{Ho}}$ are still functions of $R$, i.e. below sometimes we will use $\sigma_{\scriptscriptstyle{Hi}}(R)$ and $\sigma_{\scriptscriptstyle{Ho}}(R)$ instead, to emphasize their dependence on the shells' radius and, hence, its width.

For a later discussion of our results, let us point out that in general $E_{\gamma}\left( R-a/2 \right)\mathbf{e_{r}}+E_{\gamma}\left( R \right)\mathbf{e_{r}} \neq \vec{0}, \ \forall  R<\infty$, where the unscreened electrical field produced by cavity wall at $r=R$ is $E_{\gamma}\left( R \right) = \sigma_{o}/(\epsilon_{0}\epsilon)$. This implies a \textit{violation of the local electroneutrality condition} (VLEC) inside the shells. The VLEC is, of course, due to presence of electrical fields and ions' charge correlations. This phenomenon has been long time theoretically predicted by integral equation and Poisson-Boltzmann theories~\cite{Lozada_1984,Lozada1996,Lozada-Cassou-PRL1996,Yu_1997,Aguilar_2007,Levin_electroneutrality-2016}, and more recently corroborated by density functional theories and computer simulations~\cite{Levin_electroneutrality-2016,Levy-electroneutrality-2020,Levy-electroneutrality-PRE-2021,Keshavarzi_2020}, other theoretical approaches~\cite{Green-electroneutrality-JCP-2021,Gonzalez-Calderon-EPJ-2021,Keshavarzi_2022,malgaretti_electroneutrality-2024}, and experimental findings~\cite{Cuvillier-Rondelez-1998,Luo-electroneutrality-nature-2015}. All these theories reduce to the PB and LPB equations at the proper model parameters. Moreover, the experimental validation of the VLEC is supported by the Poisson–Boltzmann equation~\cite{Luo-electroneutrality-nature-2015}.

However, the total electroneutrality is satisfied, through \cref{Electrical-field-balance}. In addition, while for the slit-shell, for relatively small values of $R$, local electroneutrality inside the slit is numerically achieved, for the cylindrical and spherical shells this is virtually numerically impossible, due to the very large values of $R$ needed to attain this limit. Nevertheless, the $\lim_{R \to \infty} [E_{\gamma}\left( R-a/2 \right)\mathbf{e_{r}} + E_{\gamma}\left( R \right)\mathbf{e_{r}}] = \vec{0}$, as can be readily analytically demonstrated from the electrical field equations given in the next  \cref{The electical double layer}. Thus, from \cref{Electrical-field-balance}, the $\lim_{R \to \infty} [E_{\gamma}\left( R+d+a/2 \right)\mathbf{e_{r}}-E_{\gamma}\left( R+d \right) \mathbf{e_{r}}] = \vec{0}$, where the unscreened electrical field produced by cavity wall at $r=R+d$  is $E_{\gamma}\left( R+d \right)=\sigma_{\scriptscriptstyle{0}}/(\epsilon_{\scriptscriptstyle_{0}}\epsilon)$. Hence, the local electroneutrality condition is satisfied also in this limit, and \cref{Electrical-field-balance} becomes $R^{\gamma}\sigma_{\scriptscriptstyle{0}} +(R+d)^{\gamma}\sigma_{\scriptscriptstyle{0}}
+(R-a/2)^{\gamma}\sigma_{\scriptscriptstyle{0}}=(R+d+a/2)^{\gamma}\sigma_{\scriptscriptstyle{0}}$, implying that \textit{in this limit} the local electroneutrality is attained independently, inside and outside the shells.


\subsection{The Electrical Double Layer}\label{The electical double layer}

To calculate the osmotic pressure in the shell geometries, we need the electric field profile \( E(r) \) and the electrolyte concentration at contact with the inner and outer shell walls, i.e., at \( r = R - a/2 \) and \( r = R + d + a/2 \), respectively.

In this work, we focus on the analytical solutions of the linearized Poisson–Boltzmann (LPB) equation with a Stern layer correction, i.e., the ions are treated as point charges in the bulk, but finite-sized at contact with the shell walls~\cite{Verwey_TheoryStabilityLyophobicColloids_1948}. This model is valid for low surface charge densities and low electrolyte concentrations, where nonlinear and excluded volume effects are minimal.

The shell geometries we consider—slit (planar), cylindrical, and spherical—are illustrated in \cref{Geometry_twoplates,Geometry_cylinder,Geometry_sphere}. In all cases, the shells are immersed in a symmetric \( z:z \) electrolyte at the same chemical potential inside and outside the cavity. The dielectric constant \( \varepsilon \) is taken to be uniform throughout the shell and electrolyte, thus neglecting image charges.

The local charge density in the point-ion approximation is given by:
\begin{equation}
	\rho_{\text{el}}(r) = \sum_{i=1}^n e z_i \rho_{i0} \exp\left(-\beta e z_i \psi(r)\right), \label{Rho_elx2}
\end{equation}
where \( \psi(r) \) is the mean electrostatic potential (MEP), \( \rho_{i0} \) is the bulk concentration of ionic species \( i \), \( z_i \) its valence, \( e \) the elementary charge, and \( \beta = 1/(kT) \).

Substituting \cref{Rho_elx2} into the Poisson equation,
\begin{equation}
	\nabla^2 \psi(r) = -\frac{1}{\varepsilon_0 \varepsilon} \rho_{\text{el}}(r), \label{Ec.Poisson-general}
\end{equation}
yields the full Poisson–Boltzmann (PB) equation. In the linear regime, where \( \beta e z_i \psi \ll 1 \), we obtain the LPB equation:
\begin{equation}
	\nabla^2 \psi(r) = \kappa^2 \psi(r), \label{Ec.ALS_Poisson}
\end{equation}
with
\begin{equation}
	\kappa = \sqrt{\frac{2 e^2 z^2 \rho_0}{\varepsilon_0 \varepsilon k T}}, \label{Ec.kappa}
\end{equation}
where \( \rho_0 \) is the bulk ionic concentration of each species in a symmetric \( z:z \) electrolyte.

Analytical solutions of the LPB equation for the three shell geometries have been previously derived~\cite{Adrian-JML-2023}. Here we reproduce an adequate reformulation of the relevant expressions for the electrostatic potential \( \psi(r) \) and electric field \( E(r) \) in each geometry in the following subsections, where we define \( \sigma_{\text{Hi}} = \sigma(R - a/2) \) and \( \sigma_{\text{Ho}} = \sigma(R + d + a/2) \) as the effective induced charge densities at the inner and outer electrolyte contact surfaces, respectively.

The values \( \psi_{\text{H}} = \psi(R - a/2) \), \( \psi_0 = \psi(R) \), \( \varphi_0 = \psi(R + d) \), and \( \varphi_{\text{H}} = \psi(R + d + a/2) \) will be used in our discussion of the EDL structure and pressure contributions.

We emphasize that these expressions are exact within the LPB framework and that the discontinuities in the electric field at \( r = R \) and \( r = R + d \) arise from Maxwell’s boundary conditions at the dielectric interface between the electrolyte and the charged shell wall.

\subsubsection{Slit-Shell}\label{Slit-shell}

The LPB solution for a symmetric electrolyte in a slit-shell geometry (i.e., two parallel plates separated by \( 2R \)) with a Stern layer correction is given by~\cite{Adrian-JML-2023}:

\paragraph{Mean Electrostatic Potential \(\psi(r)\):}
\begin{equation}
	\psi(r=|x|) =
	\begin{cases}
		\displaystyle \frac{2\sigma_0 - \sigma_{\text{Ho}}}{\varepsilon_0 \varepsilon \kappa} \frac{\cosh[\kappa r]}{\sinh[\kappa(R - a/2)]}, & 0 \leq r \leq R - \frac{a}{2} \\[8pt]
		\displaystyle \psi_{\text{H}} + \frac{2\sigma_0 - \sigma_{\text{Ho}}}{\varepsilon_0 \varepsilon} (r - (R - a/2)), & R - \frac{a}{2} \leq r \leq R \\[8pt]
		\displaystyle \varphi_0 - \frac{\sigma_{\text{Ho}} - \sigma_0}{\varepsilon_0 \varepsilon}(r - R - d), & R \leq r \leq R + d \\[8pt]
		\displaystyle \frac{\sigma_{\text{Ho}}}{\varepsilon_0 \varepsilon \kappa}[1 - \kappa(r - R_{\text{H}})], & R + d \leq r \leq R_{\text{H}} \\[8pt]
		\displaystyle \frac{\sigma_{\text{Ho}}}{\varepsilon_0 \varepsilon \kappa} e^{-\kappa(r - R_{\text{H}})}, & R_{\text{H}} \leq r
	\end{cases}
	\label{Plates-MEP(r)}
\end{equation}

\paragraph{Electric Field \(E(r)\):}
\begin{equation}
	E(r) =
	\begin{cases}
		\displaystyle \frac{\sigma_{\text{Ho}} - 2\sigma_0}{\varepsilon_0 \varepsilon} \frac{\sinh[\kappa r]}{\sinh[\kappa(R - a/2)]}, & 0 \leq r \leq R - \frac{a}{2} \\[8pt]
		\displaystyle \frac{\sigma_{\text{Ho}} - 2\sigma_0}{\varepsilon_0 \varepsilon}, & R - \frac{a}{2} \leq r < R \\[8pt]
		\displaystyle \frac{\sigma_{\text{Ho}} - \sigma_0}{\varepsilon_0 \varepsilon}, & R < r < R + d \\[8pt]
		\displaystyle \frac{\sigma_{\text{Ho}}}{\varepsilon_0 \varepsilon}, & R + d \leq r \leq R_{\text{H}} \\[8pt]
		\displaystyle \frac{\sigma_{\text{Ho}}}{\varepsilon_0 \varepsilon} e^{-\kappa(r - R_{\text{H}})}, & R_{\text{H}} \leq r
	\end{cases}
	\label{Plates-E(r)}
\end{equation}

Here, \( R_{\text{H}} = R + d + a/2 \) is the outermost contact point of the shell and electrolyte. The surface potential values are:
\begin{align}
	\varphi_{\text{H}} &= \psi(R_{\text{H}}) = \frac{\sigma_{\text{Ho}}}{\varepsilon_0 \varepsilon \kappa}, \label{Plates-phi-H} \\
	\varphi_0 &= \psi(R + d) = \varphi_{\text{H}} + \frac{a}{2} \frac{\sigma_{\text{Ho}}}{\varepsilon_0 \varepsilon}, \label{Plates-phi-0} \\
	\psi_0 &= \psi(R) = \varphi_0 + d\left( \frac{\sigma_{\text{Ho}} - \sigma_0}{\varepsilon_0 \varepsilon} \right), \label{Plates-psi-0} \\
	\psi_{\text{H}} &= \psi(R - a/2) = \psi_0 + \frac{a}{2} \left( \frac{\sigma_{\text{Ho}} - 2\sigma_0}{\varepsilon_0 \varepsilon} \right). \label{Plates-psi-H}
\end{align}

\paragraph{Effective Surface Charge Density:}

The effective outer surface charge density \( \sigma_{\text{Ho}} \) is given by:
\begin{equation}
	\sigma_{\text{Ho}} = \left[ \frac{\kappa(a + d) + 2 \coth[\kappa(R - a/2)]}{1 + \kappa(a + d) + \coth[\kappa(R - a/2)]} \right] \sigma_0.
	\label{Plates_sigmaHO}
\end{equation}

\paragraph{Electroneutrality Condition:}

Using the electroneutrality balance for the slit-shell geometry, we obtain:
\begin{equation}
	\sigma_0 + \sigma_0 + \sigma_{\text{Hi}} = \sigma_{\text{Ho}}, \label{Slit-electroneutrality}
\end{equation}
from which \( \sigma_{\text{Hi}} \) can be readily calculated given \( \sigma_{\text{Ho}} \).

\subsubsection{Cylindrical Shell}\label{Cylindrical-shell-EDL}

The LPB solutions for a cylindrical shell geometry (infinitely long hollow cylinder) immersed in a symmetric electrolyte, with Stern correction, are given by~\cite{Adrian-JML-2023}.

\paragraph{Mean Electrostatic Potential \(\psi(r)\):}
\begin{equation}
	\psi(r) = 
	\begin{cases}
		\displaystyle -\left( \frac{R_{\text{H}} \sigma_{\text{Ho}} - (2R + d)\sigma_0}{\varepsilon_0 \varepsilon} \right) \frac{I_0(\kappa r)}{\kappa (R - \frac{a}{2}) I_1[\kappa(R - \frac{a}{2})]}, & 0 \leq r \leq R - \frac{a}{2} \\[8pt]
		\displaystyle \psi_{\text{H}} + \left( \frac{R_{\text{H}} \sigma_{\text{Ho}} - (2R + d)\sigma_0}{\varepsilon_0 \varepsilon} \right) \ln\left( \frac{R - \frac{a}{2}}{r} \right), & R - \frac{a}{2} \leq r \leq R \\[8pt]
		\displaystyle \psi_0 + \left( \frac{R_{\text{H}} \sigma_{\text{Ho}} - (R + d)\sigma_0}{\varepsilon_0 \varepsilon} \right) \ln\left( \frac{R}{r} \right), & R \leq r \leq R + d \\[8pt]
		\displaystyle \varphi_0 + \frac{\sigma_{\text{Ho}} R_{\text{H}}}{\varepsilon_0 \varepsilon} \ln\left( \frac{R + d}{r} \right), & R + d \leq r \leq R_{\text{H}} \\[8pt]
		\displaystyle \frac{\sigma_{\text{Ho}}}{\varepsilon_0 \varepsilon \kappa} \frac{K_0(\kappa r)}{K_1(\kappa R_{\text{H}})}, & R_{\text{H}} \leq r
	\end{cases}
	\label{Cylinder-MEP}
\end{equation}

\paragraph{Electric Field \(E(r)\):}
\begin{equation}
	E(r) =
	\begin{cases}
		\displaystyle \left( \frac{R_{\text{H}} \sigma_{\text{Ho}} - (2R + d)\sigma_0}{\varepsilon_0 \varepsilon} \right) \frac{I_1(\kappa r)}{(R - \frac{a}{2}) I_1[\kappa(R - \frac{a}{2})]}, & 0 \leq r \leq R - \frac{a}{2} \\[8pt]
		\displaystyle \left( \frac{R_{\text{H}} \sigma_{\text{Ho}} - (2R + d)\sigma_0}{\varepsilon_0 \varepsilon} \right) \frac{1}{r}, & R - \frac{a}{2} \leq r < R \\[8pt]
		\displaystyle \left( \frac{R_{\text{H}} \sigma_{\text{Ho}} - (R + d)\sigma_0}{\varepsilon_0 \varepsilon} \right) \frac{1}{r}, & R < r < R + d \\[8pt]
		\displaystyle \frac{\sigma_{\text{Ho}} R_{\text{H}}}{\varepsilon_0 \varepsilon r}, & R + d \leq r \leq R_{\text{H}} \\[8pt]
		\displaystyle \frac{\sigma_{\text{Ho}}}{\varepsilon_0 \varepsilon K_1(\kappa R_{\text{H}})} K_1(\kappa r), & R_{\text{H}} \leq r
	\end{cases}
	\label{Cylinder-E(r)}
\end{equation}

\paragraph{Surface Potential Values:}
\begin{align}
	\varphi_{\text{H}} &= \frac{K_0(\kappa R_{\text{H}}) \sigma_{\text{Ho}}}{\varepsilon_0 \varepsilon \kappa K_1(\kappa R_{\text{H}})} \label{Cyl-phi-H} \\
	\varphi_0 &= \varphi_{\text{H}} + \frac{R_{\text{H}} \sigma_{\text{Ho}}}{\varepsilon_0 \varepsilon} \ln\left( \frac{R_{\text{H}}}{R + d} \right) \label{Cyl-phi-0} \\
	\psi_0 &= \varphi_0 + \left( \frac{R_{\text{H}} \sigma_{\text{Ho}} - (R + d)\sigma_0}{\varepsilon_0 \varepsilon} \right) \ln\left( \frac{R + d}{R} \right) \label{Cyl-psi-0} \\
	\psi_{\text{H}} &= \psi_0 + \left( \frac{R_{\text{H}} \sigma_{\text{Ho}} - (2R + d)\sigma_0}{\varepsilon_0 \varepsilon} \right) \ln\left( \frac{R}{R - \frac{a}{2}} \right) \label{Cyl-psi-H}
\end{align}

\paragraph{Effective Surface Charge:}
\begin{equation}
	\sigma_{\text{Ho}} = \frac{L_2}{L_1} \sigma_0
	\label{Cylinder-sigmaHo}
\end{equation}
where:
\begin{align*}
	L_1 &= \frac{R_{\text{H}} I_0[\kappa(R - \frac{a}{2})]}{\kappa(R - \frac{a}{2}) I_1[\kappa(R - \frac{a}{2})]} + R_{\text{H}} \ln\left( \frac{R_{\text{H}}}{R - \frac{a}{2}} \right) + \frac{K_0(\kappa R_{\text{H}})}{\kappa K_1(\kappa R_{\text{H}})} \\
	L_2 &= \frac{(2R + d) I_0[\kappa(R - \frac{a}{2})]}{\kappa(R - \frac{a}{2}) I_1[\kappa(R - \frac{a}{2})]} + (2R + d) \ln\left( \frac{R}{R - \frac{a}{2}} \right) + (R + d) \ln\left( \frac{R + d}{R} \right)
\end{align*}

\paragraph{Electroneutrality Condition:}
\begin{equation}
	R \sigma_0 + (R + d)\sigma_0 + (R - \frac{a}{2}) \sigma_{\text{Hi}} = R_{\text{H}} \sigma_{\text{Ho}}
	\label{Cylinder-electroneutrality}
\end{equation}

From this, \( \sigma_{\text{Hi}} \) can be determined once \( \sigma_{\text{Ho}} \) is computed.

\subsubsection{Spherical Shell}\label{Spherical-shell-EDL}

The LPB analytical solution for a spherical shell immersed in a symmetric electrolyte, with Stern correction, is given by~\cite{Adrian-JML-2023}.

\paragraph{Mean Electrostatic Potential \(\psi(r)\):}
\begin{equation}
	\psi(r) = 
	\begin{cases}
		\displaystyle \frac{R_{\text{H}}^2 \sigma_{\text{Ho}} - [(R+d)^2 + R^2]\sigma_0}{\varepsilon_0 \varepsilon D(R)} \cdot \frac{\sinh(\kappa r)}{r}, & 0 \leq r \leq R - \frac{a}{2} \\[8pt]
		\displaystyle \psi_{\text{H}} + \frac{R_{\text{H}}^2 \sigma_{\text{Ho}} - [(R+d)^2 + R^2]\sigma_0}{\varepsilon_0 \varepsilon (R - \frac{a}{2})} \left( \frac{R - \frac{a}{2}}{r} - 1 \right), & R - \frac{a}{2} \leq r \leq R \\[8pt]
		\displaystyle \psi_0 + \frac{R_{\text{H}}^2 \sigma_{\text{Ho}} - (R + d)^2 \sigma_0}{\varepsilon_0 \varepsilon R} \left( \frac{R}{r} - 1 \right), & R \leq r \leq R + d \\[8pt]
		\displaystyle \varphi_0 + \frac{R_{\text{H}}^2 \sigma_{\text{Ho}}}{\varepsilon_0 \varepsilon (R + d)} \left( \frac{R + d}{r} - 1 \right), & R + d \leq r \leq R_{\text{H}} \\[8pt]
		\displaystyle \frac{R_{\text{H}}^2 \sigma_{\text{Ho}}}{\varepsilon_0 \varepsilon (1 + \kappa R_{\text{H}})} \cdot \frac{e^{-\kappa (r - R_{\text{H}})}}{r}, & R_{\text{H}} \leq r
	\end{cases}
	\label{Sphere-MEP}
\end{equation}

Here, the denominator
\[
D(R) = \sinh[\kappa (R - \frac{a}{2})] - \kappa (R - \frac{a}{2}) \cosh[\kappa (R - \frac{a}{2})]
\]

\paragraph{Electric Field \(E(r)\):}
\begin{equation}
	E(r) = 
	\begin{cases}
		\displaystyle \frac{R_{\text{H}}^2 \sigma_{\text{Ho}} - [(R+d)^2 + R^2]\sigma_0}{\varepsilon_0 \varepsilon D(R)} \cdot \frac{\sinh(\kappa r) - \kappa r \cosh(\kappa r)}{r^2}, & 0 \leq r \leq R - \frac{a}{2} \\[8pt]
		\displaystyle \frac{R_{\text{H}}^2 \sigma_{\text{Ho}} - [(R+d)^2 + R^2]\sigma_0}{\varepsilon_0 \varepsilon r^2}, & R - \frac{a}{2} \leq r < R \\[8pt]
		\displaystyle \frac{R_{\text{H}}^2 \sigma_{\text{Ho}} - (R + d)^2 \sigma_0}{\varepsilon_0 \varepsilon r^2}, & R < r < R + d \\[8pt]
		\displaystyle \frac{R_{\text{H}}^2 \sigma_{\text{Ho}}}{\varepsilon_0 \varepsilon r^2}, & R + d \leq r \leq R_{\text{H}} \\[8pt]
		\displaystyle \frac{R_{\text{H}}^2 (1 + \kappa r) \sigma_{\text{Ho}}}{\varepsilon_0 \varepsilon (1 + \kappa R_{\text{H}})} \cdot \frac{e^{-\kappa(r - R_{\text{H}})}}{r^2}, & R_{\text{H}} \leq r
	\end{cases}
	\label{Sphere-E(r)}
\end{equation}

\paragraph{Surface Potential Values:}
\begin{align}
	\varphi_{\text{H}} &= \frac{R_{\text{H}} \sigma_{\text{Ho}}}{\varepsilon_0 \varepsilon (1 + \kappa R_{\text{H}})} \label{Sph-phi-H} \\
	\varphi_0 &= \varphi_{\text{H}} + \frac{R_{\text{H}} \sigma_{\text{Ho}}}{\varepsilon_0 \varepsilon (R + d)} \cdot \frac{a}{2} \label{Sph-phi-0} \\
	\psi_0 &= \varphi_0 + \frac{[R_{\text{H}}^2 \sigma_{\text{Ho}} - (R + d)^2 \sigma_0] d}{\varepsilon_0 \varepsilon R (R + d)} \label{Sph-psi-0} \\
	\psi_{\text{H}} &= \psi_0 + \frac{[R_{\text{H}}^2 \sigma_{\text{Ho}} - (R+d)^2 - R^2]\sigma_0 \cdot \frac{a}{2}}{\varepsilon_0 \varepsilon R(R - \frac{a}{2})} \label{Sph-psi-H}
\end{align}

\paragraph{Effective Surface Charge:}
\begin{equation}
	\sigma_{\text{Ho}} = \frac{L_2}{L_1} \sigma_0
	\label{Sphere-sigmaHo}
\end{equation}

where:
\begin{align*}
	L_1 &= \frac{R_{\text{H}}}{1 + \kappa R_{\text{H}}} + \frac{\frac{a}{2} R_{\text{H}}}{R + d} + \frac{d R_{\text{H}}^2}{R(R + d)} + \frac{\frac{a}{2} R_{\text{H}}^2}{R(R - \frac{a}{2})} \\
	&\quad - \frac{\sinh[\kappa(R - \frac{a}{2})] R_{\text{H}}^2}{(R - \frac{a}{2}) D(R)} \\[5pt]
	L_2 &= \frac{d(R + d)}{R} + \frac{\frac{a}{2} [(R + d)^2 + R^2]}{R(R - \frac{a}{2})} \\
	&\quad - \frac{\sinh[\kappa(R - \frac{a}{2})] [(R + d)^2 + R^2]}{(R - \frac{a}{2}) D(R)}
\end{align*}

\paragraph{Electroneutrality Condition:}
\begin{equation}
	R^2 \sigma_0 + (R + d)^2 \sigma_0 + (R - \frac{a}{2})^2 \sigma_{\text{Hi}} = R_{\text{H}}^2 \sigma_{\text{Ho}}
	\label{Sphere-electroneutrality}
\end{equation}

This relation allows the determination of \( \sigma_{\text{Hi}} \) once \( \sigma_{\text{Ho}} \) is known.

\subsection{Electrolyte concentration contact values}\label{electrolyte concentration values}

To calculate the osmotic pressure through the contact theorems presented above, we need the steric pressure at contact with the shells walls, given by \cref{Pressure-Steric,Ions-Total-Concentration}, where within the inhomogeneous PB equation the point-ions distribution functions are

\begin{equation}
	g_{\scriptscriptstyle{+}}(r) = \exp(-\,e\,z\,\beta\psi(r)) \quad r\in[0,(R-a/2)]\cup [(R+d+a/2),\infty),
	\label{Ec.gmas}
\end{equation}

\noindent and

\begin{equation}
	g_{\scriptscriptstyle{-}}(r) =\exp(\,e\,z\,\beta\psi(r)) \quad  r\in[0,(R-a/2)]\cup [(R+d+a/2),\infty).
	\label{Ec.gmenos}
\end{equation}

\noindent Although \cref{Ec.gmas,Ec.gmenos} are symmetrically valid for positively or negatively charged shells, hereinafter we will assume a positive charge on the shells. Hence, the electrolyte's cations and anions become the shell's co-ion and counter-ion, respectively. We will refer to $r$ as the distance to the shell's geometrical center, for all shell's geometries. While for our calculations of the electric component of the pressure, $p_{\scriptscriptstyle_{E}}(R)$, we have taken the expressions given in \cref{The electical double layer} obtained from the analytical solution of the LPB, for the entropic component of the pressure, $p_{\scriptscriptstyle_{S}}(R)$, we will use directly \cref{Ec.gmas,Ec.gmenos}, since with a first order Taylor expansion of \cref{Ec.gmas,Ec.gmenos} any dependence of the steric pressure from the electrolyte concentration disappears, and for the low values of $ez_+\beta \psi_{\scriptscriptstyle{H}}$ and $ez_+\beta \varphi_{\scriptscriptstyle_{H}}$ considered in our calculations, the $p_{\scriptscriptstyle_{S}}(R)$ in this way obtained is virtually equal to that calculated with the second-order Taylor expansion of \cref{Ec.gmas,Ec.gmenos}. 

Unfortunately, although we have analytical expressions for $\psi(r)$ and $E(r)$ for the three geometries studied here, it was not possible to have closed analytical expressions for $p_{\scriptscriptstyle_{E}}(R)$ for the cylindrical and spherical geometries, hence the numerical results given below were obtained with a Python computer program.

In all our results, we consider a positively charged shell immersed in an aqueous symmetric electrolyte ($1:1$ or $2:2$), with an electrical relative permittivity, $\varepsilon$, of $78.5$, and an ion's size, $a$, of $\SI{4.25}{\angstrom}$.


\section{Results and discussion}\label{Res_Disc}

For the three shell geometries the general equation for their osmotic pressures is the sum of their steric, $p_{\scriptscriptstyle_{S}}(R)$, and electrical,  $p_{\scriptscriptstyle_{E}}(R)$, components. In particular, for the three geometries, from \cref{Osmotic-plates-Steric,Osmotic-Steric-Press-Cyl,Osmotic-Steric-Press-Sphere,Ec.gmas,Ec.gmenos}, $p_{\scriptscriptstyle_{S}}(R)$  is given by  

\begin{equation}
	\begin{split}
		p_{\scriptscriptstyle_{S}}(R)= KT[\rho_{\scriptscriptstyle{\alpha s}}(R-a/2)-\rho_{\scriptscriptstyle{\alpha s}}(R+d+a/2)]\quad\quad\\
		\\= 2KT\rho_{\scriptscriptstyle_{0}}[\cosh(ez_+\beta\psi_{\scriptscriptstyle_{H}})-\cosh(ez_+\beta \varphi_{\scriptscriptstyle_{H}})],\quad
	\end{split}
	\label{Pressure-Eq-Steric-MEP}
\end{equation}

\noindent where $\varphi_{\scriptscriptstyle{H}}=\psi(R+d+a/2)$ and $\psi_{\scriptscriptstyle{H}}=\psi(R-a/2)$  for the planar, cylindrical and spherical shells are given by \cref{Plates-phi-H,Plates-psi-H}, \cref{Cyl-phi-H,Cyl-psi-H}, and \cref{Sph-phi-H,Sph-psi-H}, respectively. Similarly, $p_{\scriptscriptstyle_{E}}(R)$, for the planar, cylindrical and spherical shells can be directly obtained from \cref{Osmotic-plates-Electric-Plates-E(r),Plates-E(r)},  Eqs. \ref{Osmotic-Net-Press-Cyl} and \ref{Cylinder-E(r)}, and Eqs. \ref{Osmotic-Press-Sph-Electric} and \ref{Sphere-E(r)}, respectively.

\begin{figure}[!htb]
	\begin{subfigure}{.5\textwidth}
		\centering
		\includegraphics[width=0.85\linewidth]{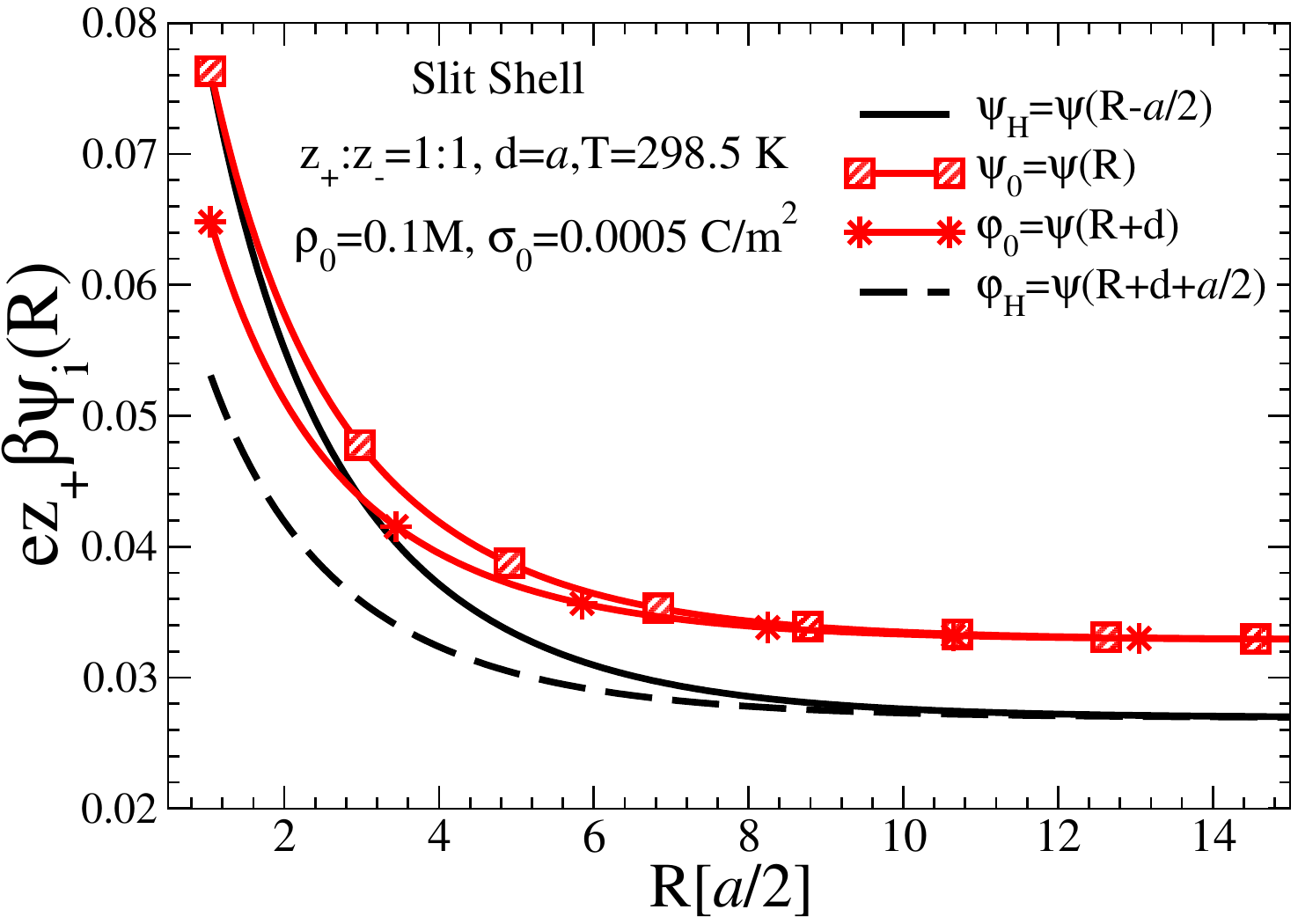}
		\caption{Low surface charge density.}
		\label{Fig.RMEP.Slit.Low.Rho}
	\end{subfigure}%
	\begin{subfigure}{.5\textwidth}
		\centering
		\includegraphics[width=0.85\linewidth]{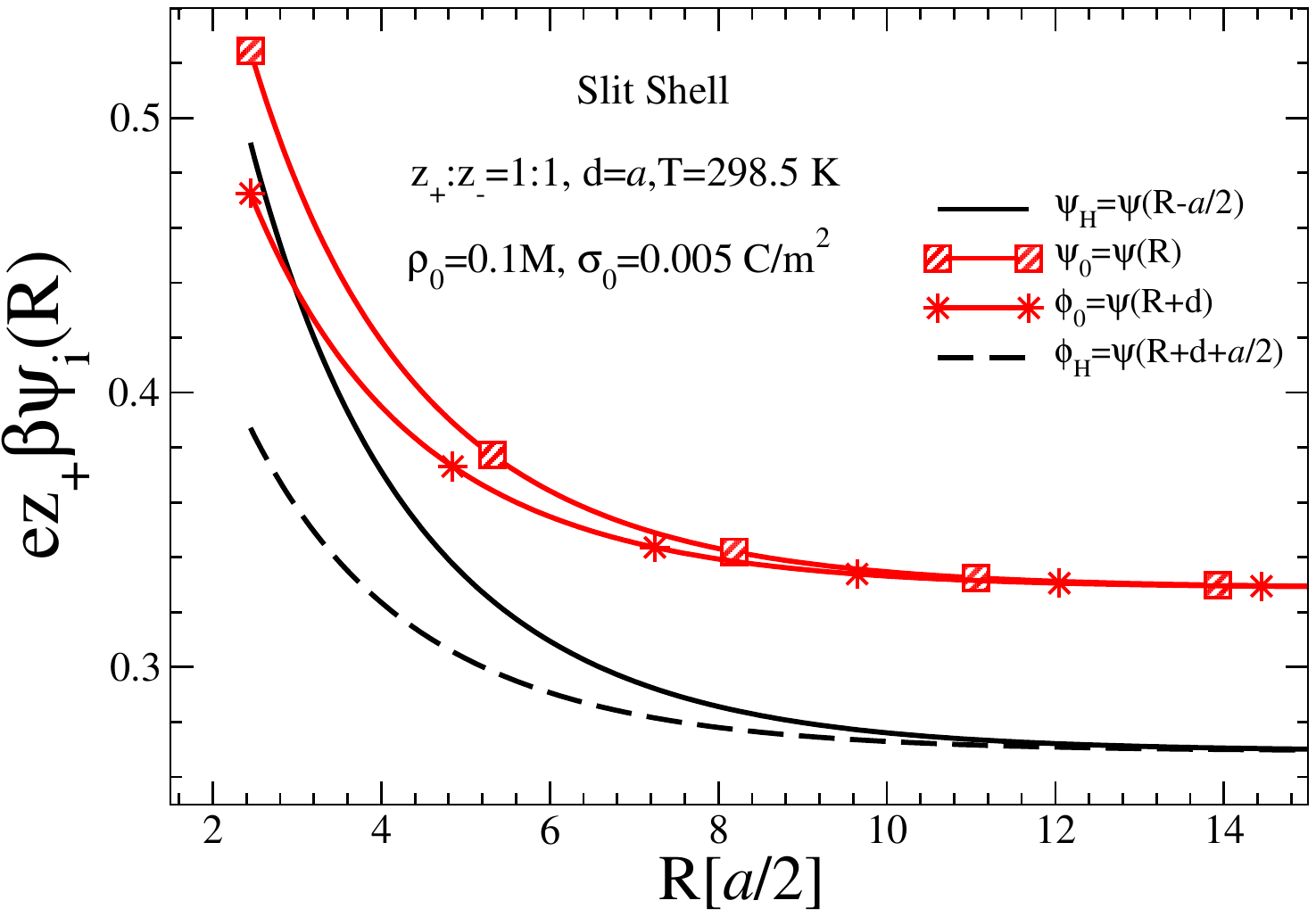}
		\caption{High charge density.}
		\label{Fig.RMEP.Slit.High.Rho}
	\end{subfigure}
	\begin{subfigure}{.5\textwidth}
		\centering
		\includegraphics[width=0.85\linewidth]{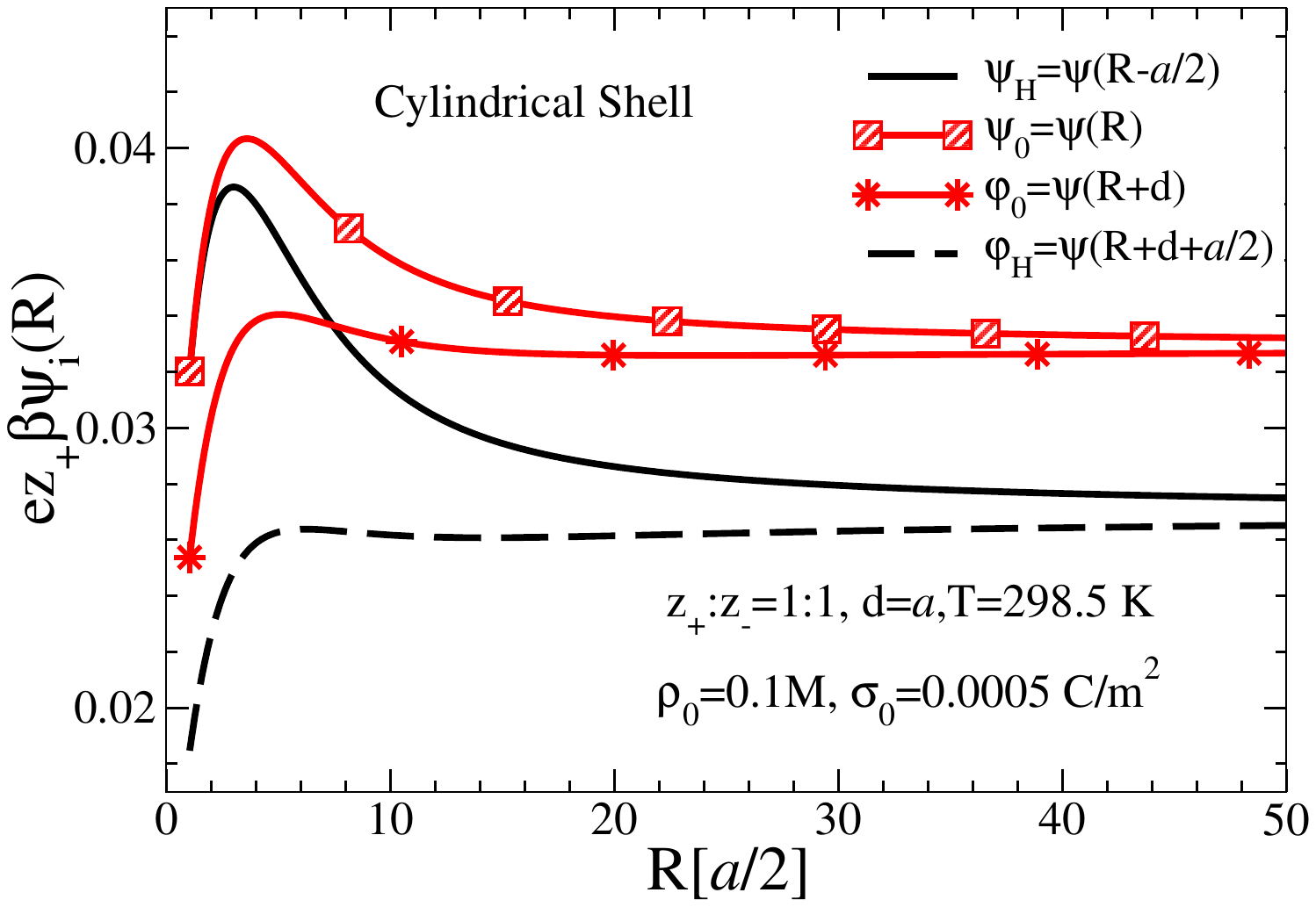}
		\caption{Low surface charge density.}
		\label{Fig.RMEP.Cyl.Low.Rho}
	\end{subfigure}%
	\begin{subfigure}{.5\textwidth}
		\centering
		\includegraphics[width=0.85\linewidth]{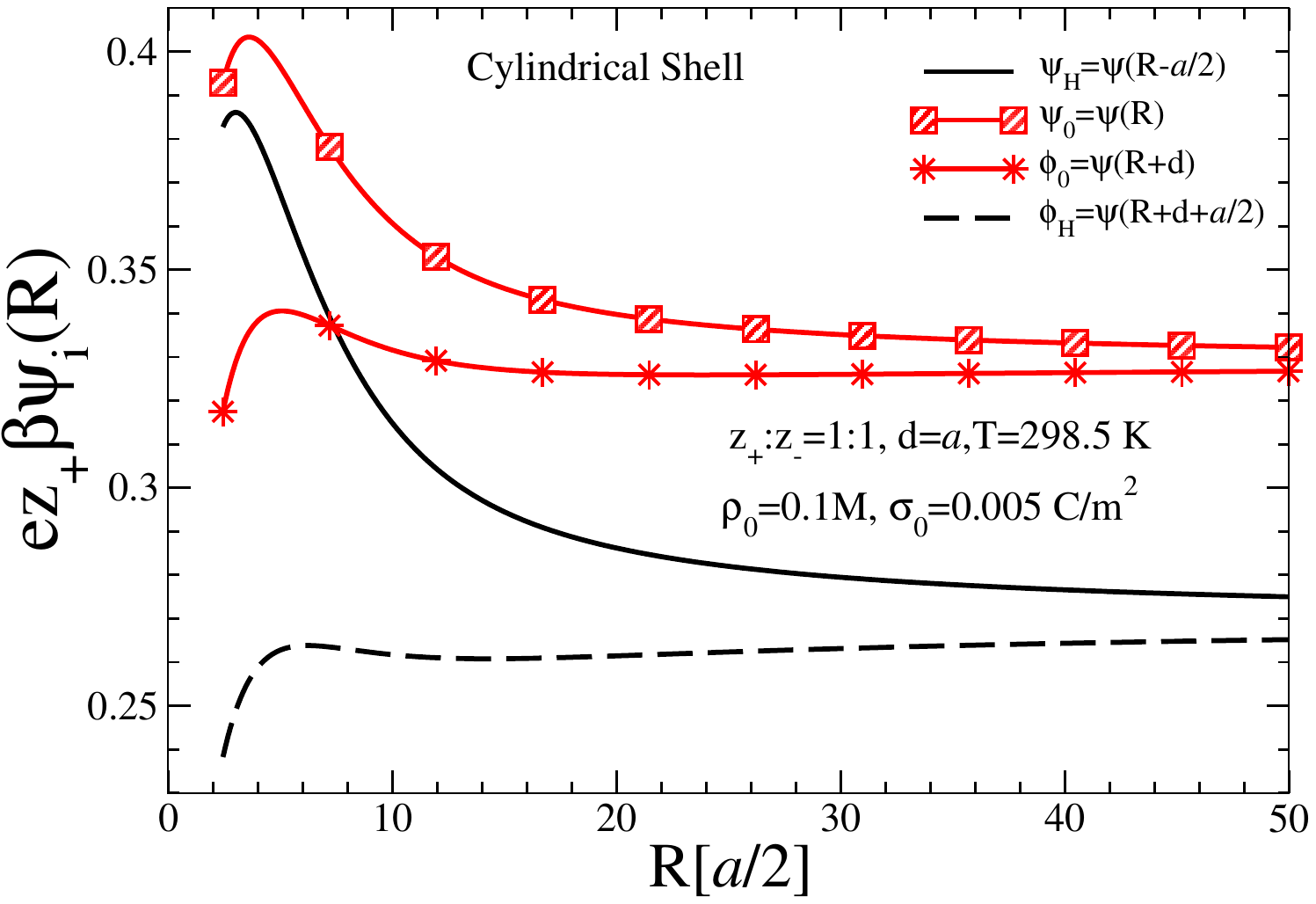}
		\caption{High surface charge density.}
		\label{Fig.RMEP.Cyl.High.Rho}
	\end{subfigure}\\
	\begin{subfigure}{.5\textwidth}
		\centering
		\includegraphics[width=0.85\linewidth]{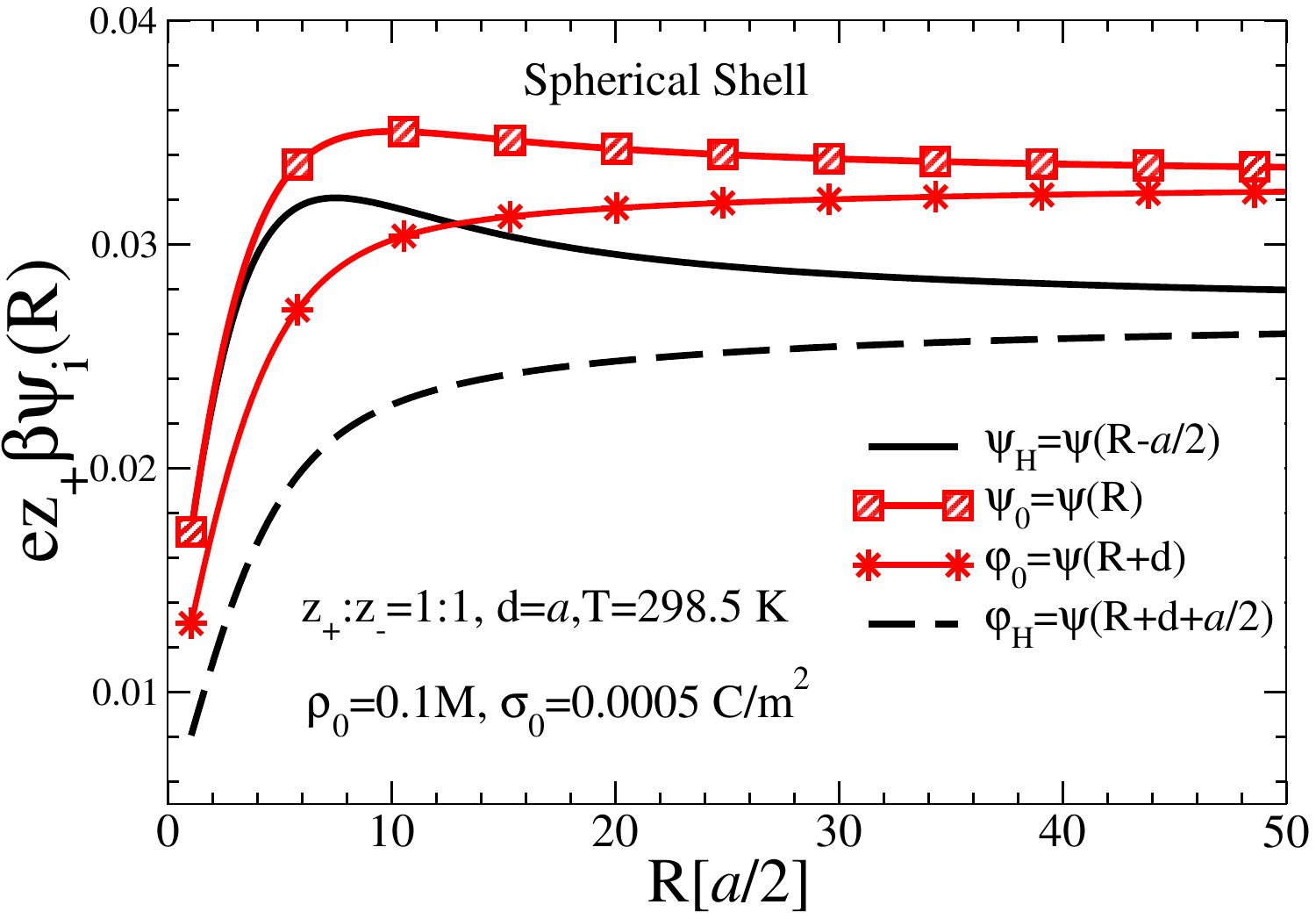}
		\caption{Low surface charge density.}
		\label{Fig.RMEP.Sphere.Low.Rho}
	\end{subfigure}
	\begin{subfigure}{.5\textwidth}
		\centering
		\includegraphics[width=0.85\linewidth]{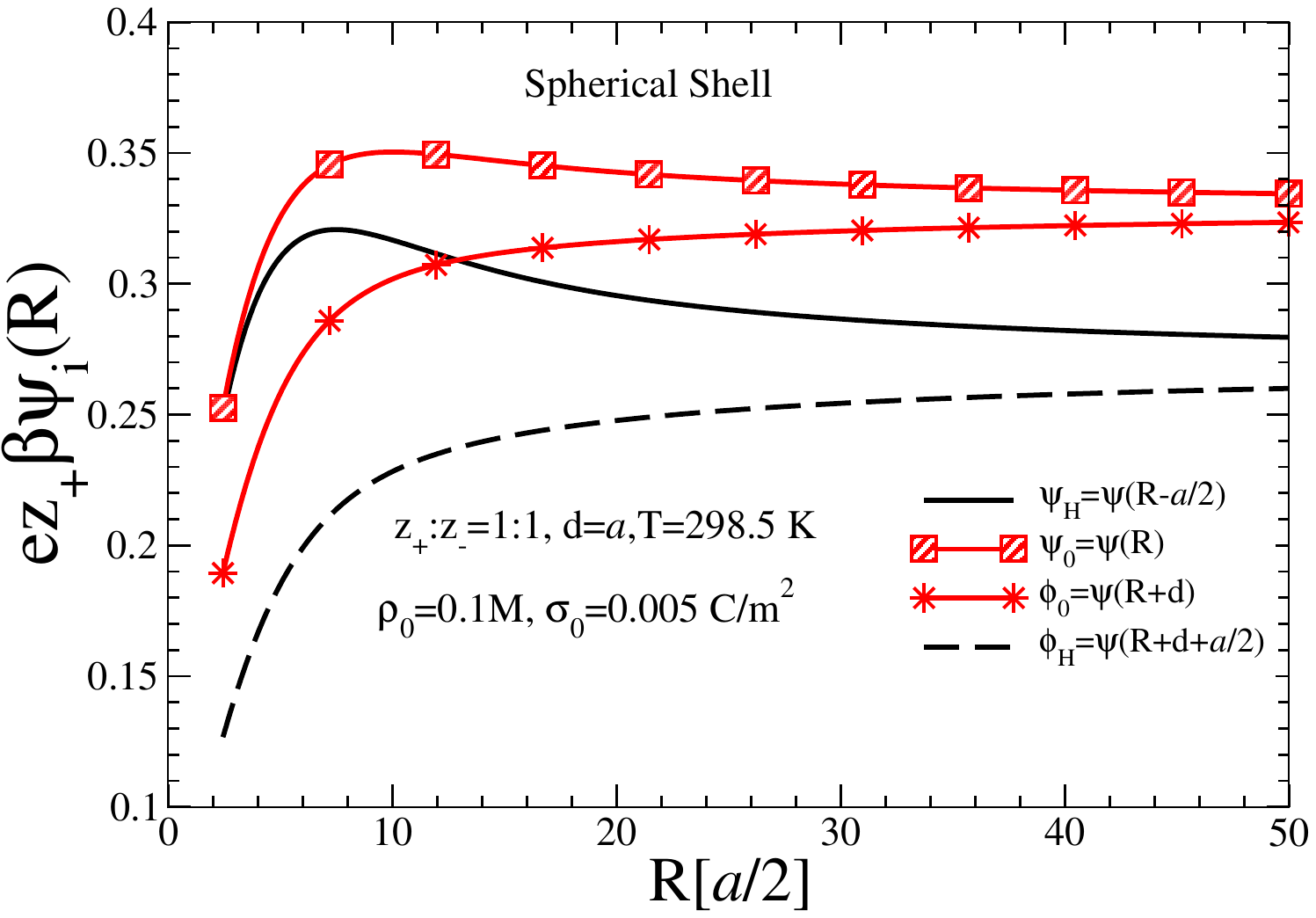}
		\caption{High surface charge density.}
		\label{Fig.RMEP.Sphere.High.Rho}
	\end{subfigure}
	\caption{Reduced mean electrostatic potential, $ez_+\beta\psi(r)$, at $r=R-a/2$, $r=R$, $r=R+d$ and $r=R+d+a/2$, defined as $ez_+\beta\psi_{\scriptscriptstyle{H}}, ez_+\beta\psi_{\scriptscriptstyle{0}}, ez_+\beta\varphi_{\scriptscriptstyle{0}}$, and $ez_+\beta\varphi_{\scriptscriptstyle{H}}$, respectively, as a function of the inner radius of the shells, $R$. In all cases, the bulk electrolyte concentration is $\rho_{\scriptscriptstyle_{0}}=0.1M$, the dielectric constant is $\varepsilon=78.5$, and $a=\SI{4.25}{\angstrom}$.}
	\label{Fig.RMEP.General}
\end{figure}
 
\subsection{The linear Poisson-Boltzmann approximation}\label{The LPB approximation}

Since our analysis relies on the linearized Poisson–Boltzmann (LPB) equation, the reduced mean electrostatic potential (MEP), $ez_+\beta \psi(r)$, must remain below unity at the shell–electrolyte interfaces, i.e., $ez_+\beta \psi_{\scriptscriptstyle{H}} < 1$ and $ez_+\beta \varphi_{\scriptscriptstyle{H}} < 1$, where $\psi_{\scriptscriptstyle{H}} = \psi(R - a/2)$ and $\varphi_{\scriptscriptstyle{H}} = \psi(R + d + a/2)$.

\Cref{Fig.RMEP.General} shows $ez_+\beta \psi_{\scriptscriptstyle{H}}$, $ez_+\beta \psi_{\scriptscriptstyle{0}}$, $ez_+\beta \varphi_{\scriptscriptstyle{0}}$, and $ez_+\beta \varphi_{\scriptscriptstyle{H}}$ as functions of shell radius $R$ for different parameter sets. In particular, \cref{Fig.RMEP.Cyl.Low.Rho,Fig.RMEP.Slit.Low.Rho,Fig.RMEP.Sphere.Low.Rho} correspond to low electrolyte concentration ($\rho_{\scriptscriptstyle{0}}=0.001\,\mathrm{M}$) and surface charge ($\sigma_{\scriptscriptstyle{0}}=0.0005\,\mathrm{C/m^2}$). Panels \cref{Fig.RMEP.Cyl.High.Rho,Fig.RMEP.Slit.High.Rho,Fig.RMEP.Sphere.High.Rho} show results for $\rho_{\scriptscriptstyle{0}} = 0.1\,\mathrm{M}$ and $\sigma_{\scriptscriptstyle{0}} = 0.005\,\mathrm{C/m^2}$.

As expected, increasing $\sigma_{\scriptscriptstyle{0}}$ proportionally increases all MEPs. Yet, even at high salt concentration and $\sigma_{\scriptscriptstyle{0}} = 0.005\,\mathrm{C/m^2}$, the LPB condition $ez_+\beta \psi_{\scriptscriptstyle{H}}<1$ and $ez_+\beta \varphi_{\scriptscriptstyle{H}} < 1$ remains satisfied.

Interestingly, the LPB approximation remains mathematically valid at molar concentrations as high as $2\,\mathrm{M}$, provided $\sigma_{\scriptscriptstyle{0}}$ is low. This is possible due to the point-ion nature of our model. Moreover, despite its simplicity, the LPB solution yields good agreement with results from integral equations for slit shells in a restricted primitive model (RPM) electrolyte~\cite{Lozada_1990-I,Lozada_1990-II}, and even for properties like the $\zeta$-potential of nano-electrodes, for $2{:}2$ electrolytes at sufficiently low $\rho_{\scriptscriptstyle{0}}$ or $\sigma_{\scriptscriptstyle{0}}$~\cite{McQuarrie_StatMech,Degreve_1993,Degreve_1995}.

In general, thinner electric double layers (EDLs)—caused by higher salt concentration or thicker shell walls—reduce $ez_+\beta \psi_{\scriptscriptstyle{H}}$ and $ez_+\beta \varphi_{\scriptscriptstyle{H}}$, improving the validity of the LPB. Conversely, lower $\rho_{\scriptscriptstyle{0}}$ or thinner walls increase EDL thickness and thus MEP values. All such effects are accounted for in our parameter scans, though not all are shown in \cref{Fig.RMEP.General}.

In our model, $\sigma_{\scriptscriptstyle{0}}$ should typically remain below $0.005\,\mathrm{C/m^2}$ for accuracy. Higher $\rho_{\scriptscriptstyle{0}}$, higher $T$, and/or thicker shells improve the LPB’s applicability. For $2{:}2$ salts, the LPB is valid for $\sigma_{\scriptscriptstyle{0}} \lesssim 0.0005\,\mathrm{C/m^2}$; at higher $\rho_{\scriptscriptstyle{0}}$, this threshold rises to $\sim 0.005\,\mathrm{C/m^2}$. Additionally, the LPB becomes more accurate with increasing $R$.

In \cref{Fig.RMEP.General}, we observe that MEPs at the shell boundaries ($\psi_{\scriptscriptstyle{H}}, \psi_{\scriptscriptstyle{0}}, \varphi_{\scriptscriptstyle{0}}, \varphi_{\scriptscriptstyle{H}}$) generally increase with decreasing $R$, before eventually declining. As expected: $\varphi_{\scriptscriptstyle{H}}(R) \leq \varphi_{\scriptscriptstyle{0}}(R) \leq \psi_{\scriptscriptstyle{0}}(R)$ and $\psi_{\scriptscriptstyle{0}}(R) \geq \psi_{\scriptscriptstyle{H}}(R) \geq \psi_{\scriptscriptstyle{d}}(R)$, where $\psi_{\scriptscriptstyle{d}} = \psi(r=0)$.

Note that all MEPs are functions of $R$. For cylindrical and spherical shells at low $\rho_{\scriptscriptstyle{0}}$, $\psi_{\scriptscriptstyle{H}}(R)$ exhibits maxima due to nonlinear field variations and confinement effects (VLEC)~\cite{Adrian-JML-2023}. In contrast, for the slit-shell, $\psi_{\scriptscriptstyle{H}}(R)$ and $\varphi_{\scriptscriptstyle{H}}(R)$ decrease monotonically with $R$.

\begin{figure}[!hbt]
	\begin{subfigure}{.5\textwidth}
		\centering
		\includegraphics[width=1.1\linewidth]{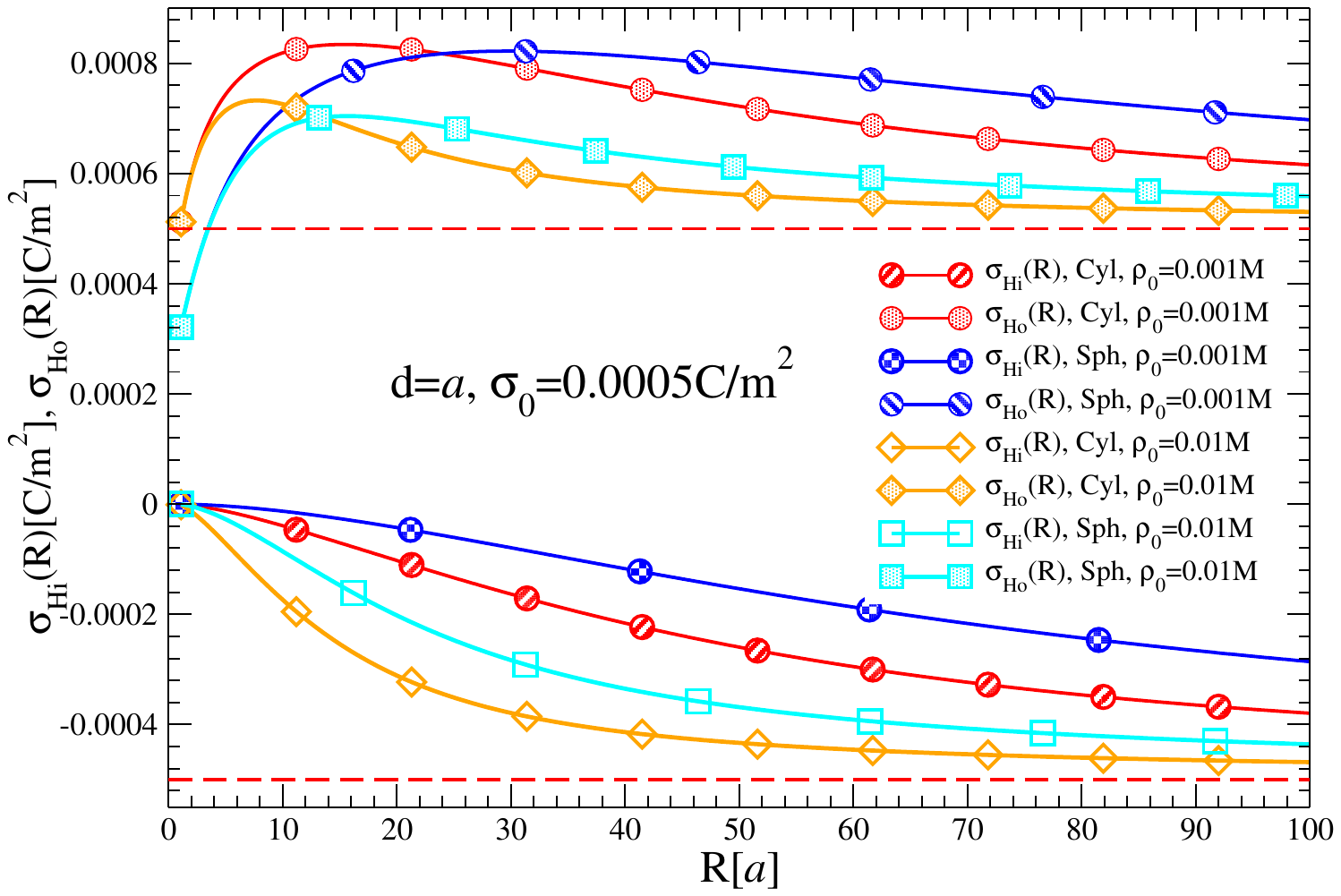}
		\caption{$\sigma_{\scriptscriptstyle{Hi}}(R)$ and $\sigma_{\scriptscriptstyle{Ho}}(R)$, at very low $\sigma_{\scriptscriptstyle{0}}$.}
		\label{Fig.sigma_Hi-sigmaHo-Cyl-Sphere-s0.0005_rho0.001_and_rho0.01}
	\end{subfigure}
	\begin{subfigure}{.5\textwidth}
		\centering
		\includegraphics[width=1.1\linewidth]{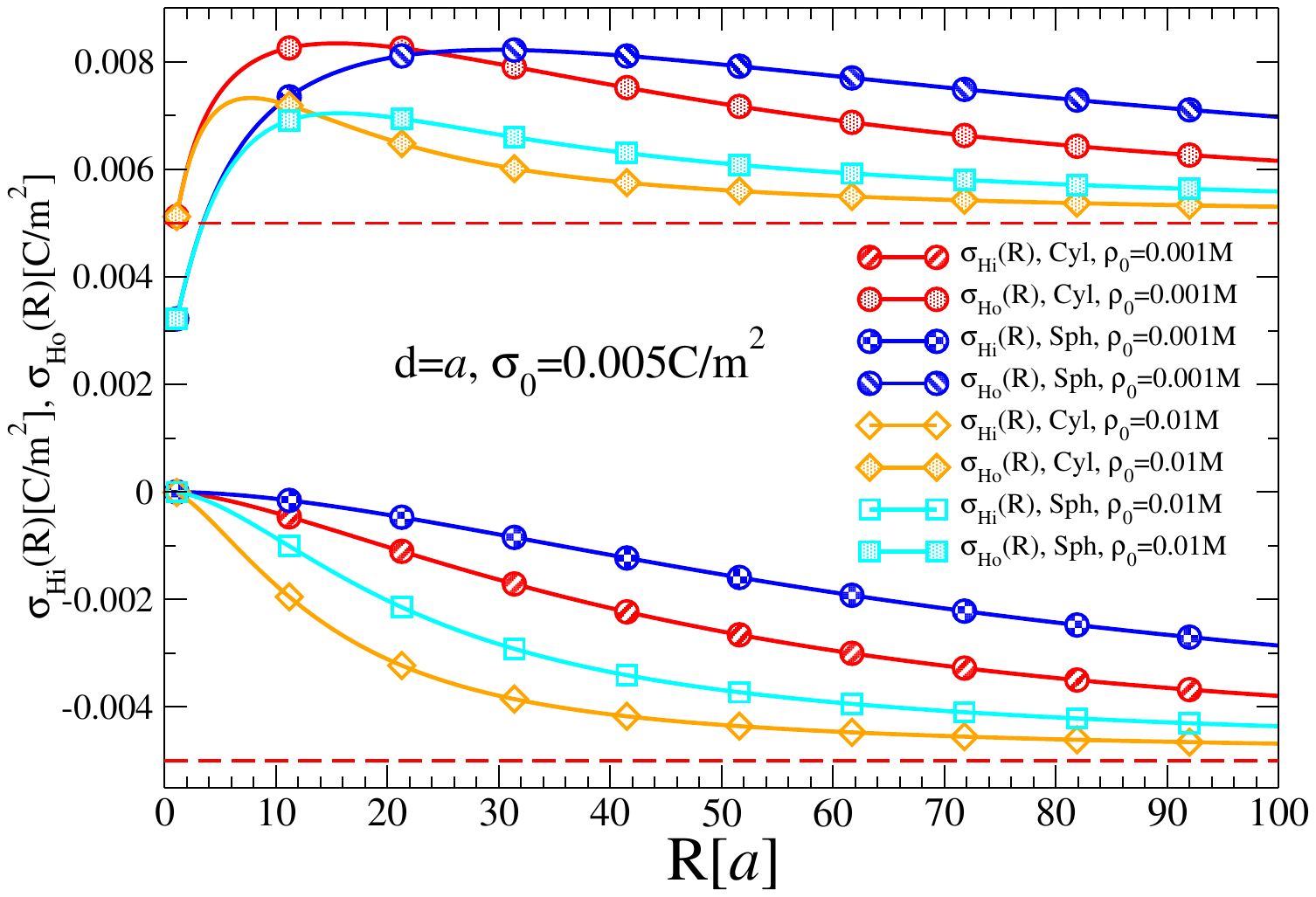}
		\caption{$\sigma_{\scriptscriptstyle{Hi}}(R)$ and $\sigma_{\scriptscriptstyle{Ho}}(R)$, at low $\sigma_{\scriptscriptstyle{0}}$.}
		\label{Fig.sigma_Hi-sigmaHo-Cyl-Sphere-s0.005_rho0.001_and_rho0.01}
	\end{subfigure}\\
	\caption{Effective surface charge densities, $\sigma_{\scriptscriptstyle{Hi}}(R)$ and $\sigma_{\scriptscriptstyle{Ho}}(R)$, at $r = R - a/2$ and $r = R + d + a/2$, respectively, for cylindrical and spherical shells. Calculations are shown for two 1:1 electrolyte concentrations: $\rho_{\scriptscriptstyle{0}} = 0.001\,\mathrm{M}$ and $0.01\,\mathrm{M}$, at $T = 298.15\,\mathrm{K}$, with $d = a$ and $\varepsilon = 78.5$. Horizontal dashed lines correspond to $\sigma_{\scriptscriptstyle{Ho}} = \sigma_{\scriptscriptstyle{0}}$ and $\sigma_{\scriptscriptstyle{Hi}} = -\sigma_{\scriptscriptstyle{0}}$.}
	\label{Fig.sigma-Hi_and_Ho}
\end{figure}

To further understand these nonlinearities, we examine the induced surface charge densities $\sigma_{\scriptscriptstyle{Hi}}(R)$ and $\sigma_{\scriptscriptstyle{Ho}}(R)$ in \cref{Fig.sigma-Hi_and_Ho}, calculated for $d = a$, $T = 298.15\,\mathrm{K}$, and two concentrations: $0.001\,\mathrm{M}$ and $0.01\,\mathrm{M}$.

In all cases, $\sigma_{\scriptscriptstyle{Hi}}(R)$ is a monotonically decreasing function of $R$, remaining above $-\sigma_{\scriptscriptstyle{0}}$. Thus, $\sigma_{\scriptscriptstyle{Hi}} + \sigma_{\scriptscriptstyle{0}} > 0$, consistent with VLEC. However, electroneutrality is always satisfied due to \cref{Electroneutrality-condition-general2}. Specifically, as $R \to \infty$, $\sigma_{\scriptscriptstyle{Hi}} \to -\sigma_{\scriptscriptstyle{0}}$, eliminating VLEC.

VLEC is more pronounced and longer-ranged for spherical shells due to their greater internal volume and weaker confinement, which reduces counter-ion adsorption at $r = R - a/2$. For the slit-shell, $\sigma_{\scriptscriptstyle{Hi}}(R)$ decreases fastest with $R$, and its VLEC vanishes rapidly. Among the three geometries, the slit-shell exhibits the highest counter-ion adsorption.

For a higher electrolyte concentration ($\rho_{\scriptscriptstyle{0}} = 0.01\,\mathrm{M}$), the electrostatic double layer (EDL) inside and outside the shells becomes narrower. As a result, the induced charge densities $\sigma_{\scriptscriptstyle{Hi}}(R)$ and $\sigma_{\scriptscriptstyle{Ho}}(R)$ become more negative and decay more rapidly with increasing $R$, compared to the case with $\rho_{\scriptscriptstyle{0}} = 0.001\,\mathrm{M}$. This behavior indicates a weaker voltage-like electrostatic confinement (VLEC) at higher electrolyte concentrations.

In contrast, increasing the bare surface charge density $\sigma_{\scriptscriptstyle{0}}$ amplifies the magnitude of $\sigma_{\scriptscriptstyle{Hi}}(R)$ and $\sigma_{\scriptscriptstyle{Ho}}(R)$ without altering their qualitative behavior, as seen in \cref{Fig.sigma_Hi-sigmaHo-Cyl-Sphere-s0.0005_rho0.001_and_rho0.01,Fig.sigma_Hi-sigmaHo-Cyl-Sphere-s0.005_rho0.001_and_rho0.01}. The function $\sigma_{\scriptscriptstyle{Ho}}(R)$ exhibits an absolute maximum for both cylindrical and spherical shells, with the location of this peak depending on the electrolyte concentration and geometry.

Notably, $\sigma_{\scriptscriptstyle{Ho}}(R)$ can significantly exceed the bare surface charge density, demonstrating the phenomenon of \textit{overcharging}, i.e., $\sigma_{\scriptscriptstyle{Ho}}(R) > \sigma_{\scriptscriptstyle{0}}$ for most $R$, except in very narrow spherical shells at high electrolyte concentrations (e.g., $\rho_{\scriptscriptstyle{0}} = 0.01\,\mathrm{M}$), as shown in \cref{Fig.sigma-Hi_and_Ho}.

Overcharging has been widely reported for planar electrodes immersed in PM electrolytes~\cite{Jimenez_2004_Feb,Guerrero_2010,Gonzalez-Calderon-EPJ-2021,Martens-PRL-2024,Samyabrata-overcharging-2024}, but is absent in RPM or point-ion models. It has also been observed in cylindrical nanopores filled with PM electrolytes, where macroions are excluded from the pore~\cite{gonzalez-tovar-2017,Gonzalez-overcharging-cyl-macroions-2022}. In such systems, the effect is attributed to the adsorption of oppositely charged macroions and their counterions, which we refer to as \textit{steric overcharging}, originating from the entropic drive for macroion adsorption.

By contrast, the overcharging observed in \cref{Fig.sigma_Hi-sigmaHo-Cyl-Sphere-s0.0005_rho0.001_and_rho0.01,Fig.sigma_Hi-sigmaHo-Cyl-Sphere-s0.005_rho0.001_and_rho0.01} is entirely due to confinement. In our point-ion model, configurational entropy effects are negligible and confined to a Stern-layer correction. As confinement is relaxed, this effect disappears: 
$\lim_{R \to \infty} \sigma_{\scriptscriptstyle{Ho}}(R) = \sigma_{\scriptscriptstyle{0}}$.

Furthermore, decreasing the ionic diameter $a$ in the Stern correction enhances overcharging, while increasing $a$ suppresses it. We term this mechanism \textit{confinement overcharging} (CO).

The maxima in $\sigma_{\scriptscriptstyle{Ho}}(R)$ result from a competition between increasing wall surface charge (since $\sigma_{\scriptscriptstyle{0}}$ is constant) and geometric effects related to the nonlinear radial dependence of the electric field $E(r)$ (see \cref{Cylindrical-shell,Spherical-shell,Electroneutrality-condition-general2,Electrical-field-balance}).

In slit-shell geometries, both $\sigma_{\scriptscriptstyle{Hi}}(R)$ and $\sigma_{\scriptscriptstyle{Ho}}(R)$ are monotonically decreasing with $R$, as shown further below. Nevertheless, $\sigma_{\scriptscriptstyle{Ho}}(R) > \sigma_{\scriptscriptstyle{0}}$ for all $R$, confirming the presence of CO. This arises because the total shell surface charge is constant, and $E(r)$ depends linearly on $\sigma_{\scriptscriptstyle{Hi}}(R)$ (see \cref{Slit-shell,Slit-electroneutrality}). Since steric overcharging is absent in RPM systems near solid electrodes, the overcharging observed here is purely due to confinement. In the large-radius limit:
\[
\lim_{R \to \infty} \sigma_{\scriptscriptstyle{Ho}}(R) = \sigma_{\scriptscriptstyle{0}}, \quad \text{and} \quad \lim_{R \to \infty} \sigma_{\scriptscriptstyle{Hi}}(R) = -\sigma_{\scriptscriptstyle{0}}.
\]

\begin{figure}[!hbt]
	\begin{subfigure}{.5\textwidth}
		\centering
		\includegraphics[width=1.1\linewidth]{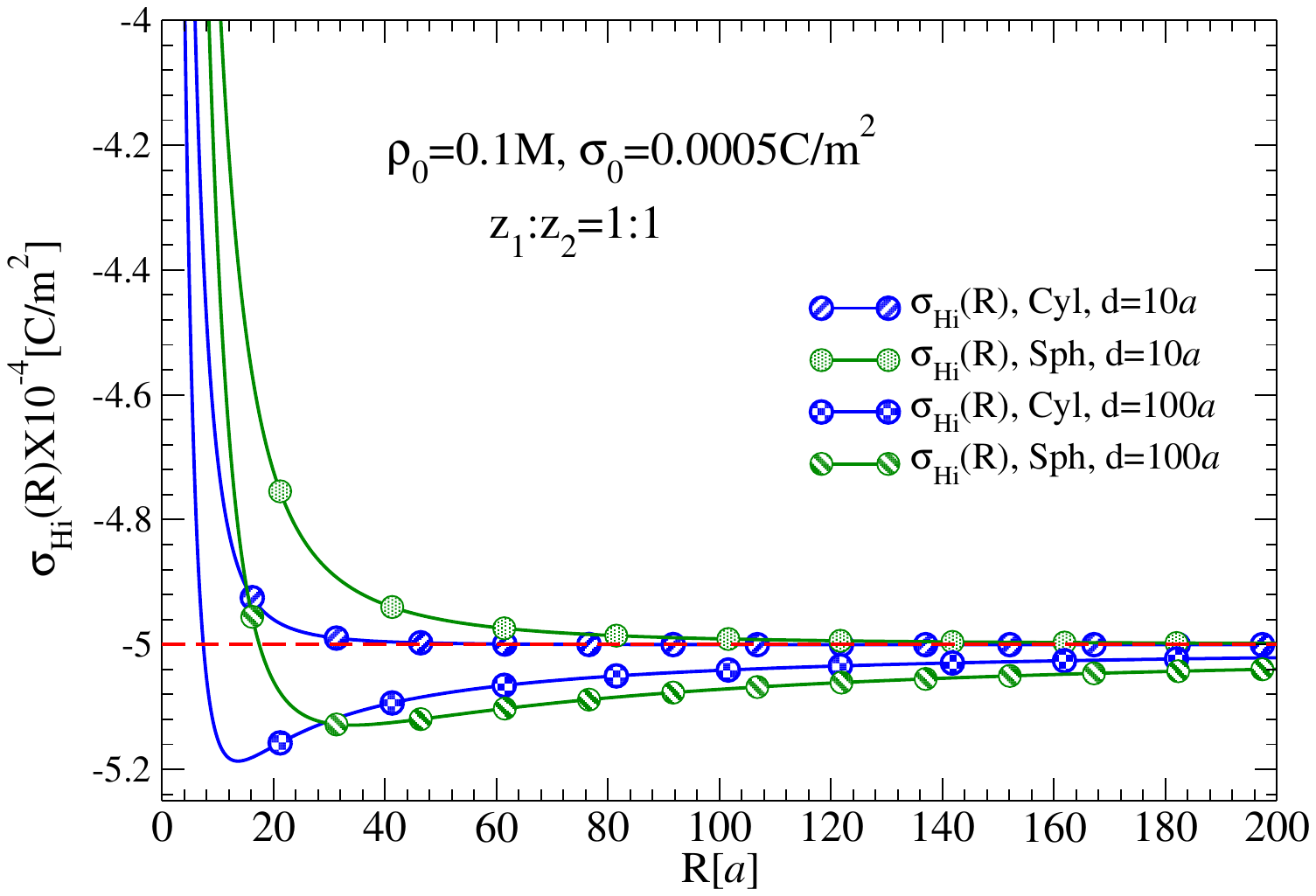}
		\caption{CCR in cylindrical and spherical shells: wall thickness dependence.}
		\label{Fig.CCR-sigma_Hi-Cyl-Sph-d10-d100-s0.0005-z1}
	\end{subfigure}
	\begin{subfigure}{.5\textwidth}
		\centering
		\includegraphics[width=1.1\linewidth]{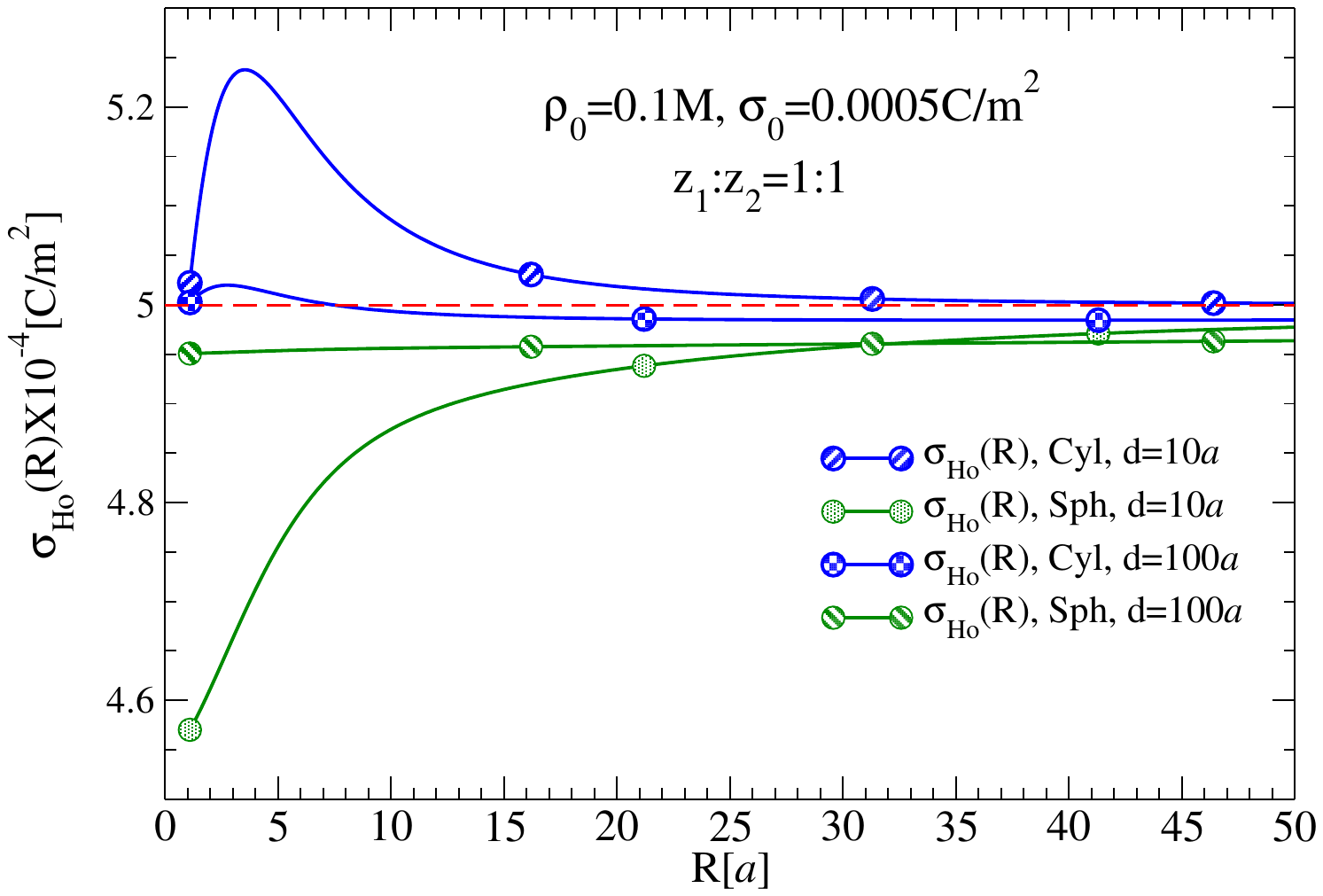}
		\caption{CO in cylindrical and spherical shells: wall thickness dependence.}
		\label{Fig.CCR-sigma_Ho-Cyl-Sph-d10-d100-s0.0005-z1}
	\end{subfigure}\\
	\begin{subfigure}{.5\textwidth}
		\centering
		\includegraphics[width=1.1\linewidth]{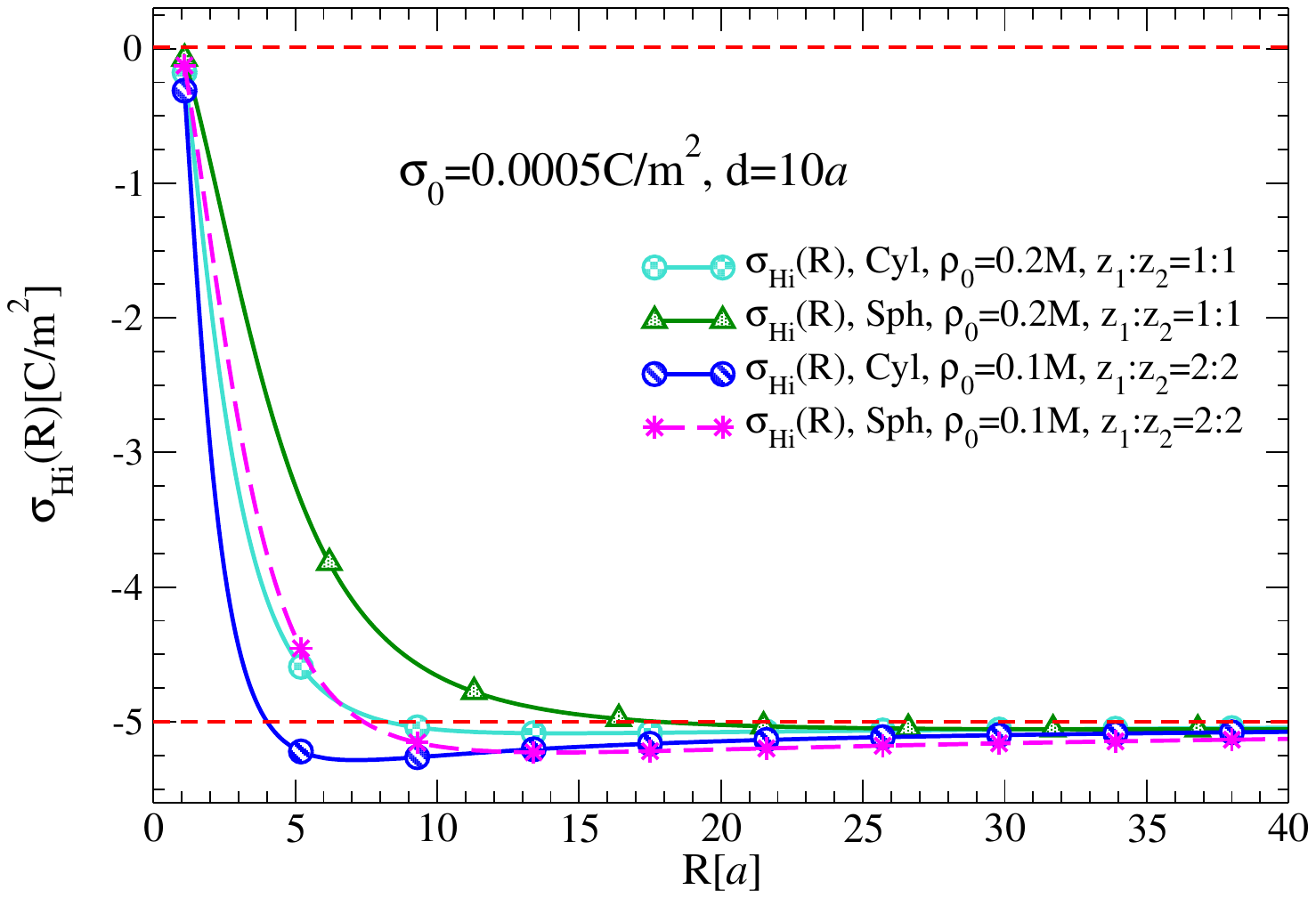}
		\caption{CCR: salt valence and concentration dependence.}
		\label{Fig.Overcharging-Charge-reversal-sigma_Hi-Cyl-Sphere-s0.0005_rho0.1-0.2_d10-z1-z2}
	\end{subfigure}
	\begin{subfigure}{.5\textwidth}
		\centering
		\includegraphics[width=1.1\linewidth]{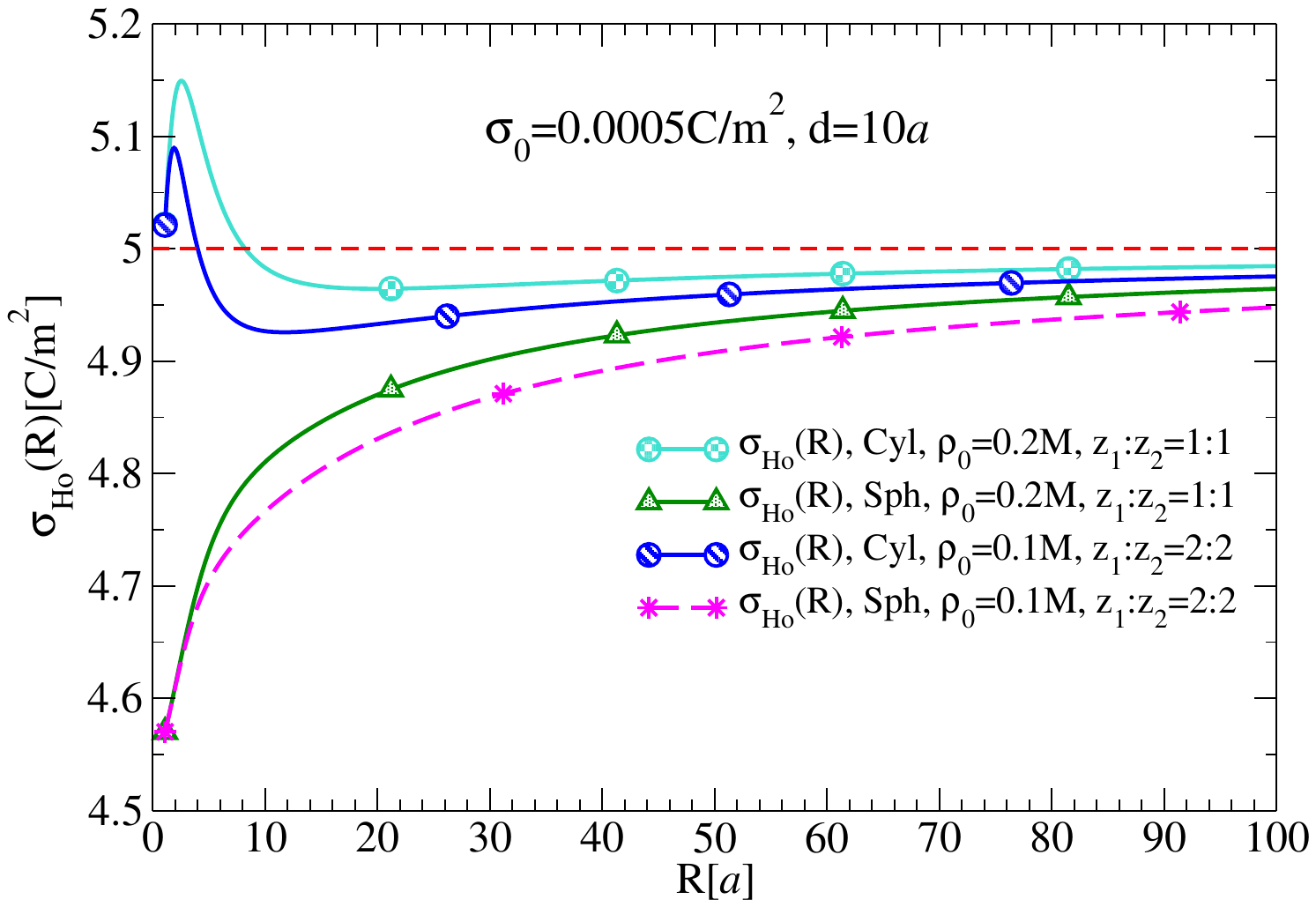}
		\caption{CO: salt valence and concentration dependence.}
		\label{Fig.Overcharging-Charge-reversal-sigma_Ho-Cyl-Sphere-s0.0005_rho0.1-0.2_d10-z1-z2}
	\end{subfigure}
	\caption{Effective surface charge densities $\sigma(r)$ at $r = R-a/2$ and $r = R+d+a/2$, i.e., $\sigma_{\scriptscriptstyle{Hi}}(R)$ and $\sigma_{\scriptscriptstyle{Ho}}(R)$, in cylindrical and spherical shells. (a) and (b): wall thickness dependence. (c) and (d): salt valence and concentration dependence. The dashed red lines correspond to $\sigma_{\scriptscriptstyle{Hi}} = -\sigma_{\scriptscriptstyle{0}}$ and $\sigma_{\scriptscriptstyle{Ho}} = \sigma_{\scriptscriptstyle{0}}$. Parameters: $T = 298.15\,\mathrm{K}$, $a = \SI{4.25}{\angstrom}$, $\varepsilon = 78.5$, $d = 10a$, $\sigma_{\scriptscriptstyle{0}} = 0.0005\,\mathrm{C/m^2}$.}
	\label{Fig.CCR_and_CO}
\end{figure}

\begin{figure}[!hbt]
	\begin{subfigure}{.5\textwidth}
		\centering
		\includegraphics[width=1.1\linewidth]{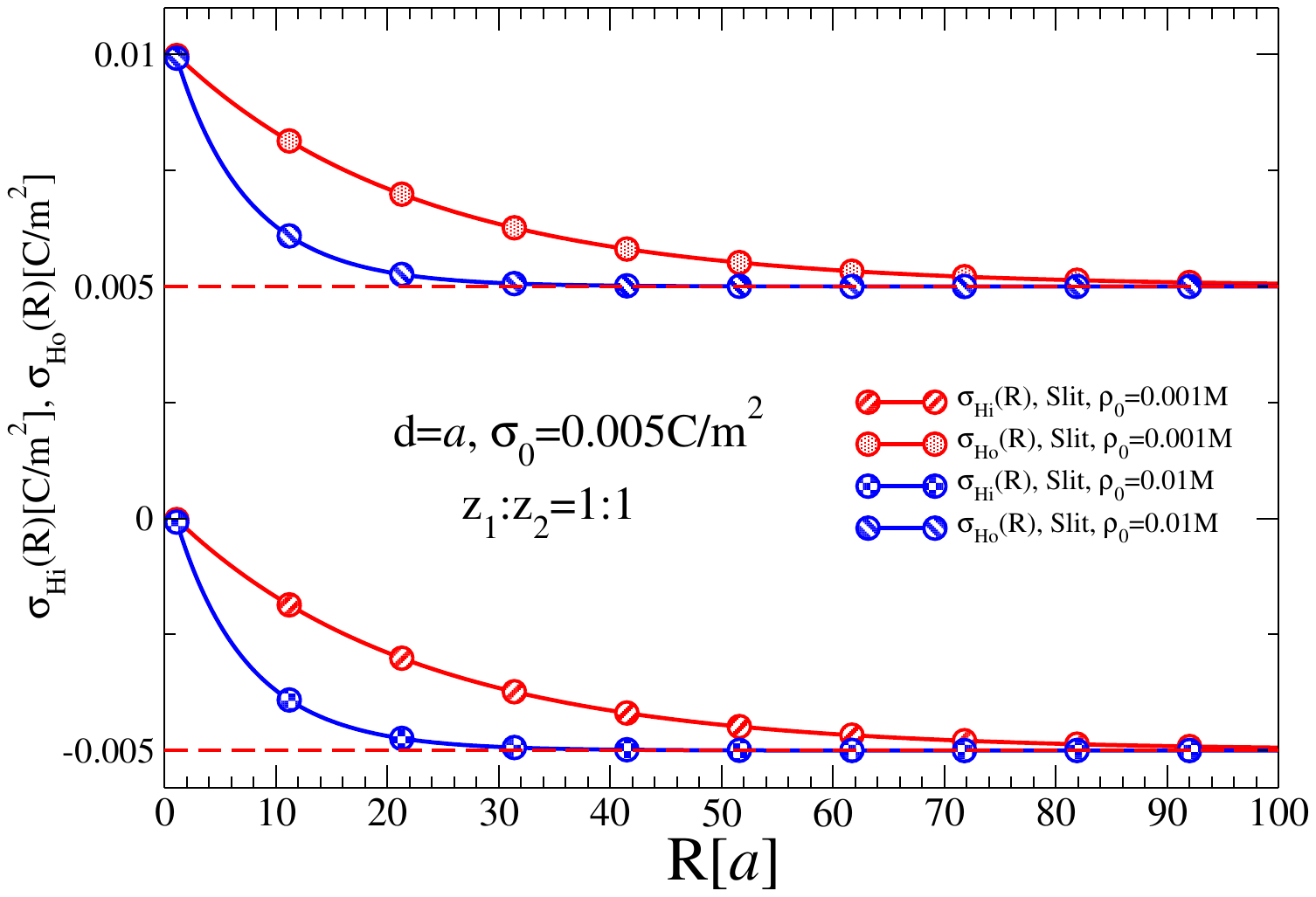}
		\caption{CO in slit shells: thin walls.}
		\label{Fig.sigma_Hi-sigmaHo-Slit-d1_s0.005_rho0.001_and_rho0.01_all-data}
	\end{subfigure}
	\begin{subfigure}{.5\textwidth}
		\centering
		\includegraphics[width=1.1\linewidth]{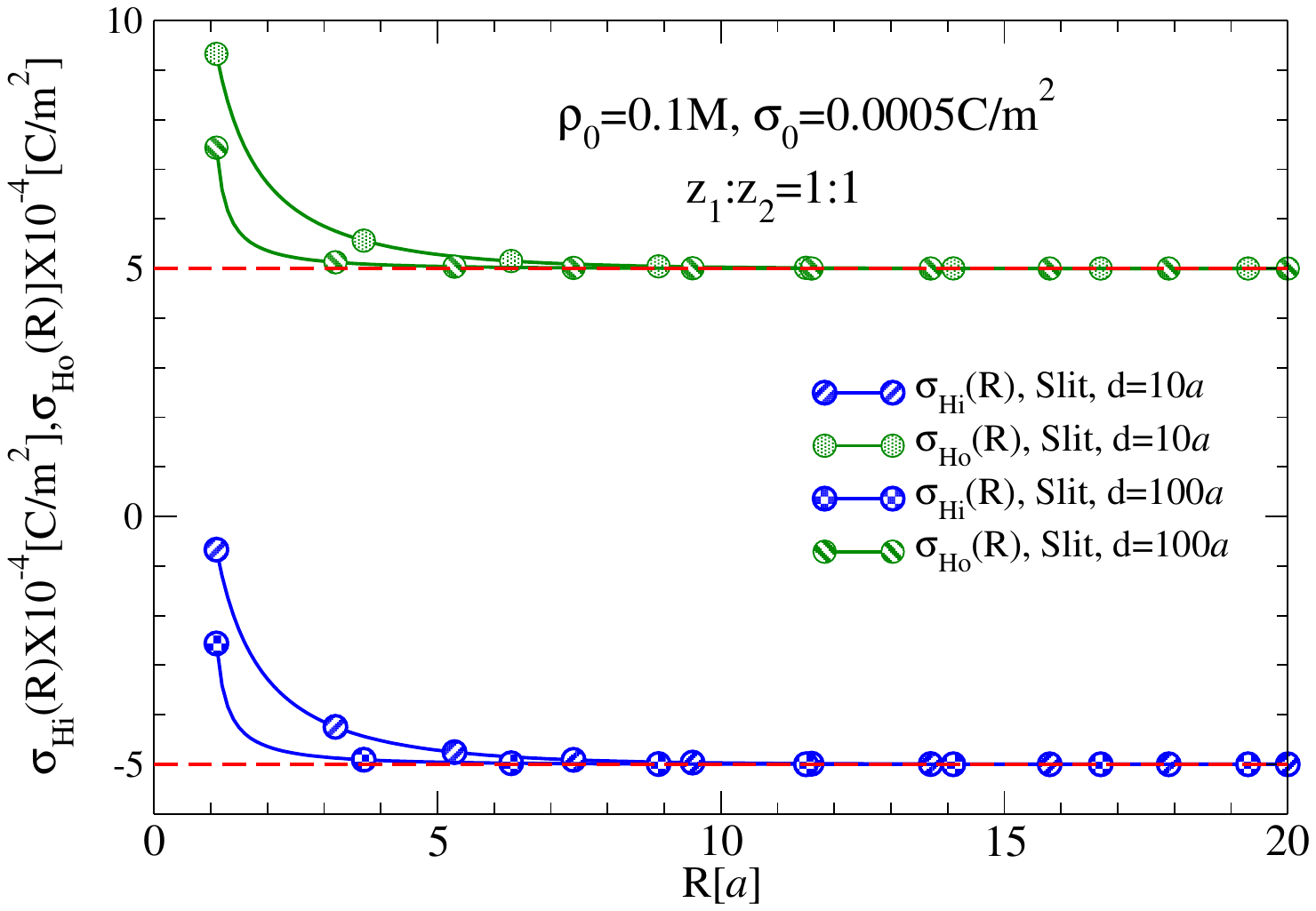}
		\caption{CO in slit shells: thicker walls.}
		\label{Fig.sigma_Hi-sigmaHo-Slit-d10-d100_z1_s0.0005_rho0.1_all-data}
	\end{subfigure}
	\caption{Confinement overcharging (CO) in slit-shell geometries. No CCR is observed. Dashed red lines indicate $\sigma_{\scriptscriptstyle{Hi}} = -\sigma_{\scriptscriptstyle{0}}$ and $\sigma_{\scriptscriptstyle{Ho}} = \sigma_{\scriptscriptstyle{0}}$. Other parameters as in \cref{Fig.CCR_and_CO}.}
	\label{Fig.CO-slit-shell}
\end{figure}

At this point, we draw attention to another confinement-induced phenomenon unique to cylindrical and spherical geometries: for certain combinations of model parameters, the induced surface charge at the inner wall, $\sigma_{\scriptscriptstyle{Hi}}(R)$, can exceed in magnitude and reverse the sign of the bare surface charge, i.e., $\sigma_{\scriptscriptstyle{Hi}}(R) + \sigma_{\scriptscriptstyle{0}} < 0$ (see \cref{Fig.CCR-sigma_Hi-Cyl-Sph-d10-d100-s0.0005-z1,Fig.Overcharging-Charge-reversal-sigma_Hi-Cyl-Sphere-s0.0005_rho0.1-0.2_d10-z1-z2}). We refer to this effect as \textit{Confinement Charge Reversal} (CCR).

Although CCR was previously observed in cylindrical shells~\cite{Gonzalez-overcharging-cyl-macroions-2022}, it was associated with steric macroion adsorption outside the shell and not recognized as a purely confinement-induced effect, as it is here. In contrast, CCR reported here arises solely from ion confinement and is a new phenomenon.

Charge reversal has been shown \emph{not} to occur in point-ion electrolytes adjacent to charged walls~\cite{Kjellander-charge-inversion-1998,Gonzalez_1985,Gonzalez_1989,gonzalez-tovar-2017,Wu-giant-charge-reversal-2017}. In previous cases involving steric overcharging, internal charge reversal was primarily driven by entropy. Here, CCR is observed under specific conditions: thick shell walls ($d \geq 5a$), electrolyte concentrations $\rho_{\scriptscriptstyle{0}} \geq 0.1\,\mathrm{M}$, and both low and high values of $\sigma_{\scriptscriptstyle{0}}$.

CCR is generally stronger in cylindrical shells at small radii and in spherical shells at larger radii. This is shown in \cref{Fig.CCR-sigma_Hi-Cyl-Sph-d10-d100-s0.0005-z1} for $d = 100a$ and in \cref{Fig.Overcharging-Charge-reversal-sigma_Hi-Cyl-Sphere-s0.0005_rho0.1-0.2_d10-z1-z2} for $d = 10a$ at higher salt concentrations or valence. Although \cref{Fig.CCR_and_CO} uses $\sigma_{\scriptscriptstyle{0}} = 0.0005\,\mathrm{C/m^2}$, CCR becomes more pronounced at higher values of $\sigma_{\scriptscriptstyle{0}}$, $\rho_{\scriptscriptstyle{0}}$, or for divalent salts.

Once CCR is triggered at a given wall thickness, it persists for larger shell radii. In the limit $R \to \infty$, however, the induced charges recover electroneutrality: $\sigma_{\scriptscriptstyle{Hi}} \to -\sigma_{\scriptscriptstyle{0}}$ and $\sigma_{\scriptscriptstyle{Ho}} \to \sigma_{\scriptscriptstyle{0}}$. CCR can be explicitly evaluated via \cref{Cylinder-sigmaHo,Sphere-sigmaHo,Electroneutrality-condition-general2}.

Importantly, CCR does not occur in slit-shell geometries for any parameter set (see \cref{Fig.CO-slit-shell}). Hence, CCR is a geometric confinement effect exclusive to curved (cylindrical and spherical) nanocavities. In contrast, Confinement Overcharging (CO) is present in slit shells across a broad parameter range, though it weakens for large shell widths.

In \cref{Fig.sigma_Hi-sigmaHo-Cyl-Sphere-s0.0005_rho0.001_and_rho0.01,Fig.sigma_Hi-sigmaHo-Cyl-Sphere-s0.005_rho0.001_and_rho0.01,Fig.CCR-sigma_Ho-Cyl-Sph-d10-d100-s0.0005-z1,Fig.Overcharging-Charge-reversal-sigma_Ho-Cyl-Sphere-s0.0005_rho0.1-0.2_d10-z1-z2,Fig.CO-slit-shell}, $\sigma_{\scriptscriptstyle{Ho}}(R)$ demonstrates the presence of CO under various parameter regimes.

\bigskip

\noindent\textbf{Summary of key trends:}
\begin{itemize}
	\item[i)] For thin walls, $\sigma_{\scriptscriptstyle{Hi}}(R)$ increases with salt concentration but decreases with stronger confinement, as in spherical shells. Consequently, $\sigma_{\scriptscriptstyle{Ho}}(R)$ is larger at low salt concentrations but increases with $R$. The crossover behavior between cylindrical and spherical shells stems from geometric differences: spherical shells adsorb fewer counterions at small $R$ but more at larger $R$ compared to cylindrical shells~\cite{Adrian-JML-2023}. However, due to the more rapid decay of the spherical electric field, $\sigma_{\scriptscriptstyle{Ho}}(R)$ is typically higher for cylindrical shells at small $R$.
	
	This crossover does not occur in slit shells. There, a greater counterion adsorption always yields a smaller $\sigma_{\scriptscriptstyle{Ho}}(R)$, resulting in a weaker CO effect (see \cref{Fig.sigma_Hi-sigmaHo-Slit-d1_s0.005_rho0.001_and_rho0.01_all-data,Fig.sigma_Hi-sigmaHo-Slit-d10-d100_z1_s0.0005_rho0.1_all-data}).
	
	\item[ii)] Increasing wall thickness, salt concentration, or valence enhances CCR while simultaneously reducing CO, as seen in \cref{Fig.sigma-Hi_and_Ho,Fig.CCR_and_CO,Fig.CO-slit-shell}.
	
	\item[iii)] CCR and CO compete: higher CCR intensity tends to suppress CO.
\end{itemize}

\bigskip

\noindent\textbf{Physical origin of CCR and CO:}

The CCR and CO effects stem from fundamental statistical mechanical principles. Since the nanocavities are immersed in a bulk electrolyte, the fluids inside and outside are at equal chemical potential and bulk concentration, satisfying global electroneutrality~\cite{Jimenez_2004_Feb}. These fluids are coupled through the cavity walls~\cite{Lozada-Cassou-PRL1996,Lozada-correlation-of-charged-fluids-1997}, and the structure of the inhomogeneous EDLs emerges from a balance between electrostatic energy and entropy.

The observed enhancement of CCR and CO for smaller ions—and their suppression for large adsorbed species—suggests these are primarily energy-driven effects.

In this work, the shell surfaces carry constant surface charge densities on both sides. If instead a fixed total surface charge were imposed, slit geometries would show little variation with $R$, whereas for curved shells, increasing $R$ would reduce $\sigma_{\scriptscriptstyle{Hi}}(R)$ and $\sigma_{\scriptscriptstyle{Ho}}(R)$ due to weaker electric fields (see \cref{Fig.sigma-Hi_and_Ho}).

Other physical scenarios, such as asymmetric surface charge distributions or variable shell thicknesses, could yield richer behavior and are left for future studies.


Beyond their theoretical relevance for understanding the entropy--energy balance in confined fluids, the CCR and CO phenomena may have important implications for ion and macroion adsorption near vesicles and biological cells. These effects could impact fields such as medicine, drug delivery, and cellular physiology.

Experimental detection of CCR and CO may require more advanced techniques. For example, the violation of the local electroneutrality condition (VLEC) was theoretically predicted in 1984~\cite{Lozada_1984}, but only experimentally confirmed in 2015~\cite{Luo-electroneutrality-nature-2015}. This experimental evidence indirectly supports the existence of CO in slit-like nanopores~\cite{Luo-electroneutrality-nature-2015}. 

Additionally, CCR and CO could be indirectly observed through osmotic pressure measurements in nanocavities of different geometries.

\bigskip

Now we return to the results for the mean electrostatic potential (MEP) at the shell boundaries, shown in \cref{Fig.RMEP.General}. From Eq.~\ref{Electroneutrality-condition-general2} and \cref{Plates-MEP(r),Cylinder-MEP,Sphere-MEP}, the potentials $\psi_{\scriptscriptstyle{H}}(R)$ and $\varphi_{\scriptscriptstyle{H}}(R)$ can be expressed analytically for each geometry. For slit-shells:

\begin{equation}
	\psi_{\scriptscriptstyle{H}} = -\frac{1}{\kappa} \coth[\kappa(R - a/2)] \frac{\sigma_{\scriptscriptstyle{Hi}}}{\varepsilon_{\scriptscriptstyle{0}} \varepsilon},
	\label{Slit-MEPsiH}
\end{equation}
\begin{equation}
	\varphi_{\scriptscriptstyle{H}} = \frac{\sigma_{\scriptscriptstyle{Ho}}}{\varepsilon_{\scriptscriptstyle{0}} \varepsilon}.
	\label{Slit-MEPhiH}
\end{equation}

For cylindrical and spherical shells:

\begin{equation}
	\begin{split}
		\psi_{\scriptscriptstyle{H}} = -\frac{1}{\kappa} \frac{I_0[\kappa(R - a/2)]}{I_1[\kappa(R - a/2)]} \frac{\sigma_{\scriptscriptstyle{Hi}}}{\varepsilon_{\scriptscriptstyle{0}} \varepsilon}, \\
		\varphi_{\scriptscriptstyle{H}} = \frac{1}{\kappa} \frac{K_0[\kappa R_{\scriptscriptstyle{H}}]}{K_1[\kappa R_{\scriptscriptstyle{H}}]} \frac{\sigma_{\scriptscriptstyle{Ho}}}{\varepsilon_{\scriptscriptstyle{0}} \varepsilon},
	\end{split}
	\label{Cylinder-MEPH}
\end{equation}

\begin{equation}
	\begin{split}
		\psi_{\scriptscriptstyle{H}} = -\frac{(R - a/2)}{\left[\kappa(R - a/2) \coth[\kappa(R - a/2)] - 1\right]} \frac{\sigma_{\scriptscriptstyle{Hi}}}{\varepsilon_{\scriptscriptstyle{0}} \varepsilon}, \\
		\varphi_{\scriptscriptstyle{H}} = \frac{R_{\scriptscriptstyle{H}}}{1 + \kappa R_{\scriptscriptstyle{H}}} \frac{\sigma_{\scriptscriptstyle{Ho}}}{\varepsilon_{\scriptscriptstyle{0}} \varepsilon},
	\end{split}
	\label{Sphere-MEPH}
\end{equation}

Here, $E(R - a/2) = \sigma_{\scriptscriptstyle{Hi}} / (\varepsilon_{\scriptscriptstyle{0}} \varepsilon)$ and $E(R_{\scriptscriptstyle{H}}) = \sigma_{\scriptscriptstyle{Ho}} / (\varepsilon_{\scriptscriptstyle{0}} \varepsilon)$ denote the electric field at the inner and outer boundaries, respectively. Therefore, for all geometries, $\psi_{\scriptscriptstyle{H}}$ and $\varphi_{\scriptscriptstyle{H}}$ are given by the electric field at the boundary multiplied by a geometry-specific factor.

In all cases, $E(R - a/2)$ becomes increasingly negative as $R$ increases due to the rising magnitude of $\sigma_{\scriptscriptstyle{Hi}}(R)$ (see \cref{Fig.sigma-Hi_and_Ho}). The geometric prefactors in \cref{Slit-MEPsiH,Cylinder-MEPH,Sphere-MEPH} are non-linear, negative, and decreasing functions of $R$, leading to maxima in $\psi_{\scriptscriptstyle{H}}(R)$ for curved geometries (\cref{Fig.RMEP.Cyl.Low.Rho,Fig.RMEP.Sphere.Low.Rho}).

On the other hand, $\sigma_{\scriptscriptstyle{Ho}}(R)$ initially increases with $R$ before decreasing back to $\sigma_{\scriptscriptstyle{0}}$, as shown in \cref{Fig.sigma-Hi_and_Ho}. Meanwhile, the geometric factors in \cref{Cylinder-MEPH,Sphere-MEPH} for $\varphi_{\scriptscriptstyle{H}}$ are increasing functions that saturate at finite constants. As a result, $\varphi_{\scriptscriptstyle{H}}(R)$ exhibits a mild maximum in the cylindrical case (\cref{Fig.RMEP.Cyl.Low.Rho}) and a monotonic increase in the spherical case (\cref{Fig.RMEP.Sphere.Low.Rho}).

For slit-shells, the geometric factor in \cref{Slit-MEPsiH} decreases sharply with $R$, while the factor in \cref{Slit-MEPhiH} remains constant. Since both $\sigma_{\scriptscriptstyle{Hi}}$ and $\sigma_{\scriptscriptstyle{Ho}}$ also decrease with $R$ in this geometry, $\psi_{\scriptscriptstyle{H}}(R)$ and $\varphi_{\scriptscriptstyle{H}}(R)$ both decrease monotonically.

Finally, in the limit $R \to \infty$, all geometrical prefactors approach constant values, while $\sigma_{\scriptscriptstyle{Hi}} \to -\sigma_{\scriptscriptstyle{0}}$ and $\sigma_{\scriptscriptstyle{Ho}} \to \sigma_{\scriptscriptstyle{0}}$. Therefore, the total electroneutrality condition holds for all $R$, as guaranteed by \cref{Electroneutrality-condition-general2,Electrical-field-balance}.

\subsection{The osmotic pressure} \label{Osmotic pressure}

We now analyze the results for the reduced osmotic pressure, defined as:
\[
\pi_{\scriptscriptstyle{N}}(R) \equiv \frac{p_{\scriptscriptstyle{N}}(R)}{kT\rho_{\scriptscriptstyle{0}}}, \quad
\pi_{\scriptscriptstyle{S}}(R) \equiv \frac{p_{\scriptscriptstyle{S}}(R)}{kT\rho_{\scriptscriptstyle{0}}}, \quad
\pi_{\scriptscriptstyle{E}}(R) \equiv \frac{p_{\scriptscriptstyle{E}}(R)}{kT\rho_{\scriptscriptstyle{0}}},
\]
\[
\pi_{\scriptscriptstyle{E1}}(R) \equiv \frac{p_{\scriptscriptstyle{E1}}(R)}{kT\rho_{\scriptscriptstyle{0}}}, \quad
\pi_{\scriptscriptstyle{E2}}(R) \equiv \frac{p_{\scriptscriptstyle{E2}}(R)}{kT\rho_{\scriptscriptstyle{0}}}, \quad
\pi_{\scriptscriptstyle{E3}}(R) \equiv \frac{p_{\scriptscriptstyle{E3}}(R)}{kT\rho_{\scriptscriptstyle{0}}}.
\]

The steric component $p_{\scriptscriptstyle{S}}(R)$ is given by \cref{Pressure-Eq-Steric-MEP}, while the dominant electrostatic term $p_{\scriptscriptstyle{E1}}(R)$ reads:

\begin{equation} \label{Osmotic-Press-E1}
	\begin{split}
		p_{\scriptscriptstyle{E1}}(R) &= -\frac{\varepsilon_{\scriptscriptstyle{0}}\varepsilon}{2} \left[ E^2(R-a/2) - E^2(R+d+a/2) \right] \\
		&= -\frac{1}{2\varepsilon_{\scriptscriptstyle{0}}\varepsilon} \left[ \sigma_{\scriptscriptstyle{Hi}}^2 - \sigma_{\scriptscriptstyle{Ho}}^2 \right].
	\end{split}
\end{equation}

For slit-shells, $p_{\scriptscriptstyle{E2}}(R) = p_{\scriptscriptstyle{E3}}(R) = 0$. For cylindrical and spherical geometries, $p_{\scriptscriptstyle{E2}}(R)$ and $p_{\scriptscriptstyle{E3}}(R)$ are the second and third terms in \cref{Osmotic-Cyl-Electric,Osmotic-Press-Sph-Electric}, respectively.

The total reduced osmotic pressure is:
\begin{equation} \label{Osmotic-Press-Net-total}
	\pi_{\scriptscriptstyle{N}}(R) = \pi_{\scriptscriptstyle{S}}(R) + \pi_{\scriptscriptstyle{E}}(R),
\end{equation}

\noindent where
\begin{equation} \label{Osmotic-Press-Reduced-MST}
	\pi_{\scriptscriptstyle{E}}(R) = \pi_{\scriptscriptstyle{E1}}(R) + \pi_{\scriptscriptstyle{E2}}(R) + \pi_{\scriptscriptstyle{E3}}(R).
\end{equation}

\begin{figure}[!htb]
	\begin{subfigure}{.5\textwidth}
		\centering
		\includegraphics[width=0.95\linewidth]{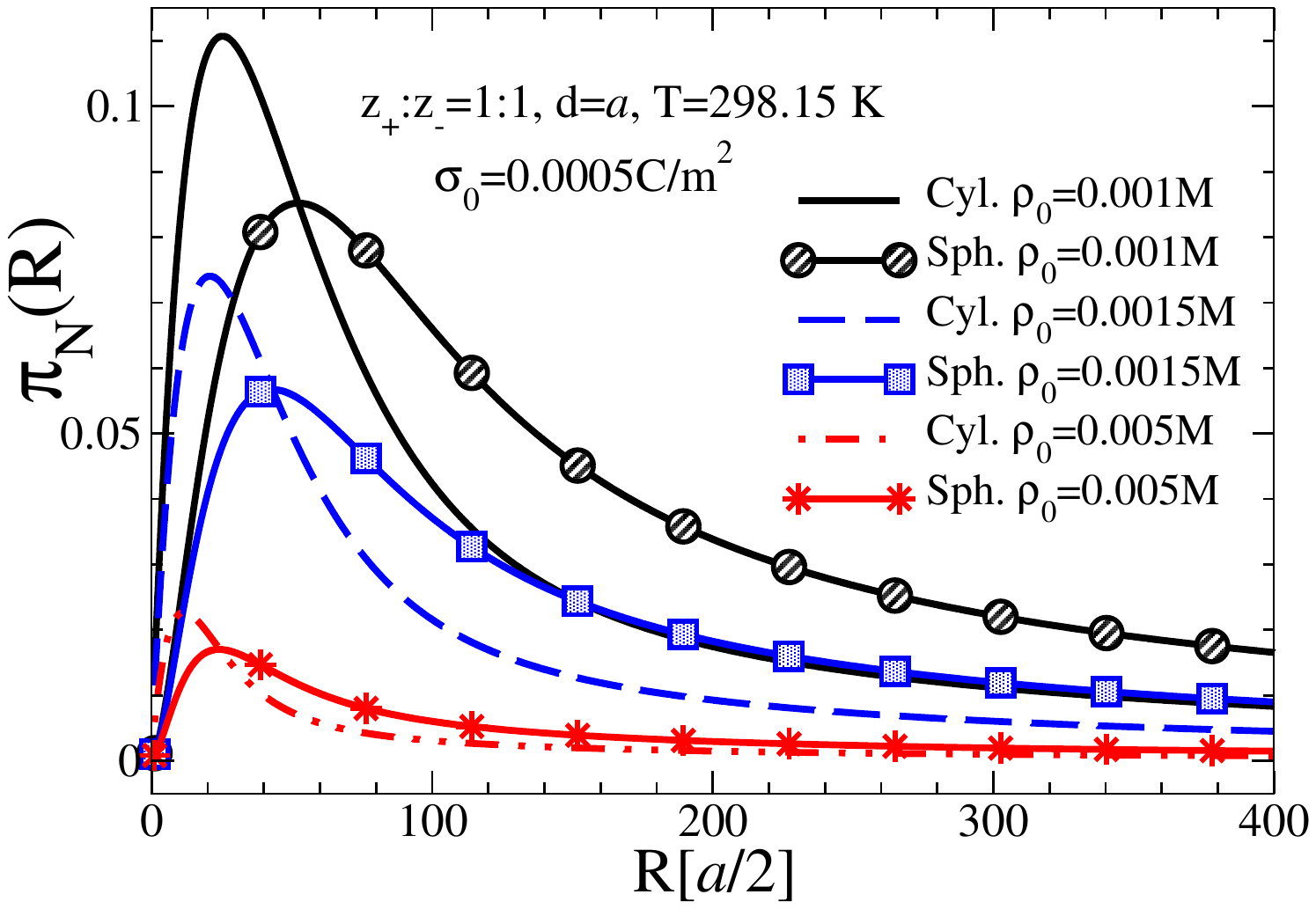}
		\caption{Electrolyte concentration dependence.}
		\label{Fig.ROP.Electrolyte.Concentration}
	\end{subfigure}
	\begin{subfigure}{.5\textwidth}
		\centering
		\includegraphics[width=0.95\linewidth]{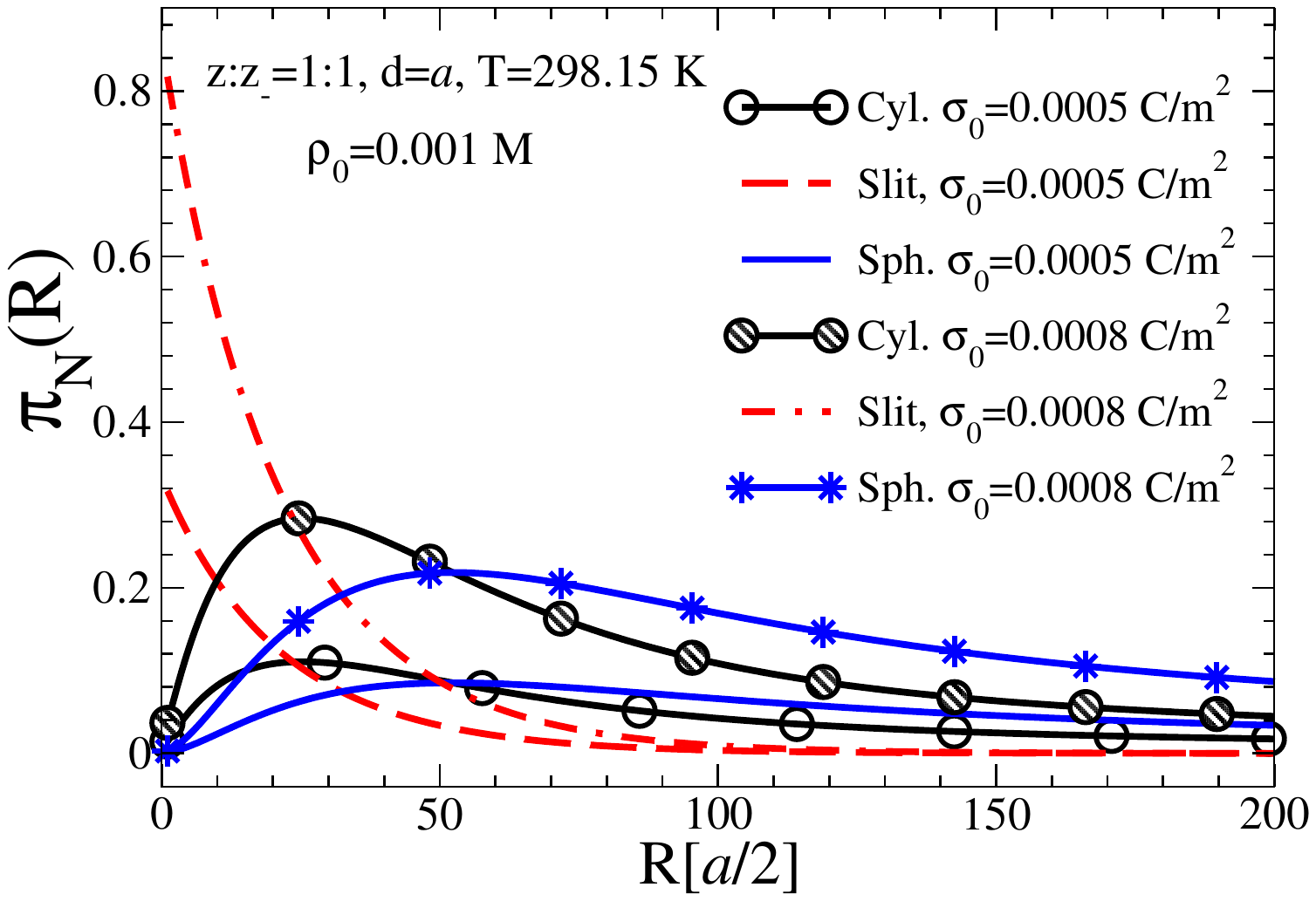}
		\caption{Surface charge density dependence.}
		\label{Fig.ROP.Surface.Charge}
	\end{subfigure}
	\begin{subfigure}{.5\textwidth}
		\centering
		\includegraphics[width=0.95\linewidth]{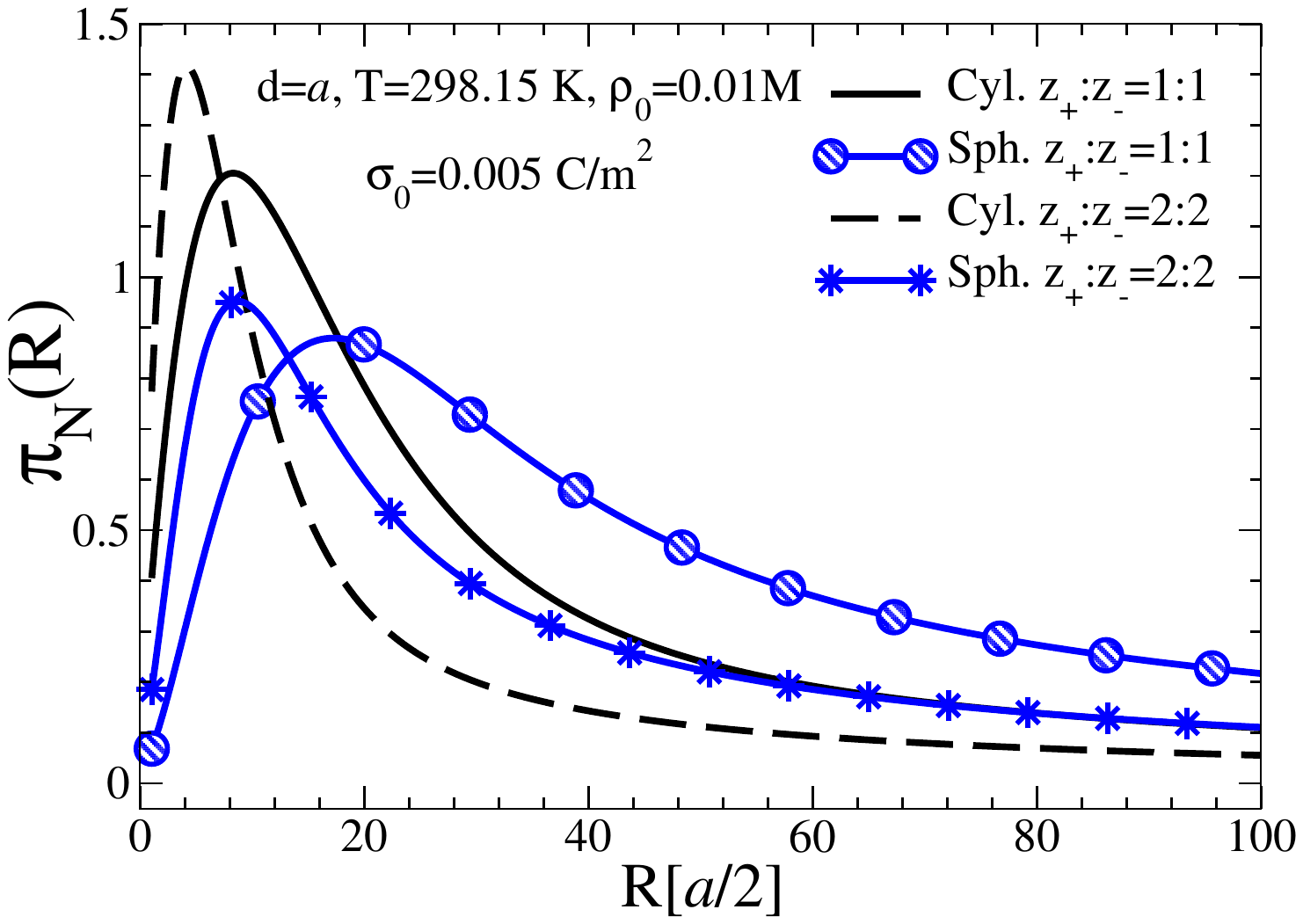}
		\caption{Electrolyte valence dependence.}
		\label{Fig.ROM.Electrolyte.Valence}
	\end{subfigure}
	\begin{subfigure}{.5\textwidth}
		\centering
		\includegraphics[width=0.95\linewidth]{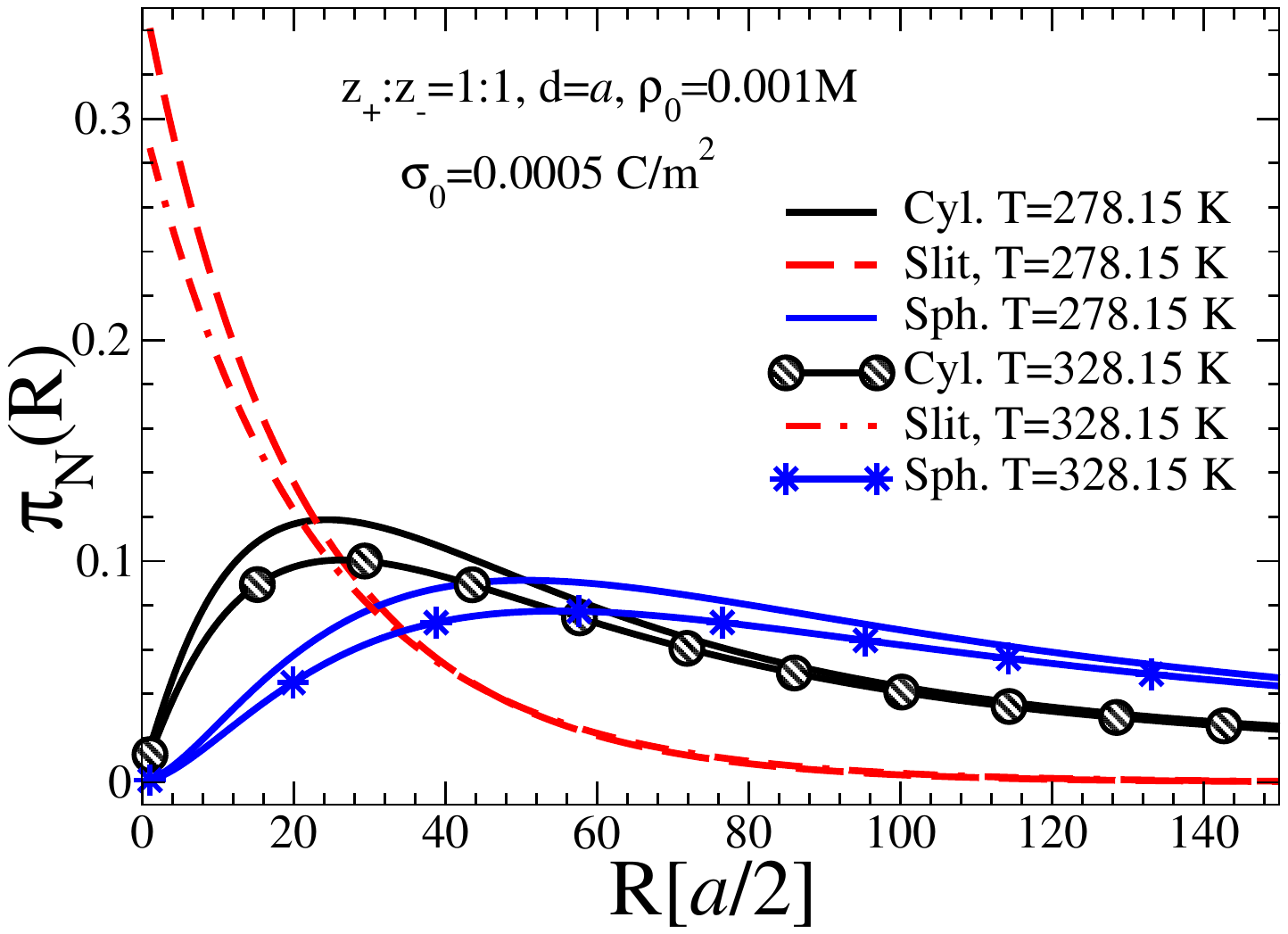}
		\caption{Temperature dependence.}
		\label{Fig.ROP.System.Temperature}
	\end{subfigure}\\
	\caption{Reduced osmotic pressure $\pi_{\scriptscriptstyle{N}}(R)$ as a function of shell radius $R$ and: (a) electrolyte concentration $\rho_{\scriptscriptstyle{0}}$; (b) surface charge density $\sigma_{\scriptscriptstyle{0}}$; (c) electrolyte valence $z_+:z_-$; and (d) temperature $T$. All cases use $\varepsilon = 78.5$ and $a = \SI{4.25}{\angstrom}$.}
	\label{Fig.ROP.General}
\end{figure}

\Cref{Fig.ROP.General} presents $\pi_{\scriptscriptstyle{N}}(R)$ as a function of shell radius $R$ for all three geometries and a variety of system parameters.

In \cref{Fig.ROP.Electrolyte.Concentration}, for cylindrical and spherical shells, $\pi_{\scriptscriptstyle{N}}(R)$ is shown for three values of $\rho_{\scriptscriptstyle{0}}$. The pressure increases with $R$, reaches a maximum at $R_{\scriptscriptstyle{Max}}$, then decays to zero as $R \to \infty$. Increasing $\rho_{\scriptscriptstyle{0}}$ lowers both the magnitude and the position of the maximum.

For small radii, $\pi_{\scriptscriptstyle{N}}(R)$ is larger in cylindrical shells than in spherical ones. However, this trend reverses beyond the spherical $R_{\scriptscriptstyle{Max}}$. This crossover is a direct consequence of the geometry-dependent electrostatics and shell-induced confinement.

The constant wall surface charge $\sigma_{\scriptscriptstyle{0}}$ implies that the induced charges satisfy $\sigma_{\scriptscriptstyle{Hi}} < 0$ and $\sigma_{\scriptscriptstyle{Ho}} > 0$. The function $\sigma_{\scriptscriptstyle{Hi}}(R)$ decreases monotonically, while $\sigma_{\scriptscriptstyle{Ho}}(R)$ shows a non-monotonic maximum (\cref{Fig.sigma-Hi_and_Ho}). For cylindrical and spherical geometries, decreasing $R$ reduces the total shell charge $Q(R)$ and $Q(R+d)$. In contrast, the planar slit-shell preserves these charges as $R$ varies.
Hence, in the limit $R \rightarrow 0$, we have $\sigma_{\scriptscriptstyle{Hi}} \rightarrow 0$ and $\sigma_{\scriptscriptstyle{Ho}} \rightarrow 0$. As $R$ increases, $\sigma_{\scriptscriptstyle{Hi}} \rightarrow -\sigma_{\scriptscriptstyle{0}}$, while $\sigma_{\scriptscriptstyle{Ho}}$ grows to a maximum and then decreases toward $\sigma_{\scriptscriptstyle{0}}$, as required by the global electroneutrality condition \cref{Electroneutrality-condition-general2}. Among the three geometries, this decay is fastest in the slit-shell, followed by the cylindrical and then the spherical shell, due to differences in the decay rates of their respective electric fields.

This interplay—an increase in surface area (and thus $Q(R)$) with a concurrent decrease in electric field strength—produces a maximum in the leading electrostatic pressure term $\pi_{\scriptscriptstyle{E1}}(R)$ (\cref{Osmotic-Press-E1}), which dominates the electrostatic contribution $\pi_{\scriptscriptstyle{E}}(R)$ (\cref{Osmotic-Press-Reduced-MST}). In contrast, for the slit-shell, $\pi_{\scriptscriptstyle{E}}(R)$ decays monotonically from a finite value to zero with increasing $R$.

In inhomogeneous fluids, higher induced surface charge densities lead to stronger counter-ion adsorption and, consequently, larger contact densities $\rho_{\scriptscriptstyle{\alpha s}}(R-a/2)$ and $\rho_{\scriptscriptstyle{\alpha s}}(R+d+a/2)$. Due to confinement, the inequality $\rho_{\scriptscriptstyle{\alpha s}}(R-a/2) - \rho_{\scriptscriptstyle{\alpha s}}(R+d+a/2) \geq 0$ holds for all $R>0$~\cite{Yu_1997,Aguilar_2007,Adrian-JML-2023}, provided there's enough room for at least one counter-ion layer inside the shell. Thus, enhanced confinement and/or larger $\sigma_{\scriptscriptstyle{0}}$ boosts inner counter-ion adsorption. However, $Q(R)$ and $Q(R+d)$ both decrease with $R$ at fixed $\sigma_{\scriptscriptstyle{0}}$, producing a non-monotonic steric pressure component (\cref{Pressure-Eq-Steric-MEP}).

As a result, both the steric and Maxwell stress contributions display maxima in the cylindrical and spherical geometries. Although these maxima are not located at exactly the same $R$, they are close enough to jointly produce the overall maxima in $\pi_{\scriptscriptstyle{N}}(R)$ shown in \cref{Fig.ROP.Electrolyte.Concentration}. The next subsections will explore their individual behaviors in more detail.

As discussed, $\pi_{\scriptscriptstyle{E1}}(R)$ is the dominant electrostatic term. It is always positive, increases with decreasing confinement, peaks at an intermediate $R$, and vanishes as $R \to \infty$, preserving electroneutrality. Its maximum is stronger and occurs at smaller radii in cylindrical shells, due to their more intense electric fields. At larger $R$, however, spherical shells yield higher and broader pressure profiles. Moreover, stronger electric fields enhance $\pi_{\scriptscriptstyle{S}}(R)$ via increased ion adsorption, further amplifying the steric contribution~\cite{Adrian-JML-2023}.

Raising the electrolyte concentration $\rho_{\scriptscriptstyle{0}}$ shifts the pressure maxima to smaller $R$, as the EDL becomes more compact. Hence, steric effects are stronger at lower $R$ and weaker at larger $R$.

\vspace{1em}

While direct experimental measurements of nanoconfined osmotic pressure are unavailable, our findings are consistent with Donnan equilibrium experiments in macroion dispersions, where osmotic pressure decreases with salt concentration~\cite{Marcia-EPL-1998,Levin_electroneutrality-2016,Jimenez_2004_Nov}. However, in our model—without macroions—the pressure arises solely from confinement, and vanishes as $R \rightarrow \infty$.

\Cref{Fig.ROP.Surface.Charge} compares two cases with $\sigma_{\scriptscriptstyle{0}} = 0.0005\,\mathrm{C/m^2}$ and $0.0008\,\mathrm{C/m^2}$, showing that a small increase in $\sigma_{\scriptscriptstyle{0}}$ yields a significantly larger osmotic pressure across all geometries. For both values, the pressure maximum occurs at the same shell size $R_{\scriptscriptstyle{Max}}$, defined as the radius where $\pi_{\scriptscriptstyle{N}}(R)$ is maximal. This reflects a mechanical balance, unchanged by the symmetric increase in wall charge. Typically, $R_{\scriptscriptstyle{Max}}$ is smaller for cylindrical shells.

A higher $\sigma_{\scriptscriptstyle{0}}$ enhances both $|\sigma_{\scriptscriptstyle{Hi}}|$ and $\sigma_{\scriptscriptstyle{Ho}}$ (\cref{Fig.sigma-Hi_and_Ho}), thereby increasing $\pi_{\scriptscriptstyle{N}}(R)$ (\cref{Osmotic-Cyl-Electric,Osmotic-Press-Sph-Electric}).

For the slit-shell, $\pi_{\scriptscriptstyle{N}}(R)$ decreases monotonically with $R$. As confinement weakens, $\sigma_{\scriptscriptstyle{Hi}}$ drops from nearly zero to $-\sigma_{\scriptscriptstyle{0}}$, while $\sigma_{\scriptscriptstyle{Ho}}$ approaches $\sigma_{\scriptscriptstyle{0}}$ (\cref{Slit-electroneutrality}). Accordingly, $\pi_{\scriptscriptstyle{E1}}(R)$ and $\pi_{\scriptscriptstyle{S}}(R)$ decrease with increasing $R$, and so does the total osmotic pressure (\cref{Osmotic-plates-Net,Fig.ROP.Surface.Charge}).

In cylindrical and spherical shells, increasing $\sigma_{\scriptscriptstyle{0}}$ intensifies ion adsorption on both walls and boosts $\pi_{\scriptscriptstyle{S}}(R)$, which can exceed $\pi_{\scriptscriptstyle{E}}(R)$ at high surface charge.

\vspace{1em}

In \cref{Fig.ROM.Electrolyte.Valence}, we show $\pi_{\scriptscriptstyle{N}}(R)$ for $1{:}1$ and $2{:}2$ electrolytes at $\rho_{\scriptscriptstyle{0}} = 0.01\,\mathrm{M}$ and $\sigma_{\scriptscriptstyle{0}} = 0.005\,\mathrm{C/m^2}$. At small $R$, the divalent electrolyte yields a higher pressure, with the maximum shifted leftward in both geometries.

Divalent counter-ions adsorb more strongly, enhancing the inner-wall ion layer. However, this stronger $\sigma_{\scriptscriptstyle{Hi}}$ also reduces $\sigma_{\scriptscriptstyle{Ho}}$—thus lowering $\pi_{\scriptscriptstyle{E1}}(R)$. Still, $\pi_{\scriptscriptstyle{N}}(R)$ is higher due to stronger steric effects. For large $R$, however, the $1{:}1$ electrolyte produces a stronger total pressure. The leftward shift in the $2{:}2$ case mimics the effect of increased salt concentration due to the role of $\kappa$ in the underlying expressions (\cref{Ec.kappa}).

\vspace{1em}

\Cref{Fig.ROP.System.Temperature} shows $\pi_{\scriptscriptstyle{N}}(R)$ at different temperatures for a $1{:}1$ electrolyte with $\rho_{\scriptscriptstyle{0}} = 0.001\,\mathrm{M}$ and $\sigma_{\scriptscriptstyle{0}} = 0.0005\,\mathrm{C/m^2}$. Temperature effects are modest but produce a rightward shift in the pressure curve, similar to that from lowering salt concentration—again reflecting the influence of $\kappa$ (\cref{Ec.kappa}).

\begin{figure}[!hbt]
	\begin{subfigure}{.5\textwidth}
		\centering
		\includegraphics[width=0.95\linewidth]{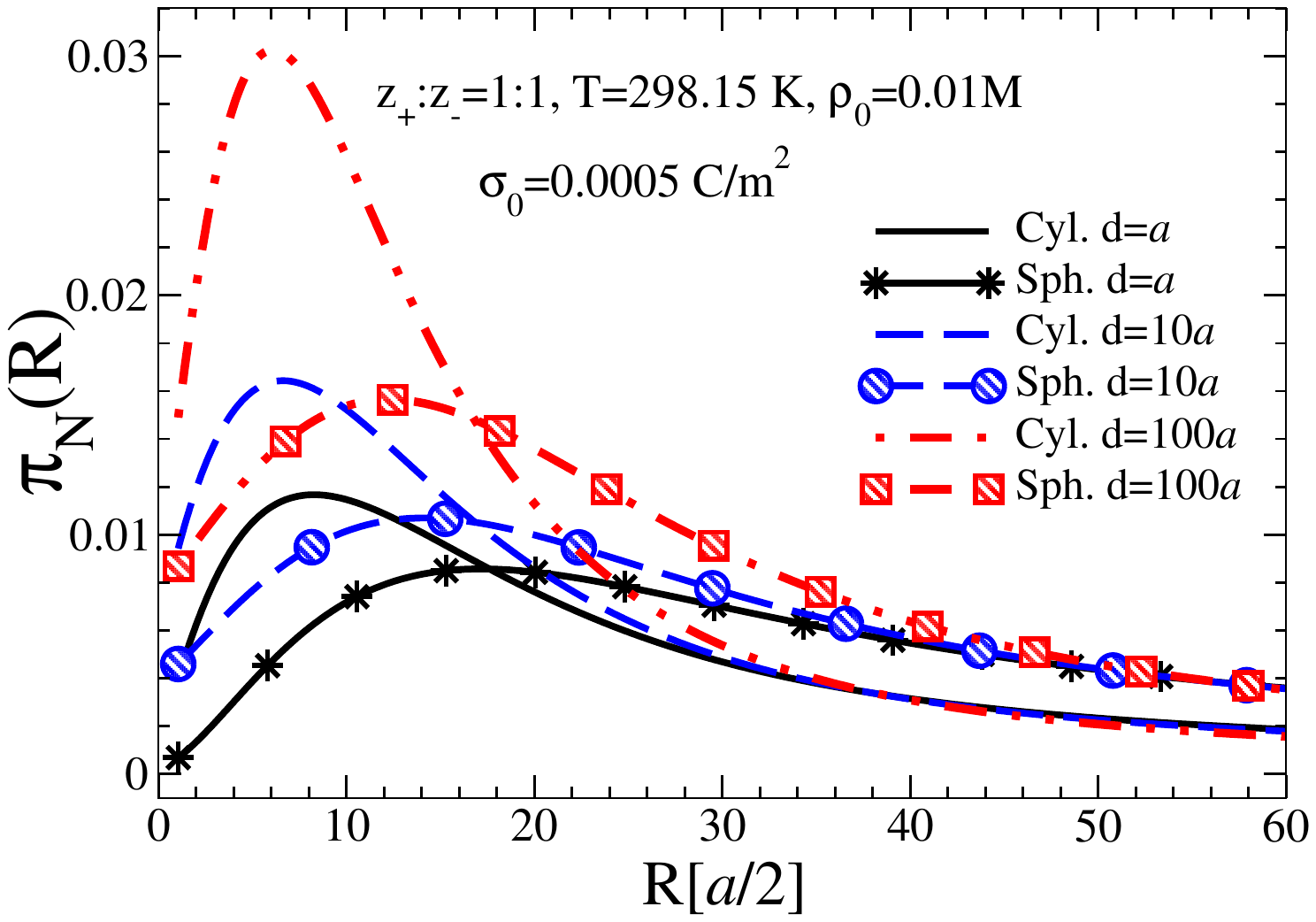}
		\caption{$\pi_{\scriptscriptstyle{N}}(R)$ vs. $d$ at low $\rho_{\scriptscriptstyle{0}}$ and $\sigma_{\scriptscriptstyle{0}}$.}
		\label{Fig.ROP.Thickness.Dependence_Low.Rho_Low.Sigma}
	\end{subfigure}
	\begin{subfigure}{.5\textwidth}
		\centering
		\includegraphics[width=0.95\linewidth]{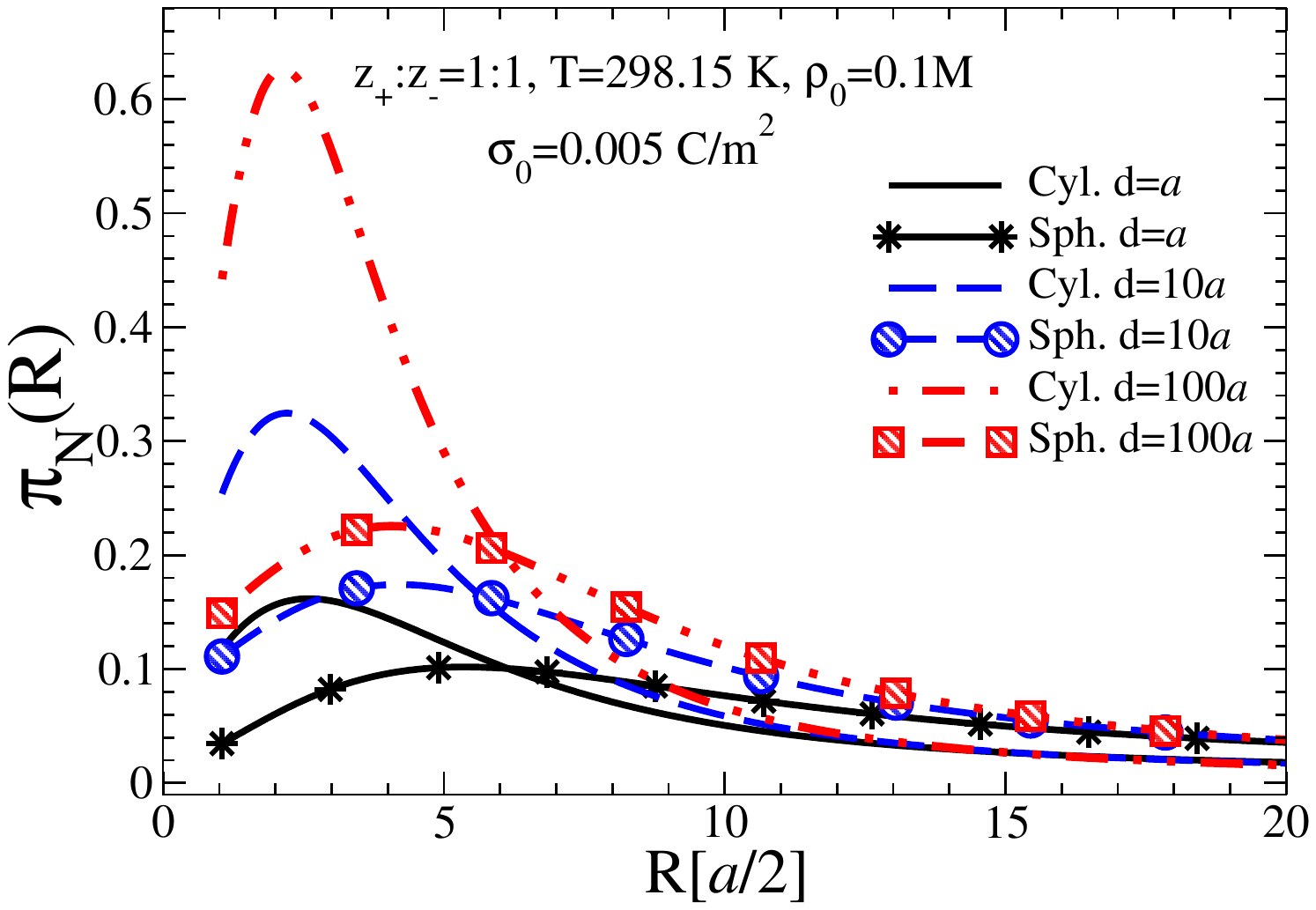}
		\caption{$\pi_{\scriptscriptstyle{N}}(R)$ vs. $d$ at high $\rho_{\scriptscriptstyle{0}}$ and $\sigma_{\scriptscriptstyle{0}}$.}
		\label{Fig.ROP.Thickness.Dependence_High_Rho.High.Sigma}
	\end{subfigure}
	\begin{subfigure}{.5\textwidth}
		\centering
		\includegraphics[width=0.95\linewidth]{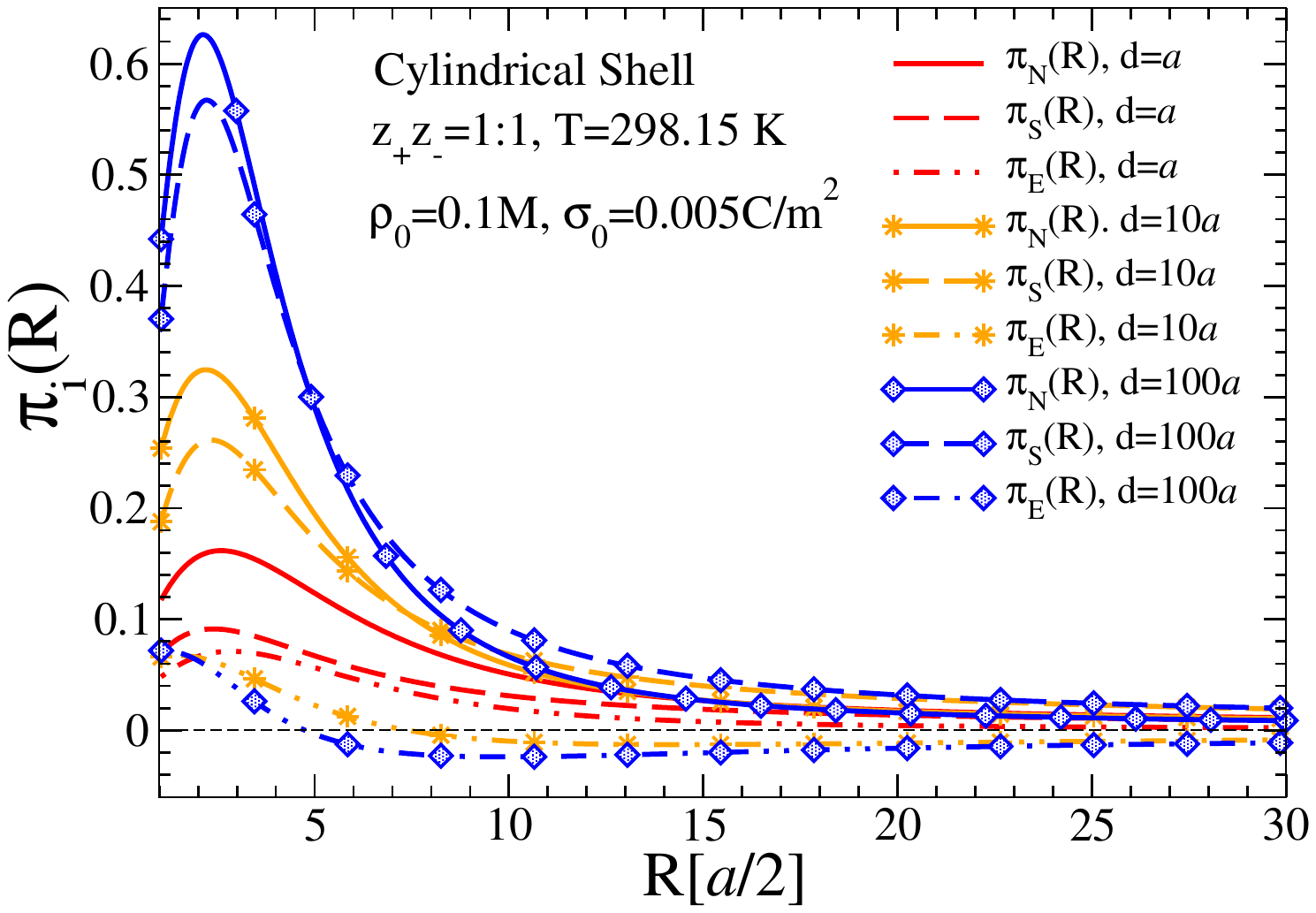}
		\caption{Cylindrical shell: $\pi_{\scriptscriptstyle{S}}(R)$ and $\pi_{\scriptscriptstyle{E}}(R)$ for varying $d$.}
		\label{cyl-PT-PS-PE-z1_d1_and_d10_d100_T298.15_s0.005_rho0.1_justpiN_piS_piE}
	\end{subfigure}
	\begin{subfigure}{.5\textwidth}
		\centering
		\includegraphics[width=0.95\linewidth]{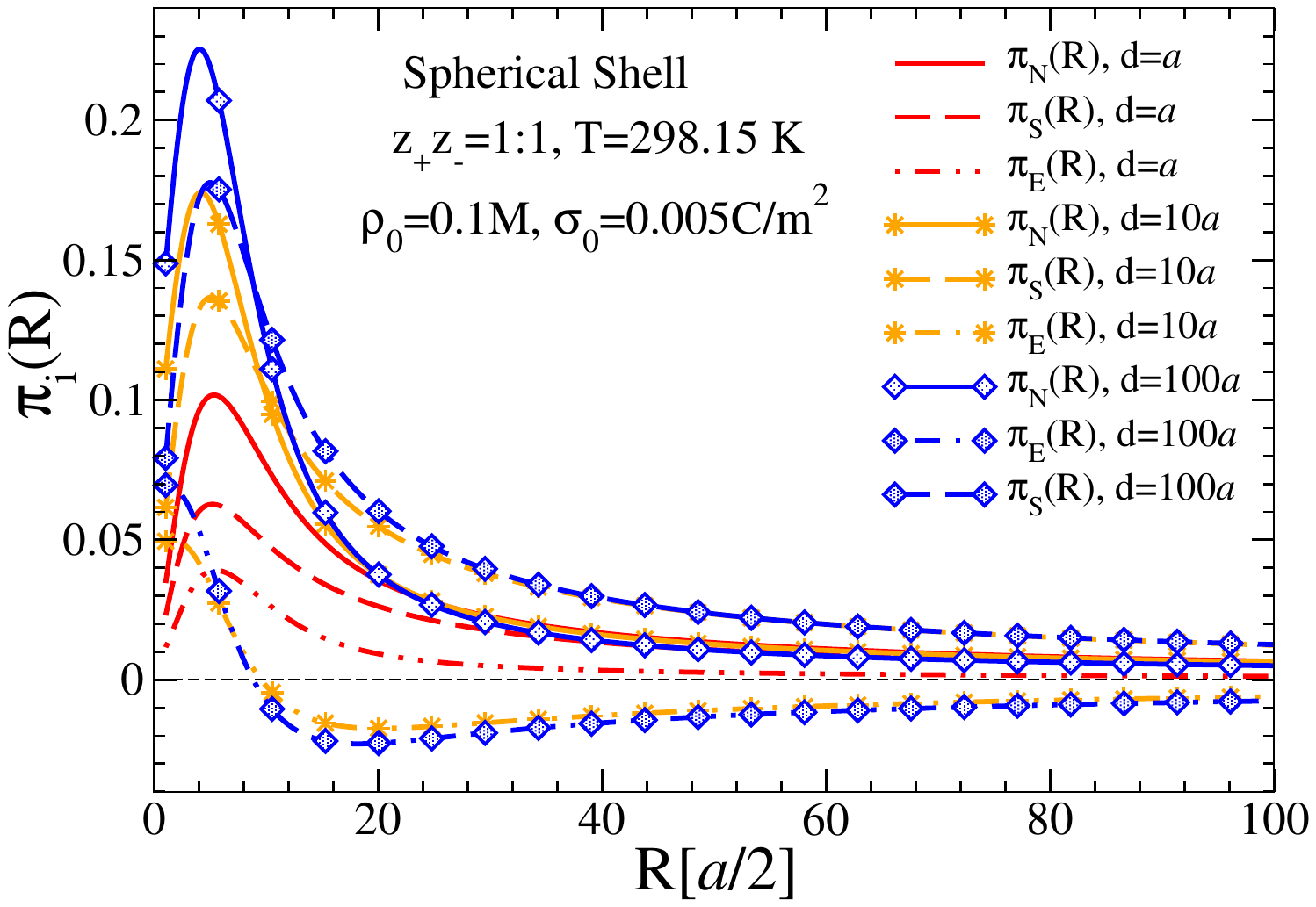}
		\caption{Spherical shell: $\pi_{\scriptscriptstyle{S}}(R)$ and $\pi_{\scriptscriptstyle{E}}(R)$ for varying $d$.}
		\label{sph-PT-PS-PE-z1_d1_and_d10_d100_T298.15_s0.005_rho0.1_justpiN_piS_piE}
	\end{subfigure}
	\caption{Reduced osmotic pressure $\pi_{\scriptscriptstyle{N}}(R)$ and its components as a function of shell radius $R$ and wall thickness $d$. (a) and (b) show full pressure curves at low and high salt/surface charge, respectively. (c) and (d) compare $\pi_{\scriptscriptstyle{S}}(R)$ and $\pi_{\scriptscriptstyle{E}}(R)$ in cylindrical and spherical geometries for $\rho_{\scriptscriptstyle{0}} = 0.1\,\mathrm{M}$ and $\sigma_{\scriptscriptstyle{0}} = 0.005\,\mathrm{C/m^2}$. $\varepsilon = 78.5$, $a = \SI{4.25}{\angstrom}$.}
	\label{Fig.ROP.Thickness_Dependence}
\end{figure}

Finally, in \cref{Fig.ROP.Thickness_Dependence}, we explore how the osmotic pressure depends on the shell wall thickness for cylindrical and spherical geometries, considering $d = a$, $10a$, and $100a$. Thicker walls result in more intense osmotic pressures. For small shell radii, cylindrical shells show higher pressure than spherical ones; however, this trend reverses at larger $R$. The crossover point does not coincide with the spherical shell’s $R_{\scriptscriptstyle{Max}}$ (see \cref{Fig.ROP.Electrolyte.Concentration}) but occurs at larger $R$. As the EDL becomes narrower with increasing $d$, the pressure maxima shift slightly to smaller radii (see \cref{Fig.ROP.Thickness.Dependence_Low.Rho_Low.Sigma,Fig.ROP.Thickness.Dependence_High_Rho.High.Sigma}).

Due to electrostatic attraction and repulsion, counter-ions are adsorbed more strongly than co-ions on both shell surfaces. However, confinement enhances adsorption at the inner wall, creating an asymmetry that significantly increases the steric contribution, $\pi_{\scriptscriptstyle{S}}(R)$, particularly in narrow shells (see \cref{cyl-PT-PS-PE-z1_d1_and_d10_d100_T298.15_s0.005_rho0.1_justpiN_piS_piE,sph-PT-PS-PE-z1_d1_and_d10_d100_T298.15_s0.005_rho0.1_justpiN_piS_piE}). As $R$ increases, this asymmetry diminishes, and $\pi_{\scriptscriptstyle{S}}(R)$ gradually vanishes.

Meanwhile, $E(R - a/2)$ becomes more negative as $R$ increases, while $E(R + d + a/2)$ first increases, then decays toward $\sigma_{\scriptscriptstyle{0}}$ (see \cref{Fig.sigma-Hi_and_Ho,Electrical-field-balance}). This non-monotonic behavior of the electric fields explains the shape of $\pi_{\scriptscriptstyle{E}}(R)$ observed in \cref{cyl-PT-PS-PE-z1_d1_and_d10_d100_T298.15_s0.005_rho0.1_justpiN_piS_piE,sph-PT-PS-PE-z1_d1_and_d10_d100_T298.15_s0.005_rho0.1_justpiN_piS_piE} (see also \cref{Osmotic-Press-Reduced-MST,Osmotic-Cyl-Electric,Osmotic-Press-Sph-Electric}).

In general, $\pi_{\scriptscriptstyle{E}}(R)$ is smaller than $\pi_{\scriptscriptstyle{S}}(R)$ and its relative contribution decreases with increasing $d$. This trend reflects the growing surface area (and charge) with thicker walls, coupled with the faster spatial decay of $E(r)$ in spherical shells. As $R$ increases, the osmotic pressures in both geometries converge and vanish in the limit $R \rightarrow \infty$.

For thick walls, $\pi_{\scriptscriptstyle{S}}(R)$ dominates and $\pi_{\scriptscriptstyle{E}}(R)$ can even become negative. Specifically, for cylindrical shells, $\pi_{\scriptscriptstyle{E}}(R) < 0$ at $R \gtrsim 7.4a/2$ for $d = 10a$, and $R \gtrsim 4.7a/2$ for $d = 100a$. For spherical shells, this occurs at $R \gtrsim 9a/2$ for both $d = 10a$ and $d = 100a$.

These negative values arise from $\abs{E(R - a/2)} > \abs{E(R + d + a/2)}$, i.e., $\sigma_{\scriptscriptstyle{Hi}} + \sigma_{\scriptscriptstyle{0}} < 0$, signaling confinement charge reversal (CCR). This amplifies the confinement field and coexists with confinement overcharging (CO), even within the point-ion model.

The total electrostatic pressure, $\pi_{\scriptscriptstyle{E}}(R)$, comprises three components: $\pi_{\scriptscriptstyle{E1}}(R)$, $\pi_{\scriptscriptstyle{E2}}(R)$, and $\pi_{\scriptscriptstyle{E3}}(R)$. All can become negative over certain $R$ ranges, reinforcing the confinement-driven origin of CCR (see \cref{Osmotic-Press-E1}).

At lower $\sigma_{\scriptscriptstyle{0}}$ and $\rho_{\scriptscriptstyle{0}}$ ($0.0005\,\mathrm{C/m^2}$, $0.001\,\mathrm{M}$), negative $\pi_{\scriptscriptstyle{E}}(R)$ appears only for thick shells. In \cref{Fig.ROP.Thickness.Dependence_Low.Rho_Low.Sigma}, this occurs at $R \gtrsim 20.4a/2$ (cylindrical) and $R \gtrsim 34a/2$ (spherical), in agreement with \cref{Fig.CCR_and_CO}. Once negative, $\pi_{\scriptscriptstyle{E}}(R)$ remains so at larger $R$. Slit-shells, by contrast, do not exhibit CCR or CO due to the absence of curvature and uniform field distribution.

\begin{figure}[!htb]
	\begin{subfigure}{.5\textwidth}
		\centering
		\includegraphics[width=0.99\linewidth]{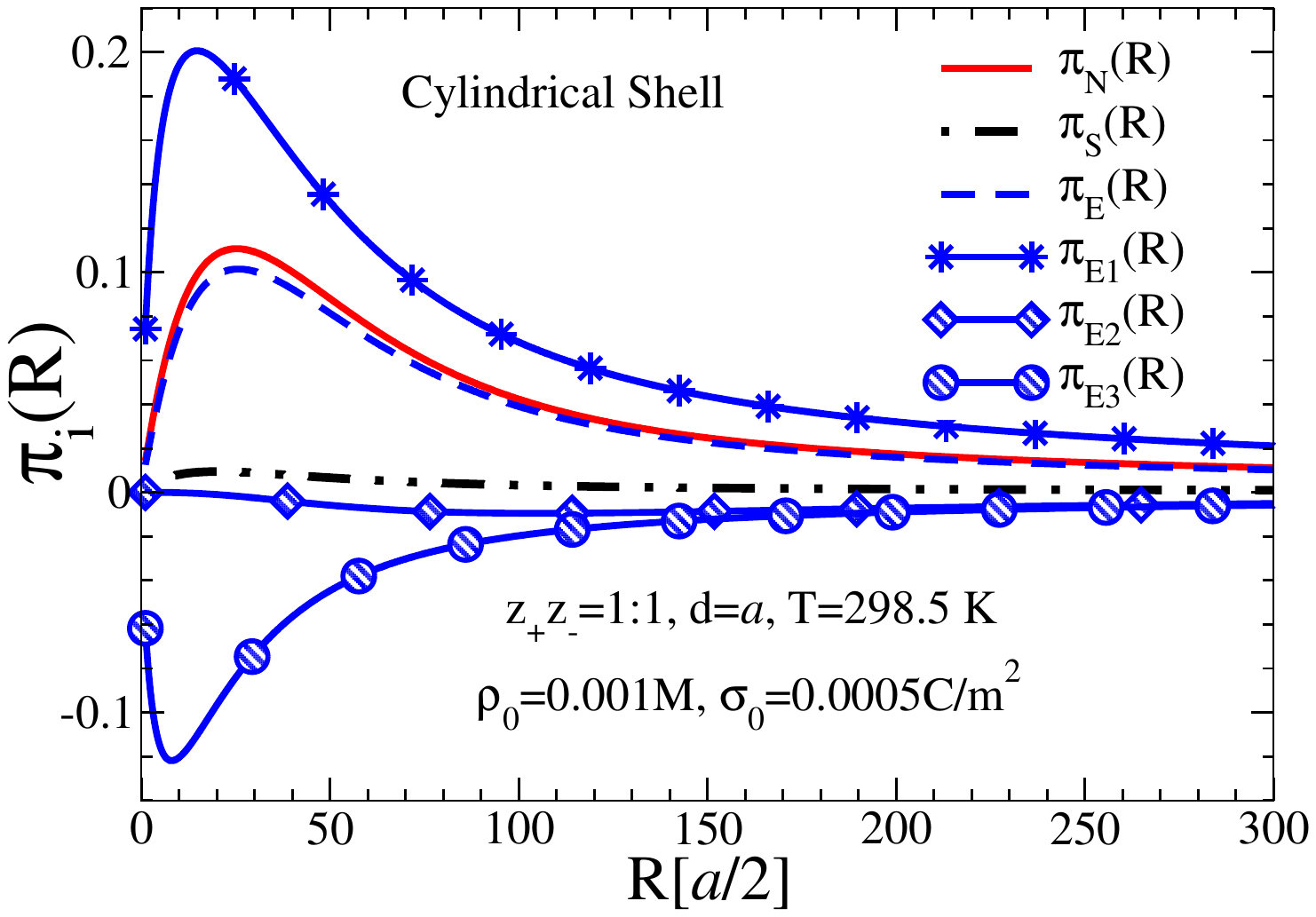}
		\caption{Cylindrical shell, low $\rho_{\scriptscriptstyle{0}}$ and $\sigma_{\scriptscriptstyle{0}}$.}
		\label{Fig.Cylinder.ROP.Components.Low.Rho.Low.Sigma}
	\end{subfigure}
	\begin{subfigure}{.5\textwidth}
		\centering
		\includegraphics[width=0.99\linewidth]{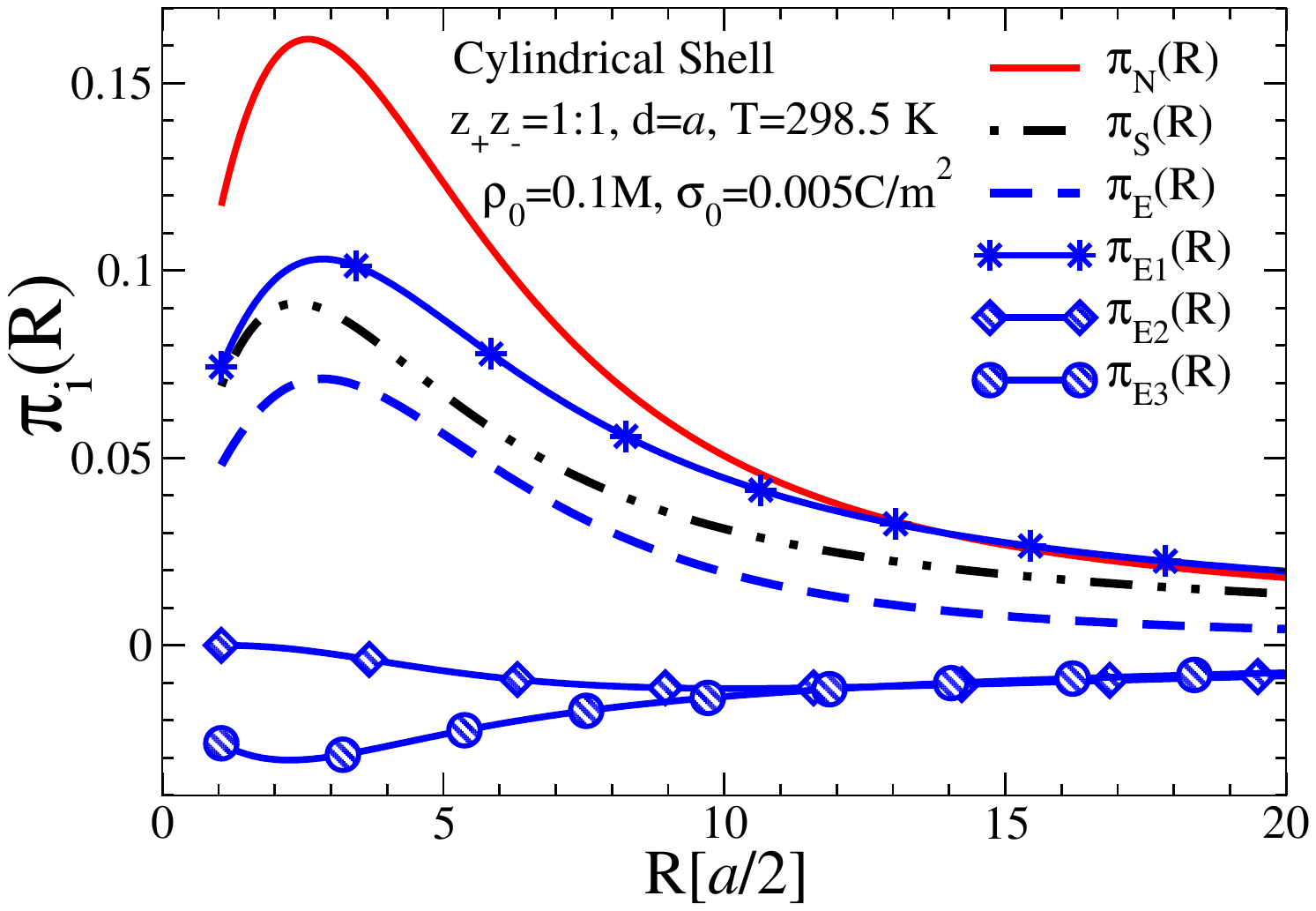}
		\caption{Cylindrical shell, high $\rho_{\scriptscriptstyle{0}}$ and $\sigma_{\scriptscriptstyle{0}}$.}
		\label{Fig.Cylinder.ROP.Components.Higer.Rho.Higher.Sigma}
	\end{subfigure}\\
	\begin{subfigure}{.5\textwidth}
		\centering
		\includegraphics[width=0.99\linewidth]{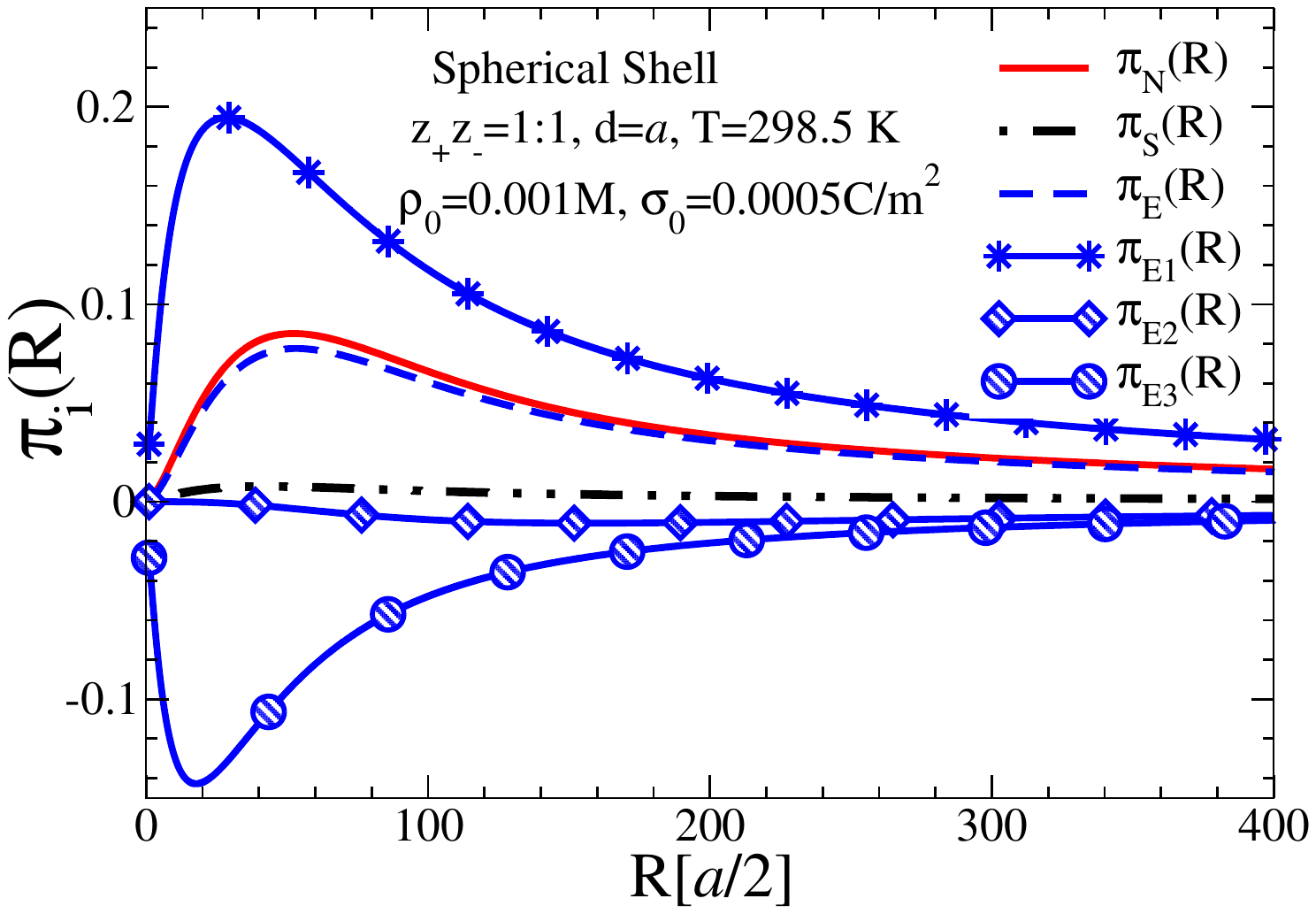}
		\caption{Spherical shell, low $\rho_{\scriptscriptstyle{0}}$ and $\sigma_{\scriptscriptstyle{0}}$.}
		\label{Fig.Sphere.ROP.Components.Low.Rho.Low.Sigma}
	\end{subfigure}
	\begin{subfigure}{.5\textwidth}
		\centering
		\includegraphics[width=0.99\linewidth]{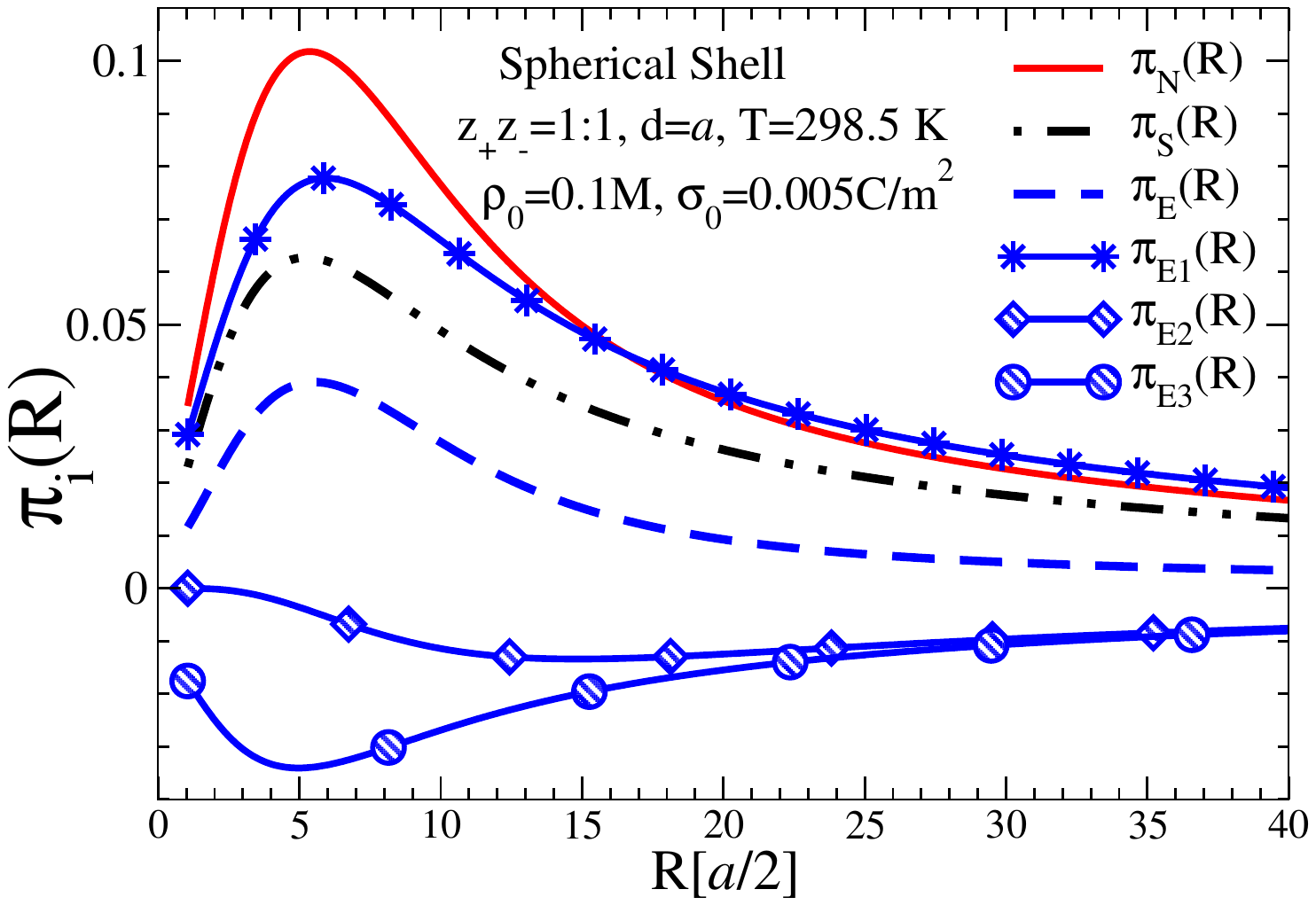}
		\caption{Spherical shell, high $\rho_{\scriptscriptstyle{0}}$ and $\sigma_{\scriptscriptstyle{0}}$.}
		\label{Fig.Sphere.ROP.Components.Higher.Rho.Higher.Sigma}
	\end{subfigure}\\
	\caption{Steric ($\pi_{\scriptscriptstyle{S}}(R)$) and electrostatic ($\pi_{\scriptscriptstyle{E}}(R)$) components of the reduced osmotic pressure $\pi_{\scriptscriptstyle{N}}(R)$ in cylindrical and spherical shells. The electrostatic term includes $\pi_{\scriptscriptstyle{E1}}$, $\pi_{\scriptscriptstyle{E2}}$, and $\pi_{\scriptscriptstyle{E3}}$ from the Maxwell stress tensor. Parameters: $\varepsilon = 78.5$, $a = \SI{4.25}{\angstrom}$.}
	\label{Fig.Cylinder.and.Sphere.ROP.Components}
\end{figure}

To summarize, \cref{Fig.Cylinder.and.Sphere.ROP.Components} decomposes $\pi_{\scriptscriptstyle{N}}(R)$ into entropic and electrostatic components for cylindrical and spherical shells, under two sets of conditions: low and high $\rho_{\scriptscriptstyle{0}}$ and $\sigma_{\scriptscriptstyle{0}}$.

In \cref{Fig.Cylinder.ROP.Components.Low.Rho.Low.Sigma}, for $(0.001\,\mathrm{M}, 0.0005\,\mathrm{C/m^2})$, a broad maximum appears at $R \approx 25.5a/2$. $\pi_{\scriptscriptstyle{E1}}(R)$ dominates but is partly canceled by $\pi_{\scriptscriptstyle{E2}}$ and $\pi_{\scriptscriptstyle{E3}}$. $\pi_{\scriptscriptstyle{S}}(R)$ is positive but modest, so the net pressure is primarily electrostatic.

In \cref{Fig.Cylinder.ROP.Components.Higer.Rho.Higher.Sigma}, for $(0.1\,\mathrm{M}, 0.005\,\mathrm{C/m^2})$, counter-ion adsorption is enhanced, decreasing $\pi_{\scriptscriptstyle{E1}}$ but greatly increasing $\pi_{\scriptscriptstyle{S}}$. The resulting pressure is $150\%$ higher, with a maximum at $R \approx 2.6a/2$, reflecting denser EDLs.

In \cref{Fig.Sphere.ROP.Components.Low.Rho.Low.Sigma}, the spherical shell with low $\rho_{\scriptscriptstyle{0}}$ and $\sigma_{\scriptscriptstyle{0}}$ exhibits a maximum at $R \approx 52a/2$, approximately double the cylindrical case, and a broader pressure profile.

Finally, \cref{Fig.Sphere.ROP.Components.Higher.Rho.Higher.Sigma} shows the high-density spherical case, where $\pi_{\scriptscriptstyle{N}}(R)$ is $120\%$ higher and peaks at $R \approx 5.4a/2$. Both $\pi_{\scriptscriptstyle{S}}(R)$ and $\pi_{\scriptscriptstyle{E}}(R)$ contribute substantially, reflecting strong confinement.

These results indicate that osmotic pressure is dominated by electrostatics in weakly confined systems and by entropic effects in highly confined, strongly charged systems.

\begin{figure}[!htb]
	\begin{subfigure}{.5\textwidth}
		\centering
		\includegraphics[width=0.99\linewidth]{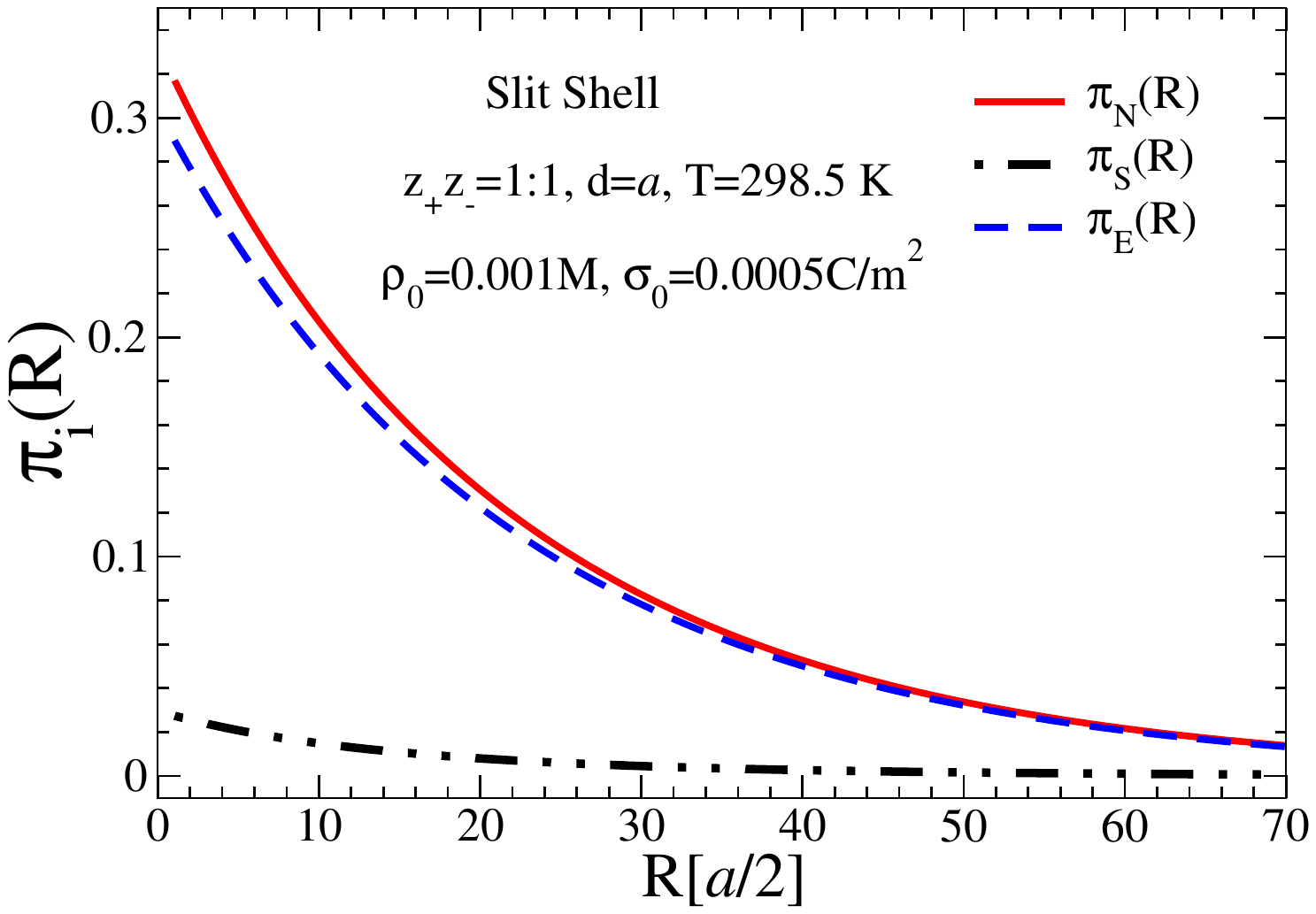}
		\caption{Slit-shell, low $\rho_{\scriptscriptstyle{0}}$ and $\sigma_{\scriptscriptstyle{0}}$.}
		\label{Fig.Slit.ROP.Components.Lowr.Rho.Low.Sigma}
	\end{subfigure}
	\begin{subfigure}{.5\textwidth}
		\centering
		\includegraphics[width=0.99\linewidth]{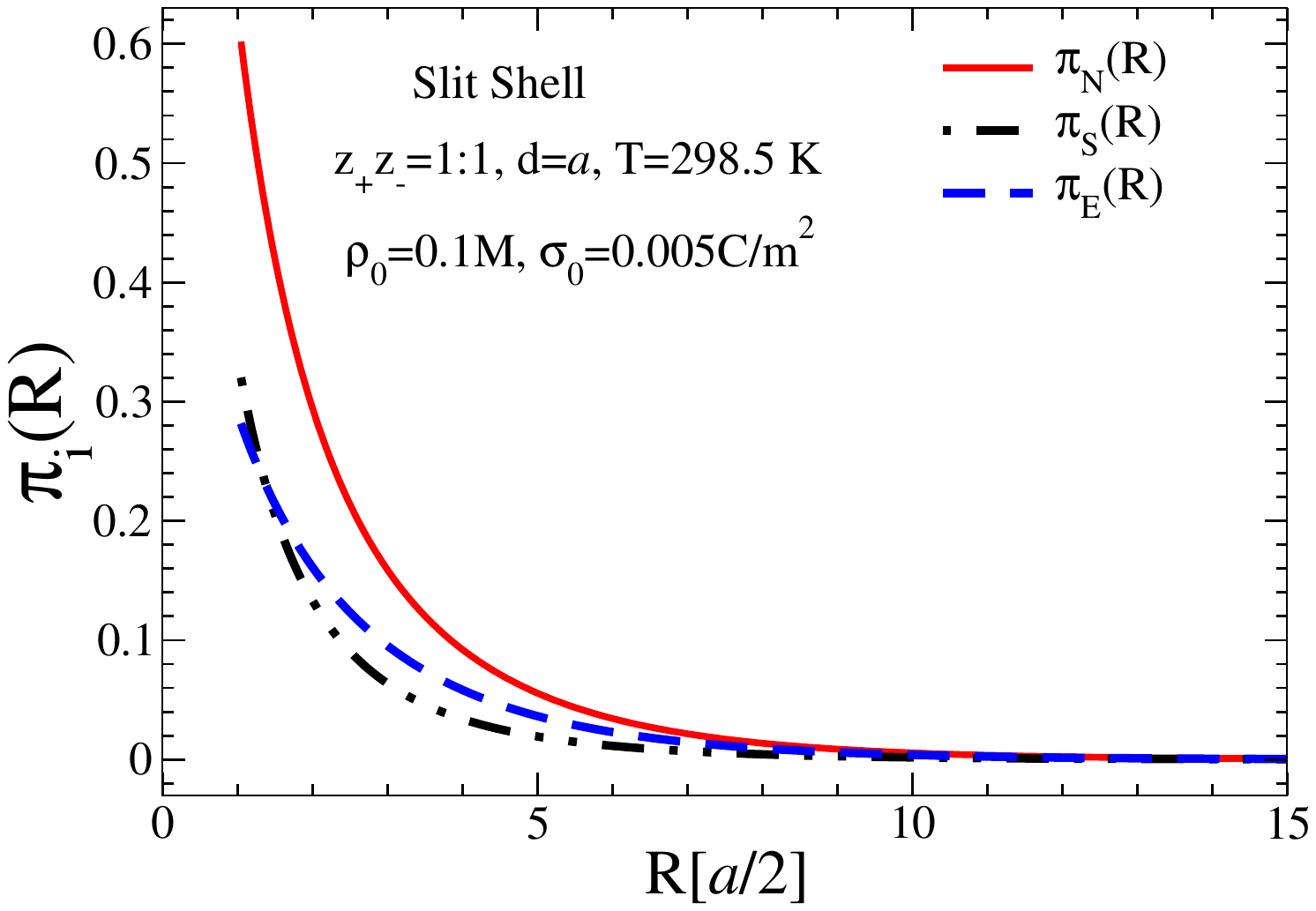}
		\caption{Slit-shell, higher $\rho_{\scriptscriptstyle{0}}$ and $\sigma_{\scriptscriptstyle{0}}$.}
		\label{Fig.Slit.ROP.Components.Higher.Rho.Higher.Sigma}
	\end{subfigure}
	\caption{Steric ($\pi_{\scriptscriptstyle{S}}$) and electrostatic ($\pi_{\scriptscriptstyle{E}}$) components of the net reduced osmotic pressure $\pi_{\scriptscriptstyle{N}}(R)$ for slit-shells. In this geometry, $\pi_{\scriptscriptstyle{E}} = \pi_{\scriptscriptstyle{E1}}$ and $\pi_{\scriptscriptstyle{E2}} = \pi_{\scriptscriptstyle{E3}} = 0$. Parameters: $\varepsilon = 78.5$, $a = \SI{4.25}{\angstrom}$.}
	\label{Fig.Slit.ROP.Components}
\end{figure}

In \cref{Fig.Slit.ROP.Components}, we examine the reduced osmotic pressure components $\pi_{\scriptscriptstyle{S}}(R)$ and $\pi_{\scriptscriptstyle{E}}(R)$ for slit-shell geometries, using two representative sets of $\rho_{\scriptscriptstyle{0}}$ and $\sigma_{\scriptscriptstyle{0}}$. In all cases, $\pi_{\scriptscriptstyle{N}}(R)$ decreases monotonically with increasing $R$ and vanishes as $R \rightarrow \infty$, reflecting the symmetric EDL structures inside and outside the slit-shell.

Although the Restricted Primitive Model (RPM) predicts non-monotonic or oscillatory $\pi_{\scriptscriptstyle{N}}(R)$ in narrow shells with high surface charge and ionic strength~\cite{Lozada_1990-I,Lozada_1990-II,Aguilar_2007}, within the parameter ranges considered here, the Poisson–Boltzmann (PB) and integral equation methods yield both qualitatively and quantitatively consistent results~\cite{Lozada_1990-I}.

In the slit-shell geometry, the electrostatic contribution is given solely by $\pi_{\scriptscriptstyle{E}}(R) = \pi_{\scriptscriptstyle{E1}}(R)$, since $\pi_{\scriptscriptstyle{E2}} = \pi_{\scriptscriptstyle{E3}} = 0$ (see \cref{Osmotic-plates-Net}). This term depends only on the difference $\sigma_{\scriptscriptstyle{Hi}}^2 - \sigma_{\scriptscriptstyle{Ho}}^2$. At low $\rho_{\scriptscriptstyle{0}}$ and $\sigma_{\scriptscriptstyle{0}}$, $\pi_{\scriptscriptstyle{E}}(R)$ dominates the osmotic pressure (\cref{Fig.Slit.ROP.Components.Lowr.Rho.Low.Sigma}). At higher charge and concentration, enhanced counter-ion adsorption increases $\pi_{\scriptscriptstyle{S}}(R)$, making it comparable to $\pi_{\scriptscriptstyle{E}}(R)$ (\cref{Fig.Slit.ROP.Components.Higher.Rho.Higher.Sigma}). Meanwhile, $\pi_{\scriptscriptstyle{E}}(R)$ decreases due to a more negative $\sigma_{\scriptscriptstyle{Hi}}$.

\

\subsection{The electrical field}\label{The electrical field}

\begin{figure}[!htb]
	\begin{subfigure}{.5\textwidth}
		\centering
		\includegraphics[width=0.99\linewidth]{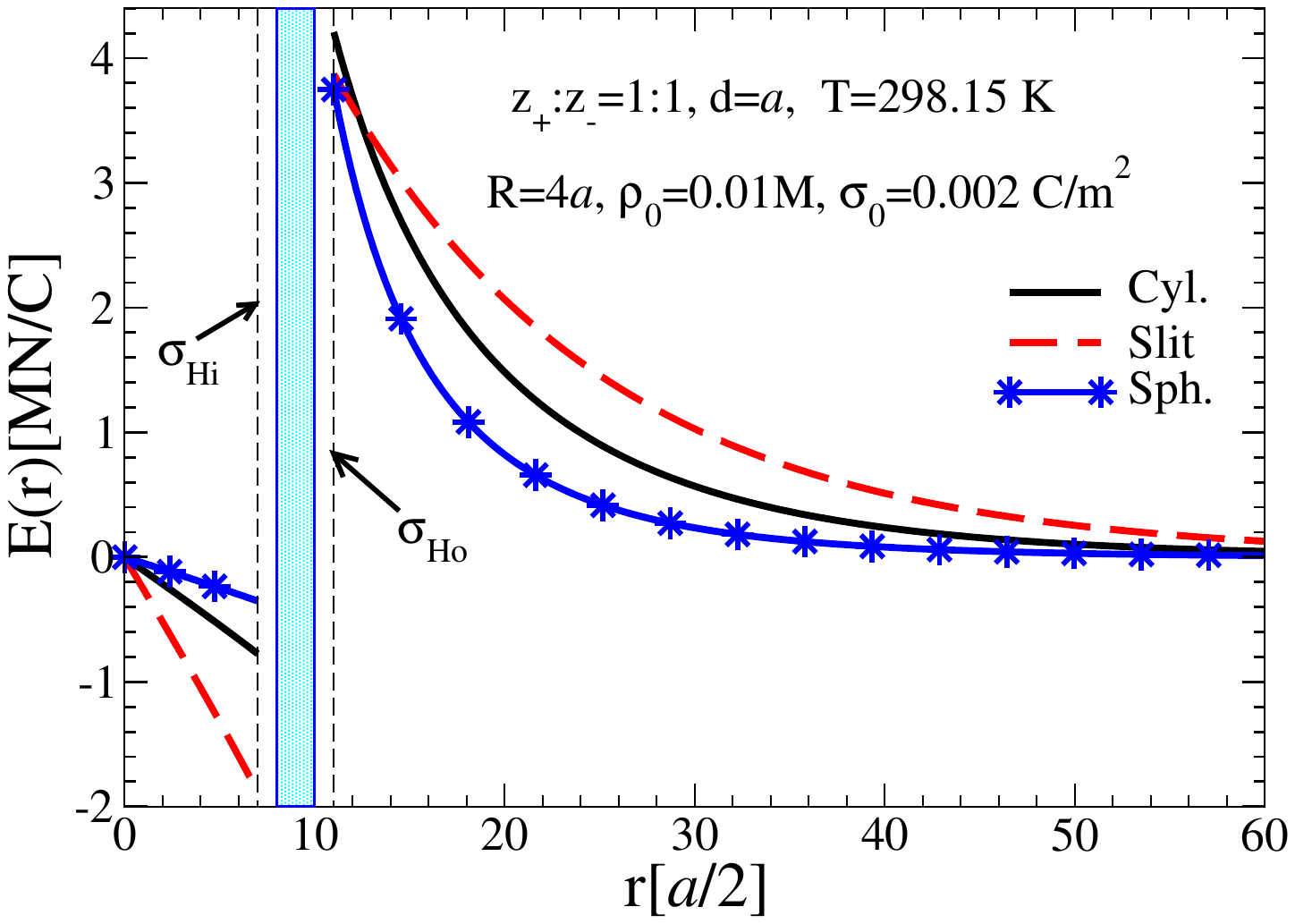}
		\caption{Electric field profile for a narrow shell, $R=4a$.}
		\label{Fig.PCS_E[r]_a4p25_R4p0_z1_d1p0_T298p0_s0p002_rho0p01}
	\end{subfigure}
	\begin{subfigure}{.5\textwidth}
		\centering
		\includegraphics[width=0.99\linewidth]{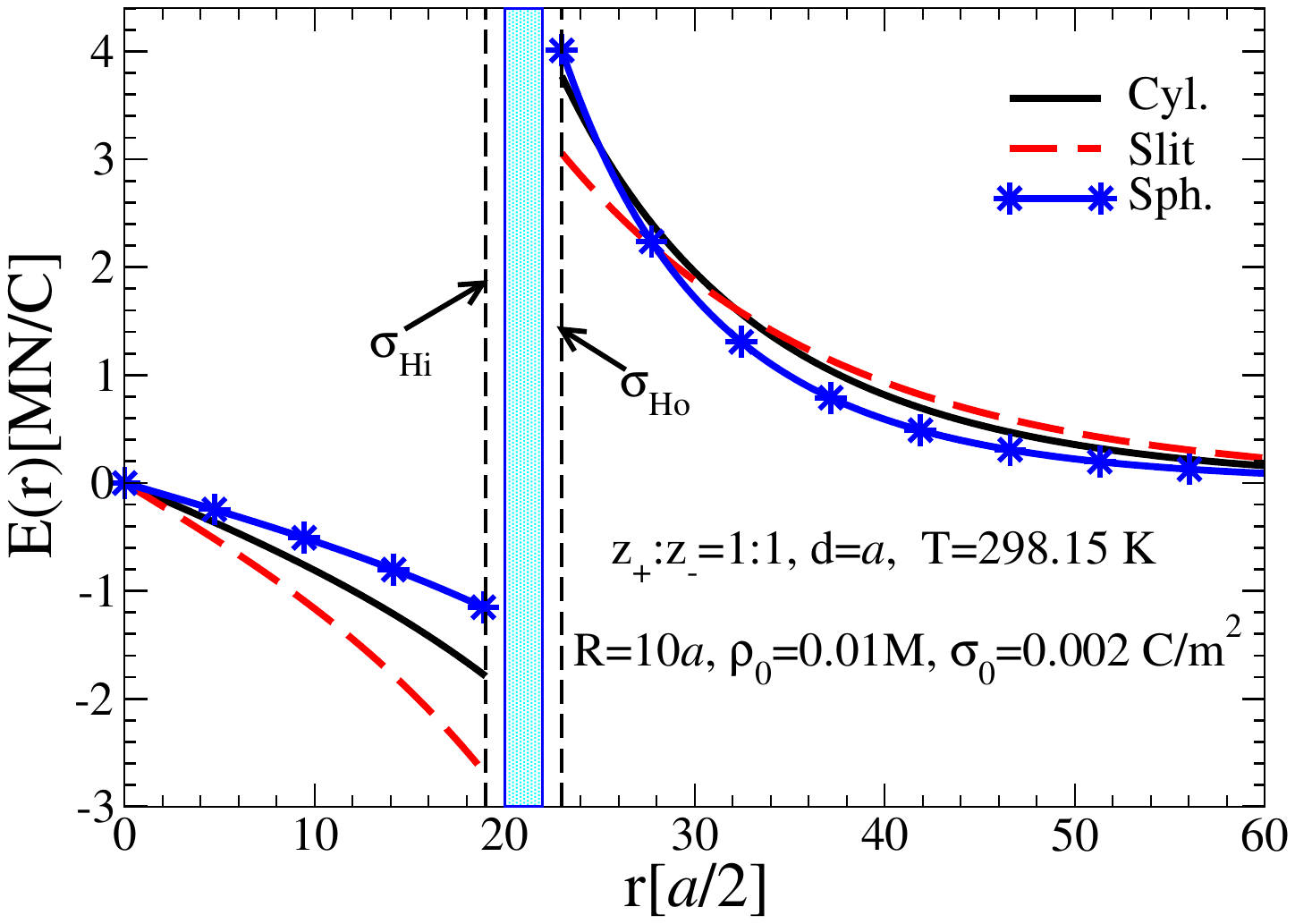}
		\caption{Electric field profile for a wider shell, $R=10a$.}
		\label{Fig.PCS_E[r]_a4p25_R10p0_z1_d1p0_T298p0_s0p002_rho0p01}
	\end{subfigure}
	\caption{Electric field $E(r)$ for planar, cylindrical, and spherical shells, for (a) $R=4a$ and (b) $R=10a$. Shell thickness is $d = a$, and vertical dashed lines mark the ionic closest approach at $r = R - a/2$ and $r = R + d + a/2$. The dielectric constant is $\varepsilon = 78.5$, and $a = \SI{4.25}{\angstrom}$.}
	\label{Fig.PCS_E[r]}
\end{figure}

\Cref{Fig.PCS_E[r]} shows the electric field profiles $E(r)$ for slit, cylindrical, and spherical shells at $\rho_{\scriptscriptstyle{0}} = 0.01\,\mathrm{M}$ and $\sigma_{\scriptscriptstyle{0}} = 0.002\,\mathrm{C/m^2}$, with shell radii $R = 4a$ and $10a$.

Inside the shells, $E(r)$ decreases monotonically, reaching $E(R-a/2) = \sigma_{\scriptscriptstyle{Hi}} / (\epsilon_{\scriptscriptstyle{0}}\epsilon)$, as expected from electrostatics (see \cref{Induced-charge-density-in}). For both radii, the field is strongest in the slit-shell, followed by the cylindrical, and weakest in the spherical shell. This ordering is purely geometric: for fixed $\sigma_{\scriptscriptstyle{0}}$, the effective area is smallest in the planar case and largest in the spherical one. As $R \rightarrow \infty$, all geometries approach the same limiting value $\sigma_{\scriptscriptstyle{0}}$ (see \cref{Induced-charge-density-in}).

Outside the shells, $E(r)$ decays fastest for the spherical shell due to curvature. Notably, a stronger internal field does not always correspond to a weaker external one. In \cref{Fig.PCS_E[r]_a4p25_R4p0_z1_d1p0_T298p0_s0p002_rho0p01}, the cylindrical shell shows a stronger $E(R+d+a/2)$ than the spherical shell, but in \cref{Fig.PCS_E[r]_a4p25_R10p0_z1_d1p0_T298p0_s0p002_rho0p01} this reverses. These differences stem from the nonlinear $r$-dependence of $E(r)$ and the variation in total surface charge $Q_{\scriptscriptstyle{0}} = A_{\scriptscriptstyle{\gamma}} \sigma_{\scriptscriptstyle{0}}$ among geometries.

As seen in \cref{Fig.Cylinder.and.Sphere.ROP.Components,Fig.Slit.ROP.Components}, the term $\pi_{\scriptscriptstyle{E1}}(R)$—proportional to $E^2(R-a/2) - E^2(R+d+a/2)$—dominates the electrostatic contribution to the net osmotic pressure. As $R$ increases, $E(R-a/2)$ becomes more negative and can exceed $-\sigma_{\scriptscriptstyle{0}}$, violating the local electroneutrality condition~\cite{Lozada_1984,Lozada1996,Lozada-Cassou-PRL1996,Levin_electroneutrality-2016,Levy-electroneutrality-2020,malgaretti_electroneutrality-2024}.

In slit-shells, $E(R-a/2)$ and $E(R+d+a/2)$ change monotonically due to constant wall charge and flat geometry. Accordingly, $\pi_{\scriptscriptstyle{E1}}(R)$, and hence $\pi_{\scriptscriptstyle{E}}(R)$, also decreases monotonically with $R$, as seen in \cref{Fig.ROP.Surface.Charge,Fig.ROP.System.Temperature,Fig.Slit.ROP.Components}.

\begin{figure}[!htb]
	\begin{subfigure}{.5\textwidth}
		\centering
		\includegraphics[width=0.99\linewidth]{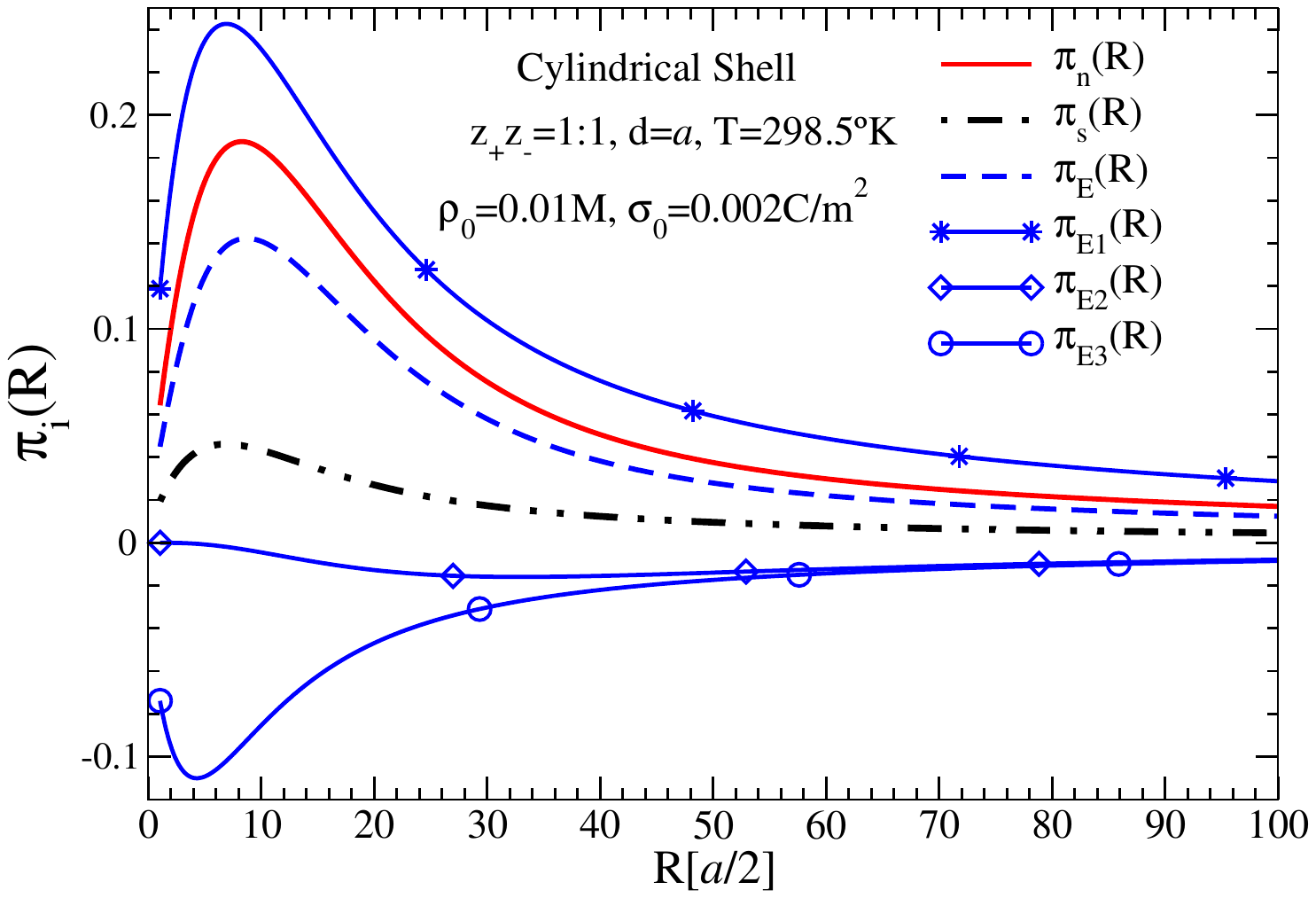}
		\caption{Cylindrical shell's osmotic pressure components.}
		\label{cyl-PT-PS-PE-z1_d1_T298.15_s0.002_rho0.01}
	\end{subfigure}
	\begin{subfigure}{.5\textwidth}
		\centering
		\includegraphics[width=0.99\linewidth]{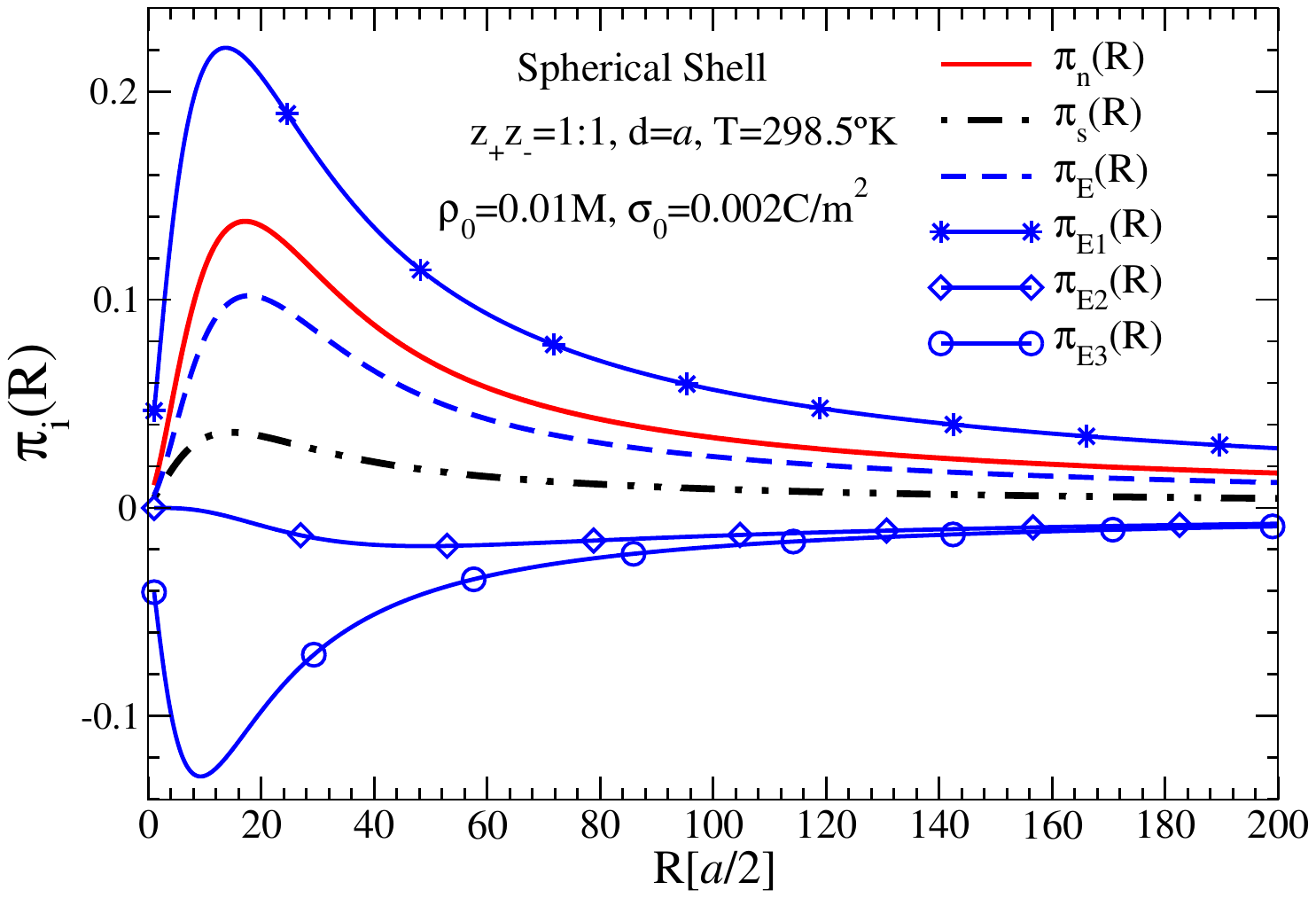}
		\caption{Spherical shell's osmotic pressure components.}
		\label{Sph-PT-PS-PE-z1_d1_T298.15_s0.002_rho0.01}
	\end{subfigure}
	\caption{Steric ($\pi_{\scriptscriptstyle{S}}$) and electrostatic ($\pi_{\scriptscriptstyle{E}}$) components of the reduced osmotic pressure, $\pi_{\scriptscriptstyle{N}}(R)$, in cylindrical and spherical shells. The electrostatic contribution is decomposed into $\pi_{\scriptscriptstyle{E1}}, \pi_{\scriptscriptstyle{E2}}$, and $\pi_{\scriptscriptstyle{E3}}$.}
	\label{Fig.Cyl-Sph-PT-PS-PE-z1_d1_T298.15_s0.002_rho0.01}
\end{figure}

By contrast, for cylindrical and spherical shells, $E(R_{\scriptscriptstyle{H}})$ first increases with $R$ before decreasing, reflecting the non-monotonic behavior of $\sigma_{\scriptscriptstyle{Ho}}(R)$ (see \cref{Fig.sigma-Hi_and_Ho}). As $R \rightarrow \infty$, the electroneutrality condition is restored: $\sigma_{\scriptscriptstyle{Hi}} \rightarrow -\sigma_{\scriptscriptstyle{0}}$, and $\sigma_{\scriptscriptstyle{Ho}} \rightarrow \sigma_{\scriptscriptstyle{0}}$ (see \cref{Electroneutrality-condition-general2}). This explains the non-monotonic profiles of $\pi_{\scriptscriptstyle{E1}}(R)$ and $\pi_{\scriptscriptstyle{E}}(R)$ in \cref{Fig.Cylinder.and.Sphere.ROP.Components,Osmotic-Press-E1}.

For example, in \cref{Fig.PCS_E[r]_a4p25_R4p0_z1_d1p0_T298p0_s0p002_rho0p01,Fig.PCS_E[r]_a4p25_R10p0_z1_d1p0_T298p0_s0p002_rho0p01}, $E(R-a/2)$ roughly doubles in magnitude as $R$ increases from $4a$ to $10a$. Meanwhile, $E(R_{\scriptscriptstyle{H}})$ increases slightly for spherical shells and decreases for cylindrical ones. As a result, $\pi_{\scriptscriptstyle{E}}(R)$ increases in the spherical case and decreases in the cylindrical case—reflected in both $\pi_{\scriptscriptstyle{S}}(R)$ and total $\pi_{\scriptscriptstyle{N}}(R)$ in \cref{Fig.Cyl-Sph-PT-PS-PE-z1_d1_T298.15_s0.002_rho0.01}. In slit geometries, all these quantities decay monotonically with $R$ due to the flat geometry and fixed wall charge.

\section{Conclusions}\label{Conclusions}

We have derived and applied contact theorems to compute the osmotic pressure in planar, cylindrical, and spherical shell geometries. Each shell is characterized by a constant, low surface charge density and immersed in a dilute symmetric electrolyte. Under these conditions, the mean electrostatic potential remains small, allowing for analytical treatment via the linearized Poisson–Boltzmann (LPB) equation. This framework yields explicit expressions for the electric double layer (EDL), electric field, and surface charge distributions inside and outside the shells, which are key to evaluating the osmotic pressure.

Our analysis reveals a strong nonlinear dependence of the osmotic pressure on the shell radius in cylindrical and spherical geometries. This behavior results from the combined effects of confinement geometry and local violations of electroneutrality. In contrast, for slit-shells, the osmotic pressure decays monotonically with radius. At small radii, slit-shells exhibit the highest osmotic pressures, followed by cylindrical and spherical shells. However, the decay of the pressure extends furthest in spherical shells, and shortest in slit-shells.

The osmotic pressure is highly sensitive to surface charge density, shell wall thickness, and electrolyte properties such as concentration and ion valence. Temperature, by contrast, plays a minor role. Thicker walls generally enhance confinement and result in more intense pressures for small shell sizes. These trends are especially relevant for biological systems operating in low-salt environments.

For example, in living cells, membranes typically restrict ion passage, creating concentration differences between the intracellular and extracellular environments~\cite{Tuszynski-Introduction-Molecular-Biophysics-book-2003}. Our model could be extended to explore such asymmetries, or cases involving charged macromolecules and other nonuniform boundary conditions.

Although we focus here on point-ion electrolytes and linear PB theory, future work will address more complex systems—including finite-sized ions, steric effects, and ion–ion correlations—using beyond-PB methods. In earlier studies, we analyzed macroion adsorption in nanocavities and next to charged interfaces~\cite{gonzalez-Advances-2019,Gonzalez-overcharging-cyl-macroions-2022,Gonzalez-Calderon-EPJ-2021,odriozola-confined-macroions-2009,Odriozola-Fortschritte-der-Physik-2017}, but osmotic pressure was not considered. We intend to bridge that gap in future research.

\vspace{0.5em}

\noindent
We have also introduced two new confinement-induced phenomena: \textit{Confinement Charge Reversal} (CCR) and \textit{Confinement Overcharging} (CO). Both effects originate from confinement energy. Entropy in our model appears only through the ideal-gas contribution and a finite-thickness Stern layer. As ions are treated as point charges, the Stern correction provides the sole mechanism for configurational entropy. Thinner Stern layers amplify CCR and CO, whereas thicker layers suppress them. Both phenomena vanish as ion size increases, highlighting their energetic (rather than entropic) nature.

Importantly, CCR and CO are not limited to the shell geometries explored here. They may also arise in fully enclosed bubble cavities, open-ended tubes, or general shell-like structures—provided confinement exists and the wall has finite width. Notably, such effects have not previously been reported for point-ion systems governed by mean-field electrostatics~\cite{Kjellander-charge-inversion-1998}. Our results thus point to a novel class of confinement-driven charge phenomena that arise independently of finite ion size or ion–ion correlations.

Our results provide a robust theoretical foundation for understanding confinement effects in nanocavities, with potential applications in nanofluidics, energy storage, and biophysics.


\section*{Acknowledgment}

The support of UNAM (PAPIIT Clave: IN108023) is acknowledged.





\end{document}